%% file: trex_overview.tex
\documentclass[preprint,10pt,trackchanges]{aastex631} 

\usepackage{grffile}  
\usepackage{url}
\let\tablenum\relax
\usepackage{siunitx}
\newcommand{\anchorfoot}[2] {\href{#1}{#2}\footnote{\url{#1}}}
\newcommand{\anchorparen}[2]{\href{#1}{#2} (\url{#1})}

\newcommand{\tablenotetextbox}[3]{\vspace{\baselineskip}#1\hspace{2ex}\parbox[t]{#2}{#3}}
\renewcommand{\S}{Section }
\newcommand{\tnm}[1]{\tablenotemark{#1}}  

\newcommand{\GHIIR}{GH{\scriptsize II}R }

\newcommand{\Spitzer} {{\em Spitzer~}}

\newcommand{\Chandra} {{\em Chandra}}
\newcommand{\ACIS}    {{ACIS}}
\newcommand{\CIAO}    {{\em CIAO}}

\newcommand{\DSnine}  {{\em DS9}}
\newcommand{\MARX}    {{\em MARX}}
\newcommand{\AEacro}  {{\em AE}}

\newcommand{\TOPCAT}    {{\em TOPCAT}}
\newcommand{\XSPEC}   {{\em XSPEC}}

\received{March 21, 2024}
\submitjournal{ApJS}
\accepted{15 April 2024}

\shorttitle{T-ReX}
\shortauthors{Townsley et al.} 

\begin{document}

\title{The Tarantula--Revealed by X-rays (T-ReX)} 

\author[0000-0001-8081-9152]{Leisa K. Townsley}
\altaffiliation{Deceased. This paper is part of a collection publishing posthumously the unfinished work of Leisa K. Townsley. Her coauthors completed the manuscript based on her initial draft, written notes and private communications.}
\affiliation{Department of Astronomy \& Astrophysics, 525 Davey Laboratory, 
Pennsylvania State University, University Park, PA 16802 USA}

\author[0000-0002-7872-2025]{Patrick S. Broos}
\affiliation{Department of Astronomy \& Astrophysics, 525 Davey Laboratory, 
Pennsylvania State University, University Park, PA 16802 USA}

\author[0000-0001-9062-3583]{Matthew S. Povich}
\affiliation{Department of Physics \& Astronomy, California State Polytechnic University, 3801 West Temple Ave, Pomona, CA 91768  USA}

\correspondingauthor{Matthew S. Povich}
\email{mspovich@cpp.edu}

\begin{abstract}

The Tarantula Nebula (30 Doradus) is the most important star-forming complex in the Local Group, offering a microscope on starburst astrophysics.  At its heart lies the exceptionally rich young stellar cluster R136, containing the most massive stars known.  Stellar winds and supernovae have carved 30 Dor into an amazing display of arcs, pillars, and bubbles.  We present first results and advanced data processing products from the 2-Ms \Chandra\ X-ray Visionary Project, ``The Tarantula -- Revealed by X-rays'' (T-ReX). The 3615 point sources in the T-ReX catalog include massive stars, compact objects, binaries, bright pre-main-sequence stars and compact young stellar (sub)clusters in 30 Dor. After removing point sources and excluding the exceptionally bright supernova remnant N157B (30 Dor B), the global diffuse X-ray maps reveal hot plasma structures resolved at 1--10~pc scales, with an absorption-corrected total-band (0.5--7~keV) X-ray luminosity of $2.110\times 10^{37}$~erg~s$^{-1}$.
Spatially-resolved spectral modeling provides evidence for emission lines enhanced by charge-exchange processes at the interfaces. We identify a candidate for the oldest X-ray pulsar detected to date in 30 Dor, PSR J0538-6902, inside a newly-resolved arctuate X-ray wind nebula, the Manta Ray. 
The long time baseline of T-ReX monitored dozens of massive stars, several showing periodic variability tied to binary orbital periods, and captured strong flares from at least three low-mass Galactic foreground stars.


\end{abstract}


\vspace{1cm}

\section{30~Doradus:  A Field of Superlatives \label{sec:30Dor}}
The Tarantula Nebula (30~Doradus, or 30~Dor), located in the Large Magellanic Cloud (LMC), is the most powerful massive star-forming region (MSFR) in the Local Group of galaxies.  30~Dor is unlike any Galactic MSFR;  with no differential galactic rotation to rip the complex apart, it persists and grows at the confluence of two supergiant shells \citep{Meaburn80} that have provided it with the fuel to power massive star formation for at least 25~Myr \citep{HTTP2}, reaching starburst proportions.  Today it is dominated by its 1--2~Myr-old central massive cluster R136, containing the richest young stellar population in the Local Group, including the most massive stars known.  Also in contrast to Galactic MSFRs, the location of 30 Dor in the LMC offers us a nearly face-on, low-metallicity starburst laboratory with low intervening absorption and a well-known distance.  
\citet{LMCdist19} refined the distance to the LMC to 49.5~kpc, but we assume a 50~kpc distance to 30 Dor to facilitate comparison with other recent studies \citep[e.g.,][]{Sabbi13,Cheng21,Chen23-30DorB}.  At this distance, the spatial scale is 0.24~pc/arcsec. 

 R136 produces ${\sim}1000$ times the ionizing radiation of Orion \citep{Conti12} from at least five times as many early O-type stars as Carina \citep{Evans11}.  It contains several of the most massive stars known \citep{Crowther10} and the full complement of massive binaries, including WR--WR, WR--O, and O--O systems \citep{CrowtherDessart98}.  In the wider 30~Dor field, the presence of $\sim$25~Myr-old evolved supergiants and embedded massive young stellar objects shows that 30~Dor is the product of multiple epochs of star formation \citep{WalbornBlades97,Whitney08}; the newest generation of embedded stars may represent triggered collapse from the effects of R136 \citep{Walborn13}.

We present the 2~Ms \Chandra\ X-ray Visionary Project, ``The Tarantula---Revealed by X-rays'' (T-ReX), an intensive imaging study of 30~Dor with the Advanced CCD Imaging Spectrometer (ACIS) instrument \citep{Garmire03} of the {\it Chandra X-ray Observatory}.  Before T-ReX, \Chandra\ had invested just 114~ks in this iconic target, revealing only the most massive stars and large-scale diffuse structures \citep{Townsley06a,Townsley06b,Townsley14}.  T-ReX leverages the unparalleled X-ray spatial resolution of \Chandra, a unique resource that will remain unmatched for another decade, revealing the X-ray properties of dozens of low-metallicity massive stars \citep[see][hereafter C22]{Crowther22}
and parsec-scale shocks from winds and supernovae that are shredding its interstellar medium \citep[ISM; the main science focus of this contribution, but see also][]{Cheng21,Chen23-30DorB}.  
This definitive \Chandra\ characterization of 30~Dor was motivated by the fundamental need to understand the complex machinery of a starburst.  Massive stars, their remnants, and the lower-mass populations that accompany them exert a profound influence on dense molecular clouds through radiation and mechanical feedback, which both perpetuates and quenches star formation.  T-ReX unveils the detailed spatial and spectral morphology of hot plasmas permeating the complex cold ISM structures, ionization fronts, and photon-dominated regions seen in optical and infrared (IR) images. 

This paper is organized as follows:
Section~\ref{sec:data} describes the T-ReX observations and data processing,  culminating in the publicly-available point-source catalog and other advanced data products. High-level descriptions and interpretation of the T-ReX images and point-source populations are presented in Sections~\ref{sec:images} and \ref{sec:ptsrcs}, respectively.  Section~\ref{sec:diffuse} presents our deeper dive into spatially-resolved spectral fitting analysis of the T-ReX diffuse emission and its physical properties. A brief summary of the anticipated legacy value of T-ReX is included in Section~\ref{sec:legacy}. Finally, two Appendices present more detailed analysis of select point-sources: Appendix~\ref{app:counterparts} details the challenges of matching T-ReX point-sources to optical/IR source catalogs and presents tables of high-reliability matching results. Appendix~\ref{app:variability} presents light-curves and basic properties of notable X-ray variable sources, including massive stellar systems with periodic variability and low-mass stars exhibiting strong X-ray flares.

\section{{\em Chandra} Observations and Data Analysis \label{sec:data}}

\subsection{Observations}

\begin{figure}[htb]
\centering
\includegraphics[width=0.49\textwidth]{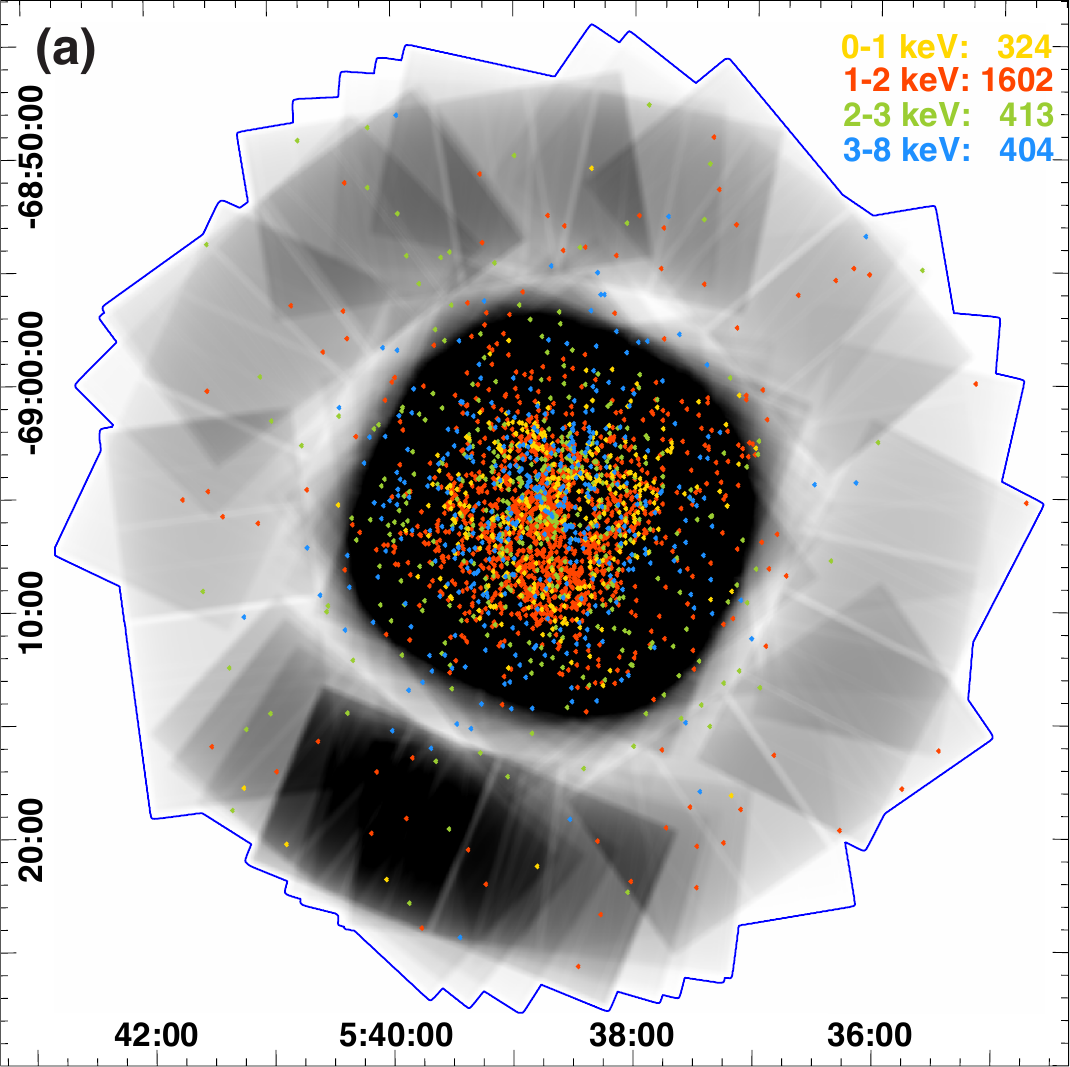}
\includegraphics[width=0.49\textwidth]{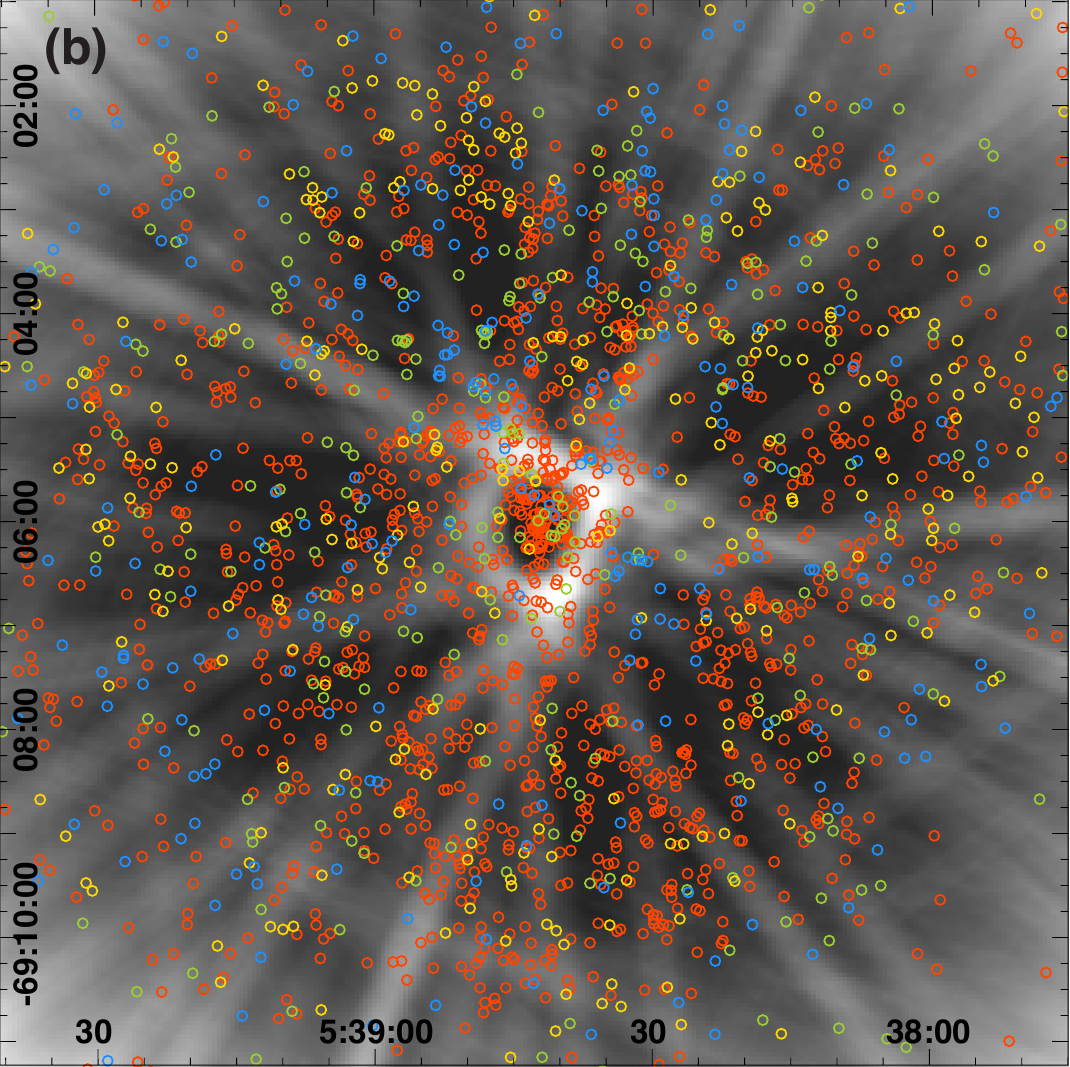}
\caption{ACIS exposure map with 2743 brighter ($\geq$5 net counts) ACIS point sources overlaid; colors denote median energy for each source.
(a) The full T-ReX field.
(b) A zoom of (a) showing the central $\sim$10$\arcmin$ of the ACIS field.
\label{emap+srcs.fig}}
\end{figure}

T-ReX obtained 51 \Chandra\ imaging observations, using the 4 ACIS-I detectors plus the two central ACIS-S detectors, with an aimpoint near R136 over the period 2014 May 3 through 2016 January 22 (Table~\ref{tbl:obslog}).
The extreme ecliptic latitude ($-87\arcdeg$) of 30~Dor limited T-ReX to 2~Ms due to spacecraft thermal constraints. Including three archival observations of 30 Dor from 2006 January brings the total exposure time to 2.05~Ms. The exposure map for all 54 observations,  with detected point sources overlaid (see Section~\ref{sec:catalog}), is presented in Figure~\ref{emap+srcs.fig}. Due to the various, arbitrary roll angles among the observations, the final exposure map resembles a daisy flower, with the two S-array detectors arranged like petals around a central disk created by rotating the four ACIS-I detectors.
Both the greatest exposure time and on-axis point-source sensitivity are hence obtained within the central portions of ACIS-I (Figure~\ref{emap+srcs.fig}b).

\input{observing_log.tex}

\clearpage
\subsection{Data Analysis}

\subsubsection{Methods, software, and procedures}

The algorithms, procedures, and tools  used for data reduction, astrometric alignment of observations, point source detection and validation, point source extraction, point source masking, construction of smoothed diffuse emission images, diffuse region extraction, and point source spectral fitting were the same as those used in the second and third installments of our Massive star-forming regions Omnibus X-ray Catalog \citep[MOXC2 and MOXC3,][]{Townsley18,Townsley19}.  Please see those papers (and their references) for explanations of our data analysis methods.
Even though all T-ReX observations have similar aimpoints, we followed our usual practice of extracting point sources and diffuse regions from each observation, and then combining those extractions and calibration products.

We employ a variety of standard and custom X-ray analysis software packages in our workflows.
\ACIS\ data are manipulated and calibrated using the \Chandra\ data analysis system, \anchorfoot{http://cxc.harvard.edu/ciao/}{\CIAO} \citep{Fruscione06}, and the \anchorfoot{http://space.mit.edu/ASC/marx}{\MARX} observatory simulator \citep{Davis12}.
Throughout our workflow, data products are visually reviewed using the \anchorfoot{http://ds9.si.edu}{{\it SAOImage DS9}} tool \citep{Joye03} and \anchorfoot{http://www.star.bris.ac.uk/~mbt/topcat/}  {\TOPCAT} \citep{Taylor05}.
Models are fit to extracted spectra using the \anchorfoot{https://heasarc.gsfc.nasa.gov/xanadu/xspec/}{\XSPEC} tool \citep{Arnaud96}.
The algorithms and scripts driving these standard tools are implemented in the  \href{http://personal.psu.edu/psb6/TARA/ae_users_guide.html}{{\em ACIS Extract}} \citep[\AEacro,][]{Broos10,AE12,AE16} software package\footnote{
The {\em ACIS Extract} software package and User's Guide are publicly available from the Astrophysics Source Code Library, from Zenodo, and at \url{http://personal.psu.edu/psb6/TARA/ae_users_guide.html}.
}
and in a set of detailed procedures we developed.
\AEacro\ and most other custom software we use are written in the \anchorfoot{www.harrisgeospatial.com/ProductsandTechnology/Software/IDL.aspx}{{\it Interactive Data Language}} (IDL).

\subsubsection{Alignment of Observations}

When \Chandra\ observations overlap, many data reduction tasks (e.g. source detection, position estimation, source extraction) are compromised if the misalignments among those observations are large (compared to the on-axis PSF).
Thus, our standard data analysis includes iterative re-alignment of the observations to each other, and to an astrometric reference catalog.
The astrometric reference catalog chosen for T-ReX consists of 20 sources\footnote{
X-ray sources p1\_1194, p1\_1234, p1\_296, p1\_1256, p1\_1186, p1\_611, p1\_830, p1\_995, p1\_682, p1\_1033, p1\_441, p1\_639, p1\_450, p1\_1391, p1\_1317, p1\_686 have likely counterparts in VMC. 
X-ray sources p1\_711, p1\_812, p1\_911, p1\_924 have likely counterparts in 2MASS. 
}
from the VMC and 2MASS catalogs that are likely counterparts to bright and un-crowded X-ray sources observed less than 5\arcmin\ off-axis.
The final shifts we applied to each observation, expressed in the  \anchorfoot{https://cxc.harvard.edu/ciao/ahelp/coords.html}{\Chandra\ SKY coordinate system} are reported in Table~\ref{tbl:obslog}.
We estimate remaining misalignment among the ObsIDs to be approximately 0.02\arcsec.

\subsection{Point Source Catalog}\label{sec:catalog}
\begin{figure}[htb]
\centering
\includegraphics[width=0.95\textwidth]{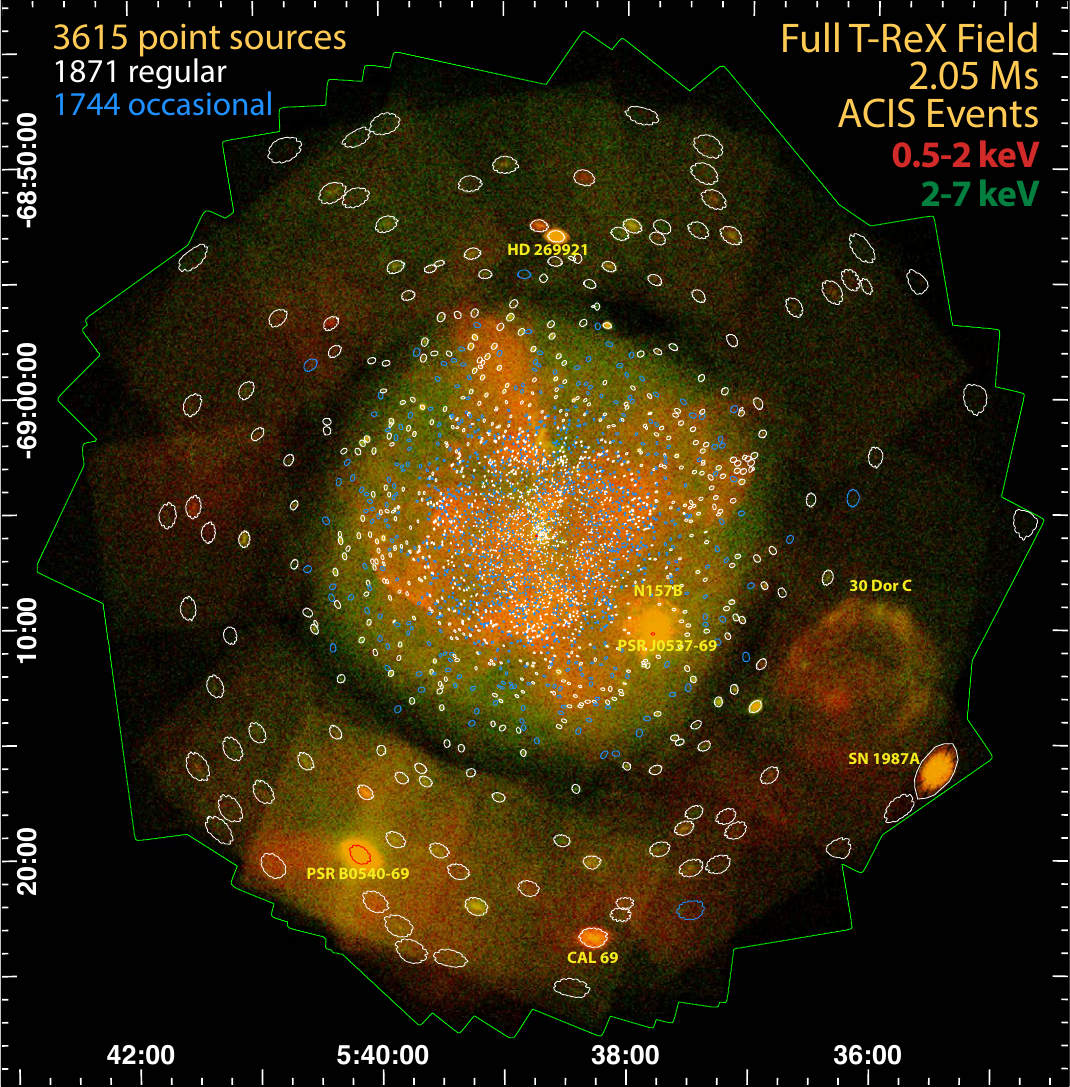}

\caption{The T-ReX event data for the full T-ReX field.  Events are binned by 4 sky pixels ($2\arcsec$).  Soft events (0.5--2~keV) are shown in red, hard events (2--7~keV) in green.  Regular point sources are outlined by blue polygons, ``occasional'' sources by white polygons. Red polygons outline sources exhibiting significant pile-up. These polygons are examples of event extraction regions, usually based on 90\% encircled energy PSF contours. Notable off-axis sources discussed in the text are labeled in yellow. 
\label{acisevents_fullfield.fig}}
\end{figure}

Extraction apertures for the 3615 point sources in the T-ReX catalog are  overlaid on an image of the full T-ReX event data in Figure~\ref{acisevents_fullfield.fig}. The catalog is published as an electronic table accompanying this article, and its columns are summarized in Table~\ref{tbl:xray_properties}. The variation in both resolution and sensitivity as a function of increasing off-axis angle $\theta$ are evident in the aperture sizes (increasing) and  point-source density (decreasing).  Both event data and point-source density decrease further in the S-array detectors farther off-axis where the exposure time is unevenly reduced compared to the field center, due to arbitrary roll angles (Figure~\ref{emap+srcs.fig}).

The extraction aperture for each observation of a point source is usually a contour of the local point spread function (PSF), chosen to enclose $\sim$90\% of the PSF for uncrowded observations or a reduced fraction to handle crowding \citep{Broos10}.   
We manually increased aperture sizes for the following very bright sources:
VFTS399 \citep[source Label p1\_610]{Clark15},
Mk34 \citep[p1\_995]{Pollock18},
R140a1/a2 (p1\_752),  
Mk39 (p1\_698),
R144 (p1\_1194),
053844.33-690554.7 (p1\_1000),
053844.12-690556.6 (p1\_979),
SN1987A (c73).
Two bright and severely crowded sources, R136a (p1\_832) and
R136c (p1\_893), required hand-drawn apertures (see Fig.~1 of C22, and related discussion of these sources as blends of multiple unresolved massive stars).


\input{xray_column_labels.tex}

Source identifiers and position quantities (\texttt{Name, RAdeg, DEdeg, PosErr, PosType}) are computed using a subset of extractions for each source chosen to minimize the position uncertainty \citep[][\S6.2 and 7.1]{Broos10}.   
Source significance quantities in each energy band ({\texttt{ProbNoSrc\_*}) are computed using a pre-defined set of ObsID combinations which do not depend on the data observed \citep[][\S2.2.3]{Townsley18}.
Because T-ReX observations have similar aimpoints, source significance ends up being calculated for each individual observation and for the combination of all observations. 
Source detection used our standard significance threshold (${<}1\%$ chance that the source is a random background fluctuation in the combination of extraction and energy band generating the highest detection significance), corresponding to  \texttt{ProbNoSrc\_MostValid}$<0.01$. Of course, identifying a multiwavelength counterpart to a faint X-ray point source (Appendix~\ref{app:counterparts}) generally increases confidence that the detection is an astrophysical object.

Variability indices (\texttt{ProbKS\_single, ProbKS\_merge,} and \texttt{ProbChisq\_PhotonFlux;} see Section~\ref{sec:var_metrics} below) are computed using all observations.
Quantities in the remaining columns of Table~\ref{tbl:xray_properties} are computed using a subset of ObsIDs, chosen independently for each source, to balance the conflicting goals of minimizing photometric uncertainty and of avoiding photometric bias \citep[][\S6.2 and 7]{Broos10}. 

As with many \ACIS\ studies, the (up to 54) extractions of a specific source in T-ReX do not share identical calibrations.
Even observations taken on the same CCD detector in the same configuration suffer from time-dependent Effective Area curves due to time-varying contamination on the \anchorfoot{http://cxc.harvard.edu/ciao/why/acisqecontamN0010.html}{\ACIS\ Optical Blocking Filters}. 
These calibration variations among source extractions introduce  biases into certain uncalibrated source properties (e.g., \texttt{MedianEnergy\_*}) and introduce source-dependent calibration errors into other source properties (e.g., \texttt{PhotonFlux, EnergyFlux, ProbKS\_merge,} and \texttt{ProbChisq\_PhotonFlux}).
Spectral fitting is needed to account for these calibration differences and to establish actual flux changes in sources that might be variable.

\subsection{Sources Flagged as ``Occasional''}\label{sec:occasional}
Nearly half ($1744/3615 = 48\%$) of sources in the catalog have the Boolean flag \texttt{IsOccasional} set (Table~\ref{tbl:xray_properties}); these sources are also identified in Figure~\ref{acisevents_fullfield.fig} and subsequent figures displaying point sources overlaid on various images.
The ``occasional'' flag, defined in \citet[][MOXC3; \S3.1]{Townsley19}, is set when ``source validation failed in all attempted multi-ObsID extraction combinations, but succeeded in one or more single-ObsID extractions.''  Since T-ReX observations have similar aimpoints, occasional sources are those that failed validation when all observations were combined, but passed validation in one or more single-ObsID extractions.  By contrast, we denote sources validated in multi-ObsID extractions as ``regular.'' 

In the MOXC3 catalog, 14\% of sources were flagged as occasional, a much lower fraction than in T-ReX. But T-ReX differs from our previous MOXC catalogs in the large number of OBsIDs targeting a single area on the sky, which increases the chances that a spurious source could be valid in a given single ObsID. This probably explains the much larger fraction of occasional sources in T-ReX.

It is important to emphasize that flagging a source \texttt{IsOccasional} is {\it not} a claim of variability, indeed, as \citet{Townsley19} noted, most occasional sources are too faint to assess X-ray variability. The majority of occasional sources in the T-ReX catalog have ${\la}12$ \texttt{NetCounts\_t}, while the large majority of regular sources exceed this value. Occasional sources also appear marginally softer than regular sources (e.g., in the \texttt{MedianEnergy\_t} distributions). As a test, we stacked the spectra of occasional sources in T-ReX and found that their total (gross) spectrum, local background spectrum, and net spectrum look very similar to each other and similar to the gross global diffuse spectrum (Section~\ref{sec:diffuse}). It is hence plausible that many occasional sources could be knots of diffuse emission, but they could also include faint astrophysical sources at the detection limit. 

We choose to include occasional sources in the catalog because (1) some of them may represent real X-ray counterparts to astrophysical objects detected at other wavelengths, and (2) they are subject to the source masking process in our diffuse emission work flow. We hence regard the 1871 ``regular'' sources as the more-reliable subset of the T-ReX catalog and those with the ``occasional'' property as a more complete set of candidate point sources.

\subsection{Variability Metrics}\label{sec:var_metrics}
The point source catalog (Table~\ref{tbl:xray_properties}) reports information that can help identify source variability on two timescales.
Short-term variability is assessed within each observation of a source.
\AEacro\ calculates a {\em p}-value for the observed arrival times of extracted (point source + background) in-band events from that observation, under the null hypothesis of a constant event rate, using a one-sample Kolmogorov-Smirnov (K--S) hypothesis test \citep[][\S7.6]{Broos10}.
If we reported all of those {\em p}-values (up to 54 in T-ReX), then variable sources within a {\em specific} observation could be identified by applying a threshold to the {\em p}-values from that observation.
However, observers are more typically interested in sources that show evidence for variability in {\em any} of the observations.
This is a classic \anchorfoot{https://en.wikipedia.org/wiki/Multiple\_comparisons\_problem}{multiple comparisons problem}, wherein any one of multiple observations can produce the ``discovery'' of short-term variability.

Every extraction that produces four or more X-ray events is tested for variability.
Instead of reporting {\em p}-values from each of those tests, we choose to report the smallest {\em p}-value produced by each source, as \texttt{ProbKS\_single} in Table~\ref{tbl:xray_properties}.
Note that the statistic \texttt{ProbKS\_single} $ = \min(P_1,P_2, \ldots P_{\rm N\_KS\_single})$ is {\em not} uniformly distributed under the null hypothesis, and is thus {\em not} a {\em p}-value.
However, a proper {\em p}-value can be calculated via 
\begin{equation}
   p = 1 - (1-\texttt{ProbKS\_single})^{\texttt{N\_KS\_single}}.
\end{equation}
%
%
If a set of $N$ sources is screened for variability by thresholding this {\em p}-value (e.g.\ ``variable'' $\Leftrightarrow p < 0.005)$, then the number of false positives expected would be approximately $N$ multiplied by the threshold (e.g.\ $N \times 0.005$).
\footnote{If instead variability were determined by thresholding the statistic \texttt{ProbKS\_single} (e.g.\ ``variable'' $\Leftrightarrow$ \texttt{ProbKS\_single}$ < 0.005)$, then the number of false positives expected would be approximately the total number of K--S tests performed in that set of sources, multiplied by the threshold (e.g.\ $\sum_{i=1}^N$\texttt{N\_KS\_single}$_{i}\times 0.005$). Such an approach would produce a much higher false-positive rate for variability, given the large number of OBsIDs in T-ReX.}
This intra-ObsID variability testing requires no calibration information, because instrument response is not expected to vary during an observation.  

To assess variability over longer timescales than a single exposure (typically 15--60~ks, Table~\ref{tbl:obslog}) one must consider all observations of a source.
We assess Inter-ObsID variability using a K--S test (\texttt{ProbKS\_merge}) that combines all the observations of the source  \citep[][\S7.6]{Broos10} and also by employing a standard $\chi^2$ test (\texttt{ProbChisq\_PhotonFlux}) on the single-ObsID measurements of \texttt{PhotonFlux\_t}.
Computation of both metrics requires calibration information. 
The instrument response for a given source will always vary across observations, due to both chromatic effects (aperture correction, mirror response, ACIS Optical Blocking Filter transmission, CCD response) and achromatic effects (dead time, exposure lost to inactive CCD pixels).
The combined energy-dependent effect of these calibration factors is represented by the Ancillary Response File (ARF) calculated for each source extraction.
For all multi-ObsID projects, the accuracy of the two long-term variability metrics produced by \AEacro\ (\texttt{ProbKS\_merge} and \texttt{ProbChisq\_PhotonFlux}) is limited by their assumption that the single-ObsID ARFs have the same {\em shape} (i.e. differ only in normalization).
In T-ReX the calibration factor that most strongly invalidates this assumption is the time-dependent variation in the ACIS Optical Blocking Filter transmission.

Thus in T-ReX and many previous projects \citep{Broos11,Townsley14,Townsley18,Townsley19} thresholding 
\texttt{ProbKS\_merge} or \texttt{ProbChisq\_PhotonFlux} should be viewed as a {\em screening} test for possible variability, rather than as a calibrated null hypothesis test using an accurate {\em p}-value; false-positives are expected. 
The only accurate method for measuring source variability across all observations would be simultaneous fitting of all single-ObsID spectra, which is beyond the scope of this paper. 

We do not attempt to flag all candidate variable sources in the T-ReX point source catalog, but in 
Appendix~\ref{app:variability} we discuss seven examples of high-quality periodic X-ray variables and two large X-ray flares associated with foreground Galactic stars in the T-ReX field.

\subsection{Sources Suffering Photon Pile-up \label{sec:pileup}}
Three point sources (Mk34, R140a, PSR~J0537-6910) have extractions that are significantly impacted by an instrumental non-linearity known as \anchorfoot{http://cxc.harvard.edu/ciao/why/pileup_intro.html}{{\em photon pile-up}}, which is described in the three MOXC papers \citep[and references therein]{Townsley14,Townsley18,Townsley19}.  When possible, we model the piled-up spectra in these extractions \citep{Broos11} and use that model to estimate a ``reconstructed'' ACIS spectrum free from pile-up effects.  
The metric we choose to describe the severity of pile-up is the ratio of the pile-up-free count rate to the observed (piled-up) count rate in the total (0.5--8~keV) energy band, which we call the Rate Correction.

Rate Corrections for the colliding-wind Wolf-Rayet binary system Mk34 (053844.25-690605.9 = p1\_995, VFTS~770) range from 1.176 when the count rate was highest (ObsID 17312) to 1.0 (no pile-up) when the count rate was lowest. 
\citet[][Figure~6]{Pollock18} compare spectral fits to the piled and corrected spectra.

The nearly constant colliding-wind Wolf-Rayet binary system R140a 
(053841.59-690513.4 = p1\_752, see C22) produced moderate Rate Corrections ($\sim$1.07) in all observations.

Although the pulsar PSR~J0537-6910 (053747.40-691020.2 = p1\_255) has count rates about 5 times larger than Mk34 and R140a, it has similarly moderate Rate Corrections because its photons were distributed over a much larger PSF (6\arcmin\ off-axis).

\subsection{Archive of Reduced Data Products \label{sec:repository}}
The Zenodo archive \anchorfoot{https://zenodo.org/doi/10.5281/zenodo.8372420}{\citep{Townsley23}}  
contains data products and plots that may be of use to other investigators who wish to use the T-ReX data set; see the file \textsf{README.TXT} there for details.


\section{An Imaging Tour of T-ReX \label{sec:images}}

To emphasize various morphological features of the T-ReX diffuse X-ray emission, we present in this section several visualizations of the imaging data with no overlays. Soft diffuse emission, divided into three energy bands, across the full T-ReX field is shown in Figure~\ref{3soft_fullfield.fig}. This field is  $40\arcmin \approx 580$~pc in diameter. The image has been smoothed to achieve a roughly constant SNR and showcases the dramatic improvement in spatial resolution afforded by the increased SNR of the 2~Ms T-ReX exposure compared to previously published images \citep{Townsley06a,Townsley11c}. 

\begin{figure}[htb]
\centering
\includegraphics[width=0.99\textwidth]{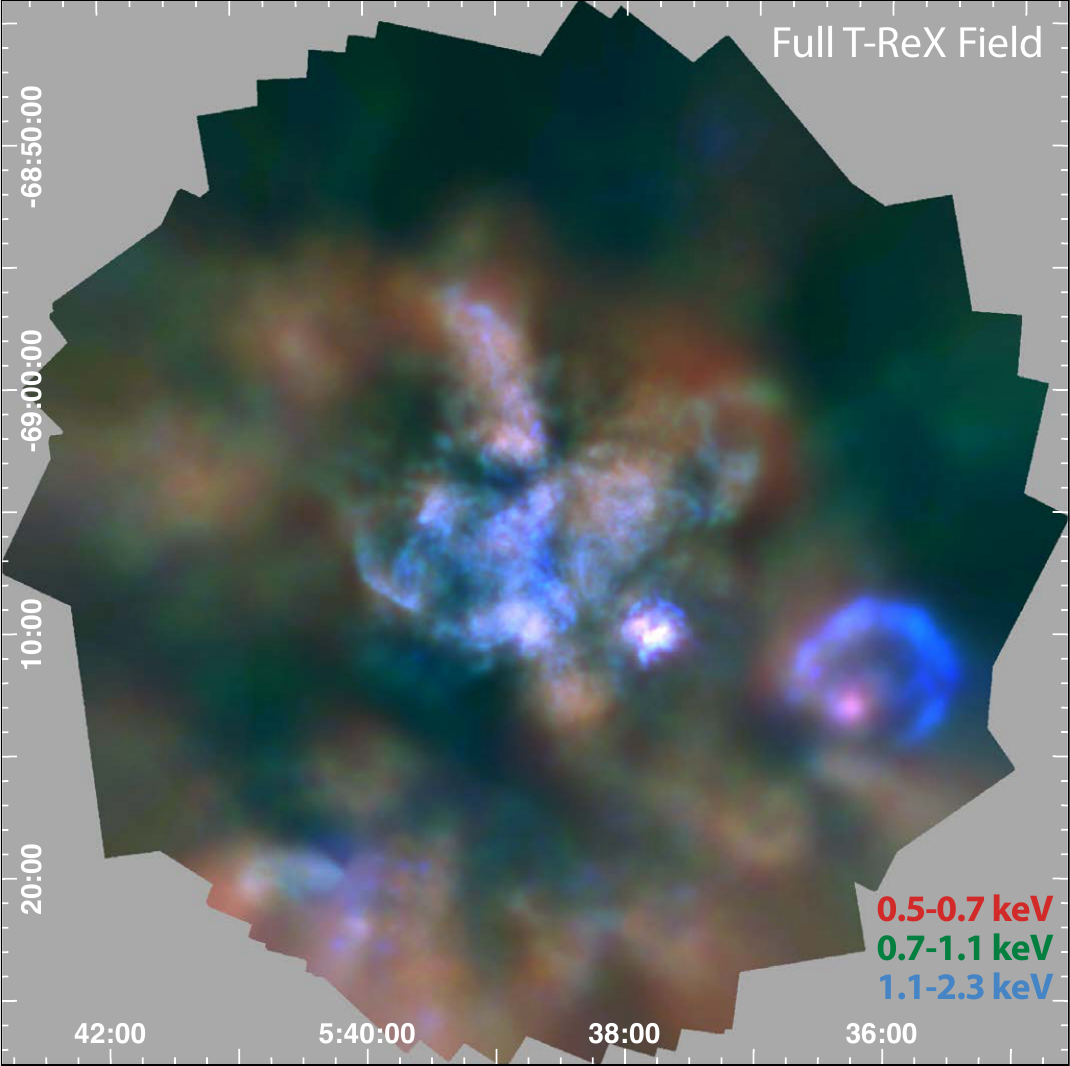}
\caption{T-ReX diffuse emission in three soft X-ray bands, smoothed with adaptive kernels to achieve SNR = 15.  All 3615 point sources (seen in Figure~\ref{acisevents_fullfield.fig} above) were excised before smoothing. Axes are in celestial (J2000) coordinates.
\label{3soft_fullfield.fig}}
\end{figure}

Based on earlier, shallower and hence blurrier images, 30~Dor appeared to exhibit at least five plasma-filled superbubbles \citep{Townsley06a}, products of strong OB winds and multiple supernovae.  Because of their proximity and the low absorbing column to the LMC, they provide the best opportunity to study the X-ray spectrum and structure of any superbubble system.
\Chandra\ has already shown \citep{Townsley11c,Lopez11,Pellegrini11} that they are spatially complex X-ray structures with a range of X-ray plasma temperatures, ionization timescales, absorptions, and luminosities.  The T-ReX images reveal that the ISM in 30 Dor, as in Carina and other MSFRs, forms a complex labyrinth of low-density cells and channels filled with hot plasma, surrounded by walls of colder material. The boundaries apparently separating the superbubble lobes are much less distinct in the deeper observations that more effectively penetrate absorbing columns of the cold, dense ISM walls, shells, and filaments.

\begin{figure}[htb]
\centering
\includegraphics[width=0.99\textwidth]{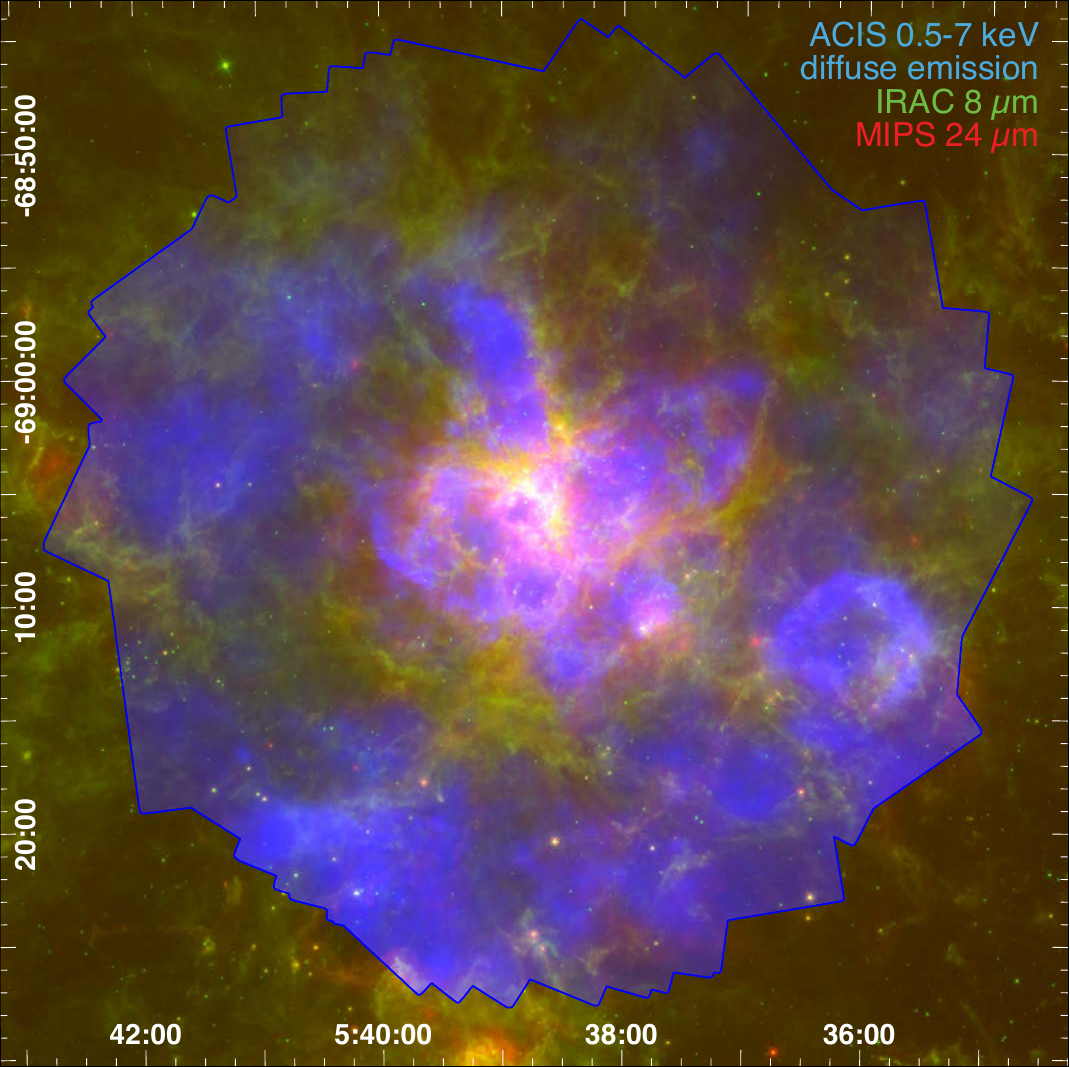}
\caption{ T-ReX diffuse emission in the \Spitzer context.  \Spitzer images are from the Surveying the Agents of Galaxy Evolution (SAGE) survey \citep{SAGE}.
\label{context.fig}}
\end{figure}

The outer regions of the T-ReX field are suffused with fainter, softer (red-green in in Figure~\ref{3soft_fullfield.fig}) emission that also shows rich spatial structure. Much of this spatial complexity could be explained by the imprint of spatially-varying absorption (see Figure~\ref{context.fig}}) on background of X-ray emission that is intrinsically softer than 30 Dor itself. Some of this emission could be hot plasma leaking from the 30 Dor superbubbles.

\begin{figure}[htb]
\centering
\includegraphics[width=0.49\textwidth]{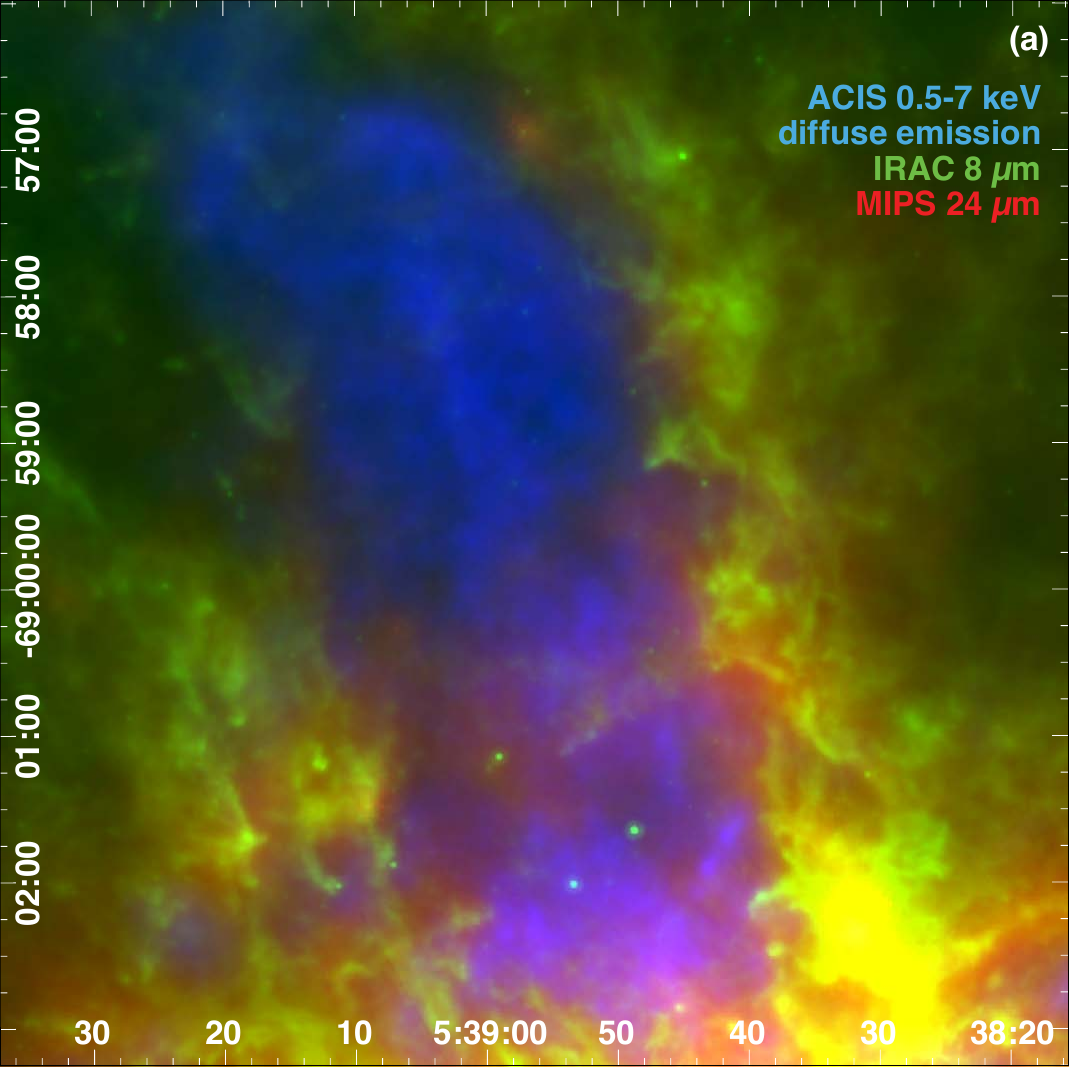}
\includegraphics[width=0.49\textwidth]{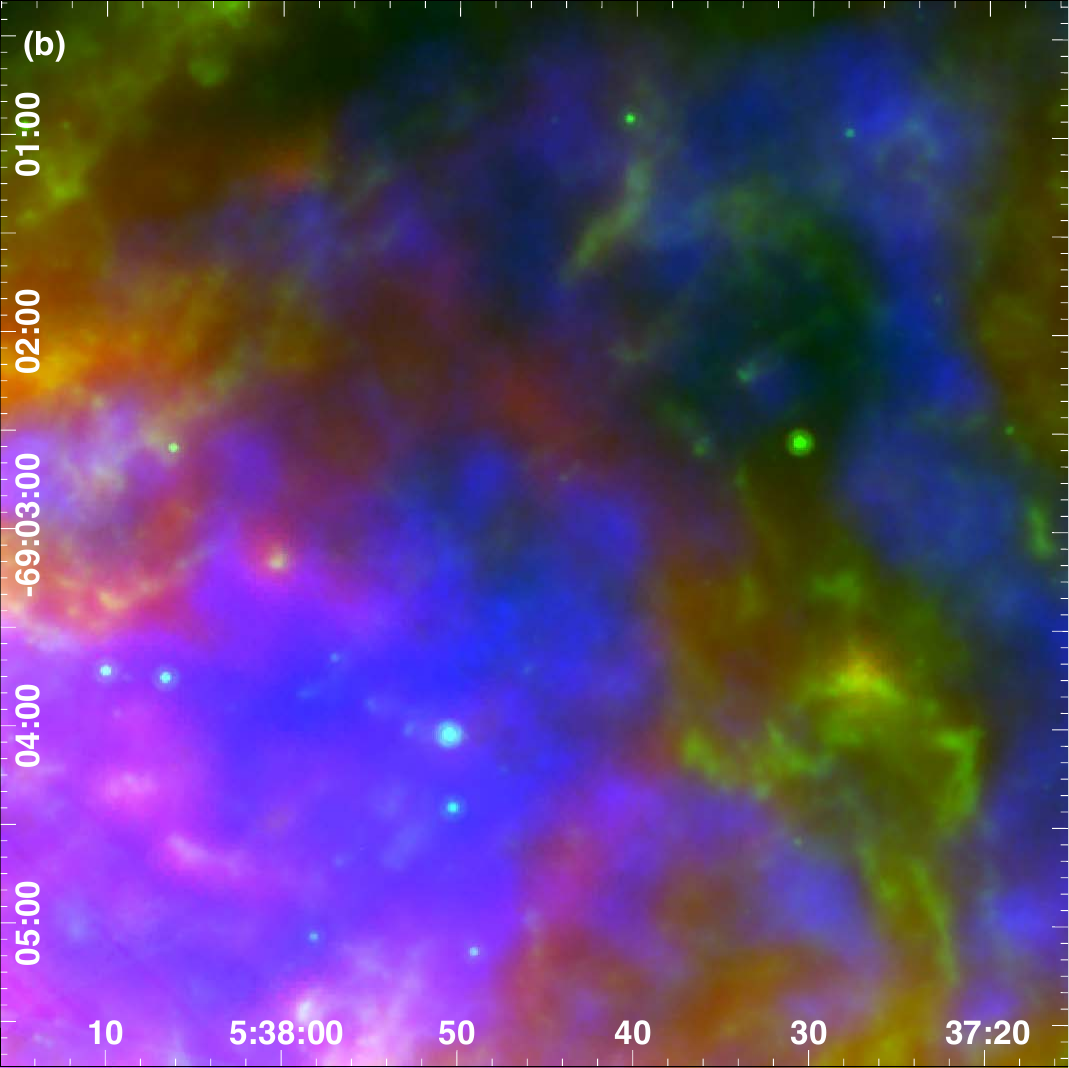}
\includegraphics[width=0.49\textwidth]{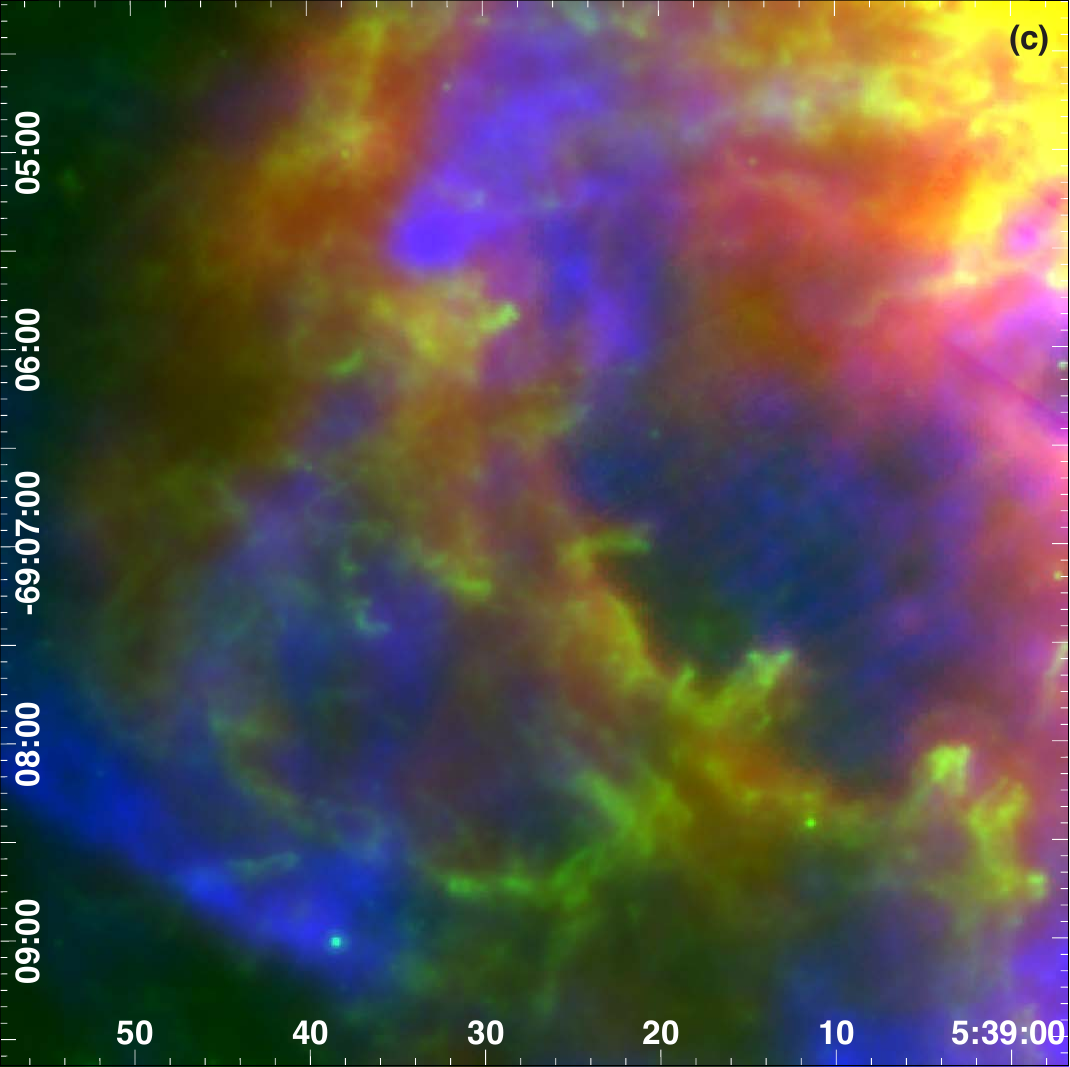}
\includegraphics[width=0.49\textwidth]{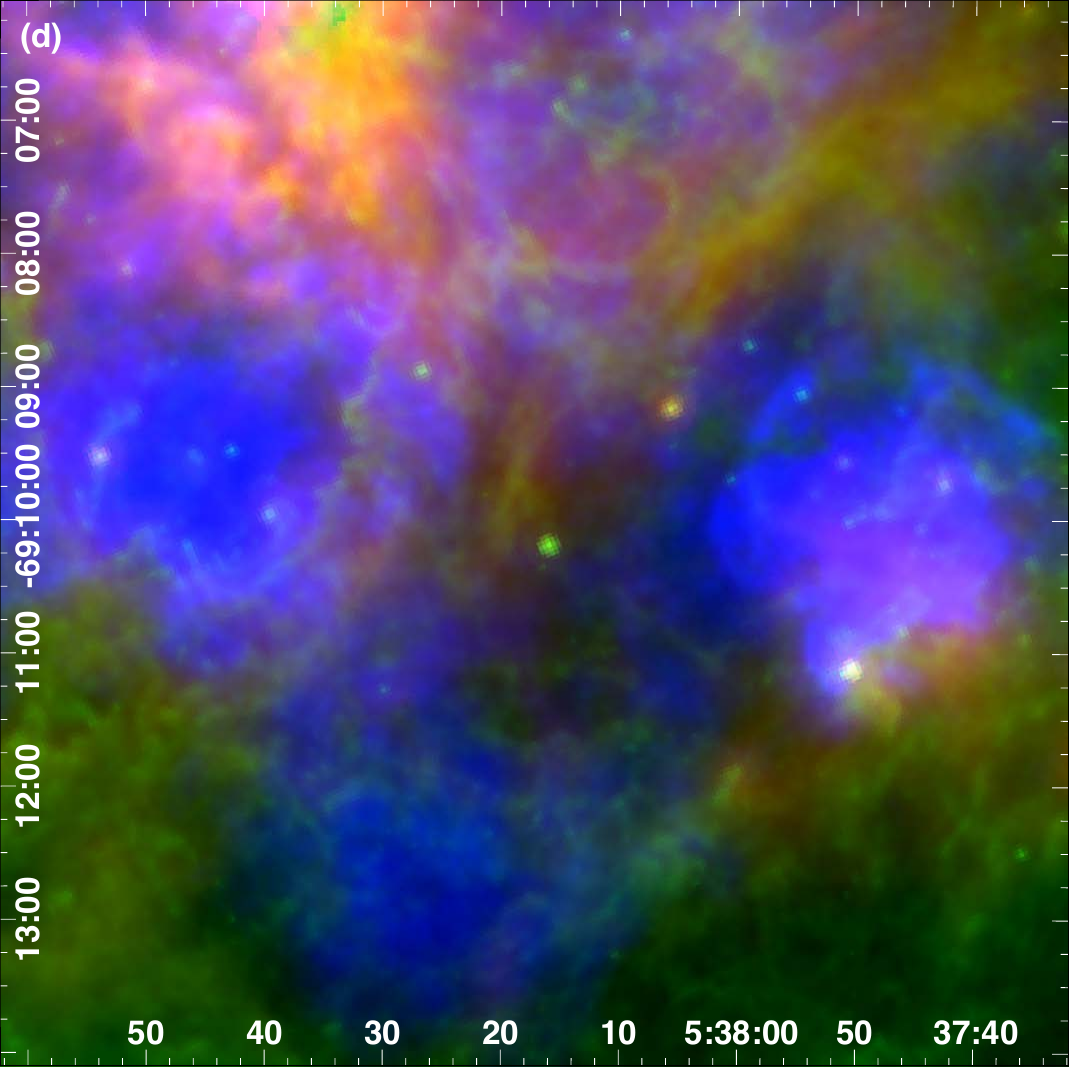}
\caption{ Light and shadow:  zoomed and rescaled regions from Figure~\ref{context.fig}.  These images illustrate the complex interfaces between the hot and cold ISM of 30~Dor.  With respect to R136, these regions are located:
(a) northeast;
(b) northwest;
(c) southeast;
(d) southwest.
\label{interfaces.fig}}
\end{figure}

Three neighboring structures, labeled in Figure~\ref{acisevents_fullfield.fig}, stand out in the 1.1--2.3~keV band against the softer background.
The ${\sim}5$-Myr-old star-forming region NGC~2060 \citep{Sabbi16}, located ${\sim}6\arcmin$ southwest of R136,
is dominated by the young SNR N157B (also known as 30 Dor B), containing the 16-ms PSR~J0537-6910 \citep{Chen06,Chen23-30DorB}. N157B and its prominent cometary pulsar wind nebula
contributed over half of the total X-ray counts in prior \Chandra\ studies of 30 Dor \citep{Townsley06a}. West of N157B lies the 30 Dor C superbubble, a large shell of non-thermal X-ray emission, which appears as an elongated, broken blue ring surrounding (in projection) a compact, softer source that has been identified as MCSNR J0536--6913 \citep{Kavanagh15}.  
A hard-edged artifact produced by smoothing over the mask for the exceptionally bright PSR B0540-69 appears in the S-array mosaic SE of 30 Dor. Enhanced hard diffuse emission in the vicinity is likely associated with the young, oxygen-rich SNR reported by \citet{Park10} using on-axis \Chandra\ observations.


In Figures~\ref{context.fig} and \ref{interfaces.fig} we highlight the complex morphological interplay between the hot X-ray emitting plasma, and the cool, dusty ISM revealed in \Spitzer images.\footnote{Our T-ReX diffuse map, combined with {\it JWST} images of the central portions of 30~Dor, was released publicly \\ by the \Chandra\ X-Ray Center in 2022 October, see \url{https://chandra.si.edu/photo/2023/30dor/index.html}.} In these images, the {\it Spitzer}/IRAC 8~$\mu$m band (green) shows primarily emission from polycyclic aromatic hydrocarbons (PAHs) that produce a strong 7.7~$\mu$m fluorescence feature when exposed to UV irradiation \citep{Povich07}, making it an excellent tracer of locations where the surfaces of cold, molecular clouds are exposed to ionizing radiation. The {\it Spitzer}/MIPS 24~$\mu$m band (red) traces warm dust, which emits most strongly where dust grains are intermixed with ionized plasma and becomes stochastically heated by the hard radiation field \citep{Watson08,EverettChurchwell10}. The full ACIS 0.5--7~keV emission (blue) tends to fill in the cavities and voids in the mid-IR nebula. Regions where the diffuse X-rays are especially faint, notably a broad region bordering the East-Southeast half of 30~Dor (Figure~\ref{3soft_fullfield.fig}), tend to be filled in by mid-IR cloud structures (Figure~\ref{context.fig}), indicating that the cold clouds are located in front of the hot plasma in these locations and absorb the soft X-rays. In other regions, diffuse X-rays appear to surround or pass in front of bright mid-IR filaments (many examples are evident in Figure~\ref{interfaces.fig}, especially panels c and d), revealing interfaces between the hot and cold ISM.

\section{Point Sources \label{sec:ptsrcs}}

\begin{figure}[htb]
\centering
\includegraphics[width=0.99\textwidth]{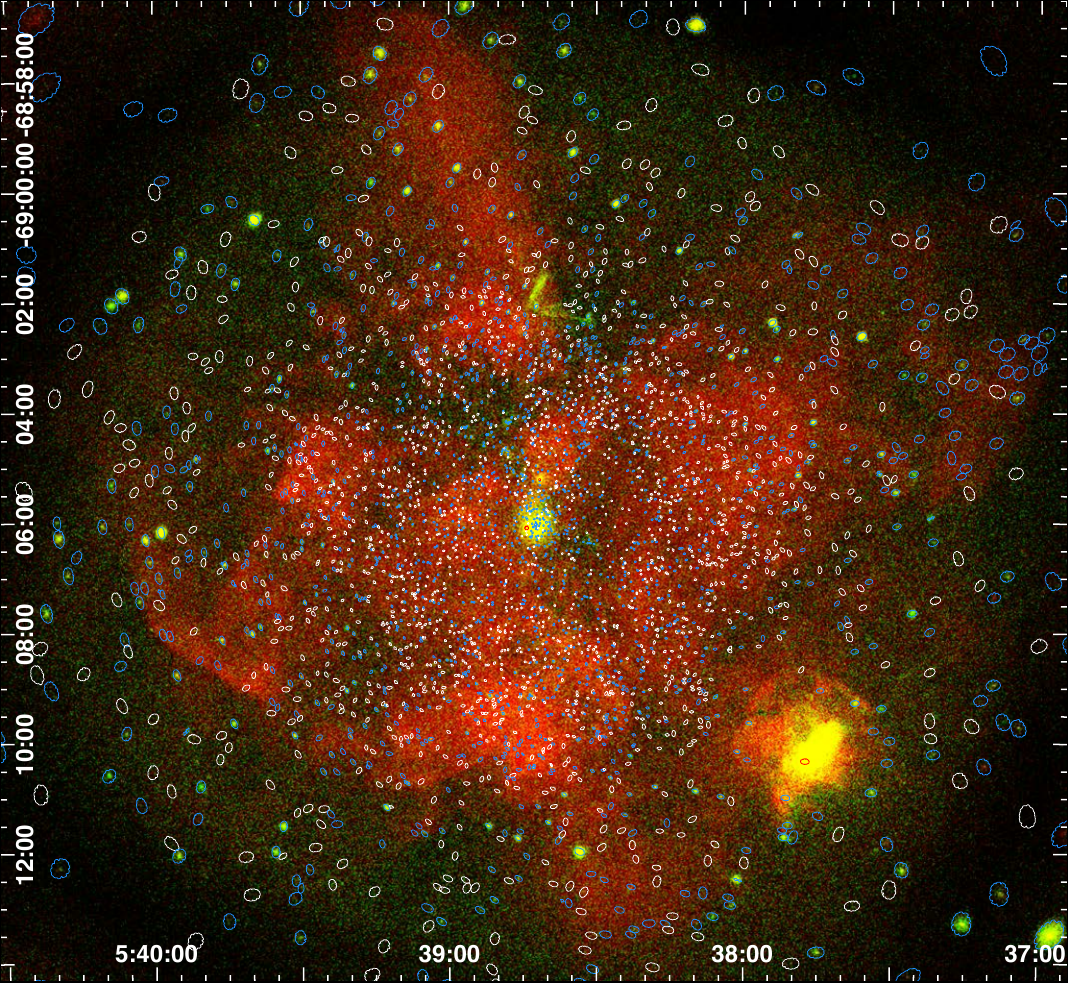}
\caption{The T-ReX event data with point source extraction apertures overlaid, as in Figure~\ref{acisevents_fullfield.fig}, now zooming in on 30~Dor itself.  Events are binned by 2 sky pixels.
\label{acisevents2.fig}}
\end{figure}

\begin{figure}[htb]
\centering
\includegraphics[width=0.99\textwidth]{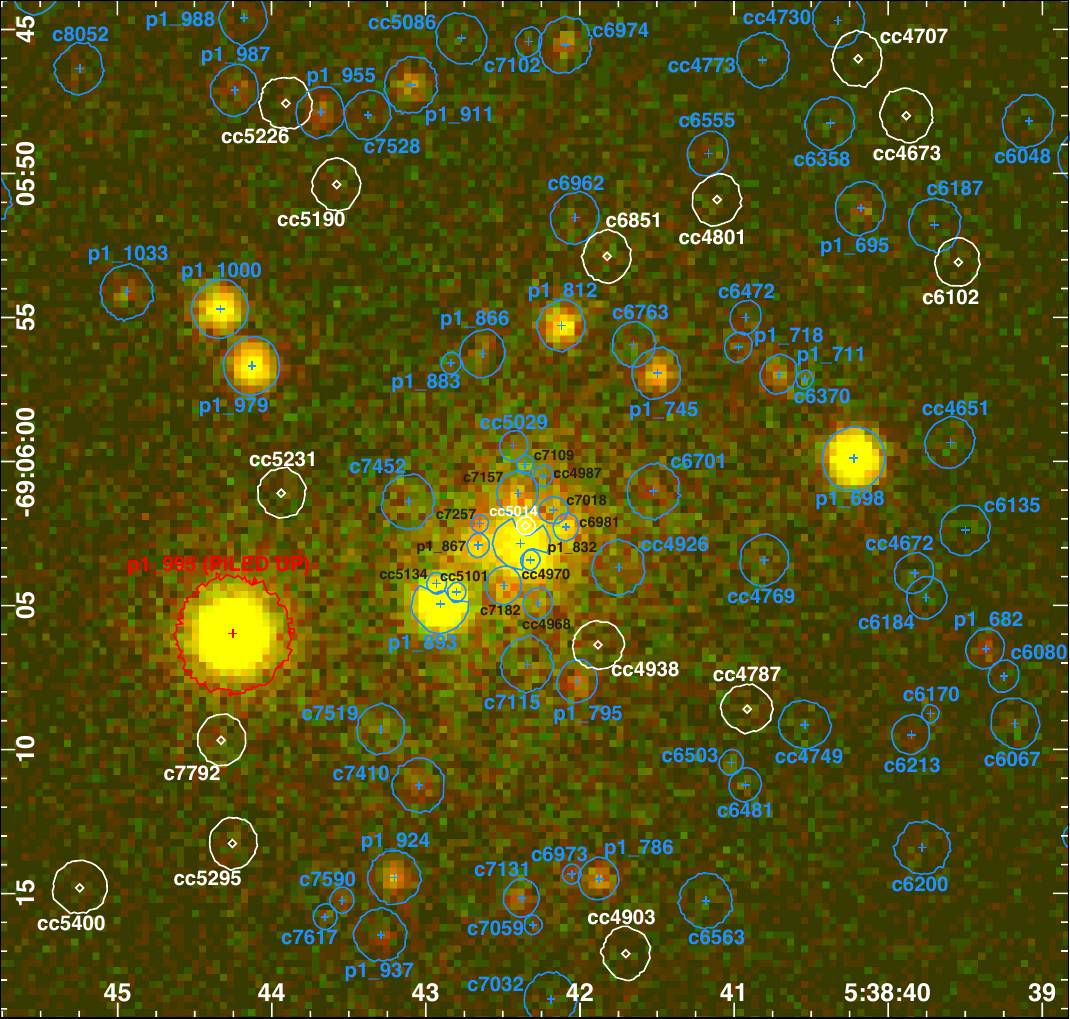}
\caption{T-ReX event data (as in Figures~\ref{acisevents_fullfield.fig} and \ref{acisevents2.fig}) zoomed in to
show the inner ${\sim}35\arcsec = 8.4$~pc of the field, centered on the T-ReX aimpoint (R136).  Events are binned by 0.5 sky pixels.  Source positions are shown as blue diamonds for occasional sources and as white crosses for regular sources, with their ACIS source labels (Table~\ref{tbl:xray_properties}; regular source labels in the most crowded central region are shown in black, with reduced size, for clarity).  The piled-up WR star Mk34 (p1\_995) is noted in red.
\label{acisevents_R136.fig}}
\end{figure}

In this and the following section we focus on the central region of the T-ReX field containing 30 Dor itself, imaged by the ACIS I-array (Figure~\ref{acisevents2.fig}). Point sources in the T-ReX catalog are distributed widely throughout this field, except in the vicinity of N157B, where bright diffuse emission prevented detection of point sources associated with NGC 2060 other than the extraordinarily bright, piled source PSR J0537--69 (source p1\_255 with $3.4\times 10^5$ net counts). 
The field center contains the R136 cluster, and we zoom in on this region in Figure~\ref{acisevents_R136.fig} to better display its dense concentration of X-ray events and cluster of detected point-sources. The spatial distribution of regular sources exhibits more structure compared to occasional sources; besides R136 there are several smaller clusters of X-ray sources that appear to be associated with embedded subclusters \citep{Walborn13,Sabbi16} in the molecular cloud to the north and west.

After the pulsar, most of the brightest point sources in the 30 Dor field are massive stars studied in detail by C22, including two Wolf-Rayet systems (Mk 34 and R140a) and the candidate high mass X-ray binary VFTS 399 \citep{Clark15}. We present the cross-matching of T-ReX point sources with massive stellar counterpart catalogs in Appendix~\ref{app:counterparts}; these matching lists, including blended sources in R136 (Figure~\ref{acisevents_R136.fig}) that could represent multiple OB stars, were used to construct the C22 sample. C22 found that the majority of Wolf-Rayet and Of/WN systems in 30 Dor were detected in X-rays (70.0\% and 85.7\%, respectively), but for O-type stars in general a similar majority went undetected, and the fraction of detected B-type stars was very low. The absorption-corrected X-ray luminosity ($L_X$) from spectral fitting of detected massive stars and upper limits on undetected massive stars indicate that the T-ReX massive star sample is highly incomplete for $\log{L_X}<32~{\rm erg s}^{-1}$.

We can also estimate the sensitivity of T-ReX to lower-mass, pre-main sequence stars. Among the regular sources in the T-ReX catalog, 1501/1871 (80\%) were detected with ${>}12$~net counts. Assuming a typical T-Tauri X-ray spectrum and absorption equivalent to $A_V=0.5$~mag, 12 counts on-axis corresponds to $\log{L_X}>31.5~{\rm erg s}^{-1}$. A calculation using the absorption-corrected surface brightness of the global diffuse emission (see Section~\ref{sec:diffuse} below) yields a very similar luminosity for a circular region of diffuse background emission enclosed by an on-axis point-source extraction aperture. Diffuse X-ray emission limits point-source sensitivity throughout the T-ReX field, not only in the brightest regions.
This luminosity threshold suggests that T Tauri stars detected in T-ReX are likely to be in a mega-flaring state \citep{GetmanFeigelson21}.
The \Chandra\ Carina Complex Project \citep[CCCP,][]{Townsley11a} provides a useful comparison to put this source luminosity in context, as it is a Galactic starburst region containing multiple stellar clusters with different ages, similar to 30 Dor \citep{HTTP2,Povich19}. \citet{XIMPS} performed X-ray spectral fitting on 370 bright CCCP stellar sources that were not associated with OB stars. Imposing the T-ReX luminosity cut on the CCCP sample would exclude all but a small number (${\la}20$) of flaring, low-mass T Tauri stars ($M<2~M_{\Sun}$). About one-third of  intermediate-mass, pre-main sequence stars (IMPS; 2--4~$M_{\Sun}$) in CCCP would be detectable in T-ReX, but with the caveat that their X-ray luminosity fades rapidly after ${\sim}1$~Myr \citep{Gregory16,XIMPS}. X-ray activity for pre-main sequence stars of all masses decays after ${\sim}7$~Myr \citep{Getman22}.

Considering the above arguments, T-ReX is sensitive primarily to early O stars and convective (flaring) IMPS, tracing the youngest stellar populations in 30 Dor, primarily R136. This explains why the older cluster Hodge 301 \citep[20~Myr;][]{Sabbi16} is not detected in T-ReX. The great density of R136 itself means that the cluster core is not resolvable by ACIS, and (probably hundreds) of unresolved stellar sources contribute to elevated diffuse X-ray backgrounds that further reduce point-source sensitivity in the vicinity (Figure~\ref{acisevents_R136.fig}). 
R136 contains the majority of stellar sources that should be detectable in principle, based on X-ray luminosity, but in reality this cluster also exhibits extreme source crowding and elevated diffuse background levels that dramatically reduced point-source sensitivity compared to the T-ReX field overall. 

Beyond R136 and the neighboring smaller, embedded clusters, the point source population in the wider T-ReX field appears broadly consistent with a population of AGN and other background extragalactic sources, as indicated by the general lack of spatial clustering and the dearth of multiwavelength counterparts to T-ReX sources (see Appendix~\ref{app:counterparts}). 
Using the 95\% completeness limits of the 2-Ms {\it Chandra} Deep Field North catalog \citep[CDF-N][]{CDF-N2Ms, CHAMP04}\footnote{The CDF-S 2 Ms catalog \citep{CDF-S2MS} would yield very similar numbers.} as a control for the T-ReX field if the LMC were removed, we would predict 720 sources with soft-band (0.5--2~keV) flux ${>}8\times 10^{-17}$~erg~s$^{-1}$~cm$^{-2}$ and 480 sources with hard-band (2--8~keV) flux ${>}4\times 10^{-16}$~erg~s$^{-1}$~cm$^{-2}$. Assuming an average absorbing column through 30 Dor of $N_H=2\times 10^{22}$~cm$^{-2}$ and a characteristic photon index of 1.7, \anchorfoot{https://cxc.harvard.edu/toolkit/pimms.jsp}{PIMMS} shows that the limiting observed fluxes would be reduced to ${\sim}1.2\times 10^{-17}$~erg~s$^{-1}$~cm$^{-2}$ in the soft band. The T-ReX catalog contains ${\sim}1600$ regular sources above this soft-band limit, more than double the expected background contamination rate. By contrast, T-ReX contains only 525 hard-band sources brighter than the CDF-N 95\% completeness limit (with absorption producing a negligible correction), hence the large majority of T-ReX sources detected in the hard band are consistent with background contaminants.

\clearpage

\section{Diffuse X-ray Emission \label{sec:diffuse}}

\begin{figure}[htb]
\centering
\includegraphics[width=0.9\textwidth]{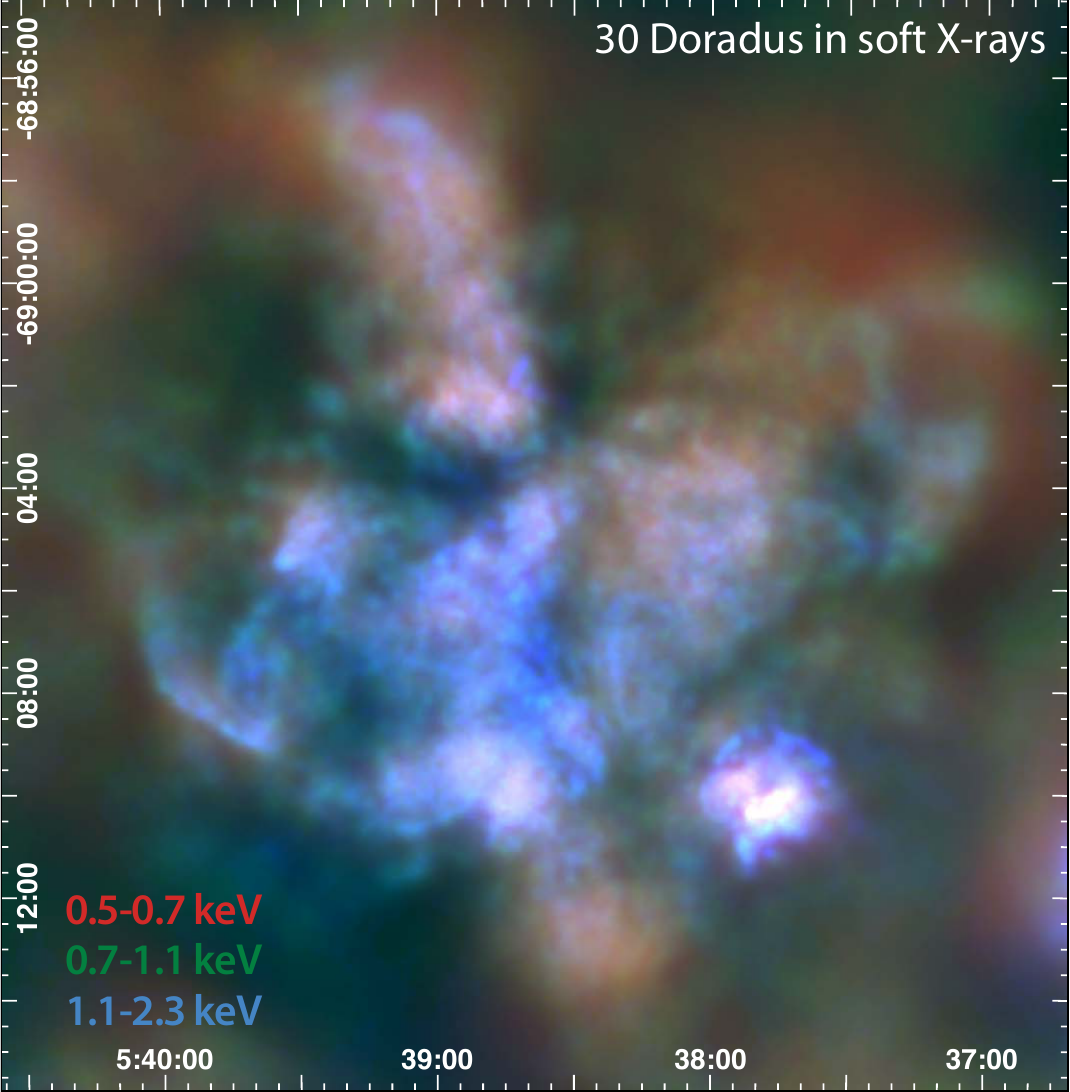}
\caption{30~Dor diffuse emission in three soft X-ray bands.  This is a zoom of Figure~\ref{3soft_fullfield.fig}, cropped to a similar field as shown in Figure~\ref{acisevents2.fig}.
\label{3soft_30dor.fig}}
\end{figure}

30 Dor offers us a microscope on starburst astrophysics, having endured 25 Myr of the birth and death of the most massive stars known.  Across its 250-pc extent, stellar winds and supernovae have carved the ISM within 30 Dor into an amazing display of arcs, pillars, and bubbles (Figures~\ref{context.fig} and \ref{interfaces.fig}). 
For over 40 years, we have also known that 30 Dor is a bright X-ray emitter, so its familiar stars and cold ISM structures suffer irradiation by multi-million-degree plasmas.  
Here we focus on the diffuse X-ray emission revealed by T-ReX, highlighting hot/cold interfaces (Figure~\ref{interfaces.fig}) and other hot plasma structures resolved to 1--10~pc scales across the entire 30 Dor complex (Figures~\ref{3soft_fullfield.fig} and \ref{3soft_30dor.fig}).  Massive star winds and cavity supernovae over the millenia have contributed to a broad mix of X-ray-emitting plasmas and absorbing columns, and the hot ISM within 30 Dor exhibits similar complexity as seen at colder temperatures.
In Figure~\ref{3soft_30dor.fig} we present a square approximately the size of the ACIS I-array ($17\arcmin = 245$~pc on a side), where the sensitivity and spatial resolution are maximized. In this section we present spectral fitting of the diffuse emission using extraction regions from within this image.

 \citet{Townsley11b,Townsley11c} provided a detailed description of our general spectral modeling strategy as applied to the Carina Nebula, other Galactic massive star-forming complexes, and the 30 Dor observations predating T-ReX. The spectral model consists of  three shocked thermal plasma components implemented in \XSPEC\ {(\em vpshock)}. Components 1 and 2 consist of cooler plasmas ($0.25~{\rm keV}\la kT \la 0.35$~keV) with different cooling timescales; the high timescales for component 2 (e.g. $\log{(\rm \tau)}=13.7/[{\rm s~cm}^{-3}$]) imply collisional ionization equilibrium (CIE) while the lower timescales for component 1 suggest the presence of non-equilibrium ionization (NEI) plasma. Component 3 consists of hotter plasmas ($0.7~{\rm keV} \la kT \la 1.3$~keV) exhibiting a mixture of CIE and NEI. Each spectral fit plot presented below (Figures~\ref{N157Bspectra.fig}--\ref{tessellates.fig}) labels the {\em vpshock} parameters for the various employed components: absorbing column $N_H$ is labeled as NH[123], plasma temperature $kT$ as kT[123], timescale $\tau$ as tau[123] and absorption-corrected surface brightness as SB[123].

We assumed that absorbing material in both the LMC and foreground Milky Way have relative abundances equal to those of the Sun, but that the LMC abundances are scaled down from the Milky Way (by about 50\%). Thus in \XSPEC\ we chose to model total absorption with a single {\em TBabs} model that uses solar abundances. Abundances for most elements in the {\em vpshock} models were frozen at $0.4 Z_{\odot}$, appropriate for the low-metallicity environment of the LMC \citep{RD90}, but we found that enhanced abundances of Ne ($0.59 Z_{\odot}$), Mg $1.06 Z_{\odot}$, and Si ($0.75 Z_{\odot}$) were needed to account for other line features in the spectrum.
The NH[123] values labeled in Figures~\ref{N157Bspectra.fig}--\ref{tessellates.fig} may be doubled to correct for LMC abundances. 


Dotted Gaussians representing unmodeled lines were added because they substantially improve the fit \citep{Townsley11b,Townsley11c}. The central line energies, widths, and normalizations were all free parameters, and yet, as we demonstrate below, the multiple Gaussian components typically land on or near the energies of real X-ray emission lines. They may represent charge exchange \citep[CX, see][for a recent review]{Gu+Shah23} or some other physics not captured by the current plasma models.  

We also allow a contribution from unresolved pre-MS stars (2-temperature {\em apec} model for collisionally-ionized thermal plasma) and an absorbed power law to account for a hard tail in the spectrum due to a background of unresolved AGN and/or nonthermal emission in 30 Dor.

\subsection{N157B SNR and Pulsar Wind Nebula \label{sec:evolved}}

The N157B SNR (also known as 30 Dor B) and the associated wind nebula produced by its pulsar, PSR J0537 \citep{Marshall98,Wang01}, produce the brightest diffuse X-ray emission by far in the T-ReX field, with (absorption-corrected) surface brightnesses exceeding the typical values of 30 Dor itself by one to two orders of magnitude. It was therefore critical to remove these diffuse sources prior to modeling the global diffuse emission of 30 Dor.

\begin{figure}[htb]
\centering
\includegraphics[width=0.49\textwidth]{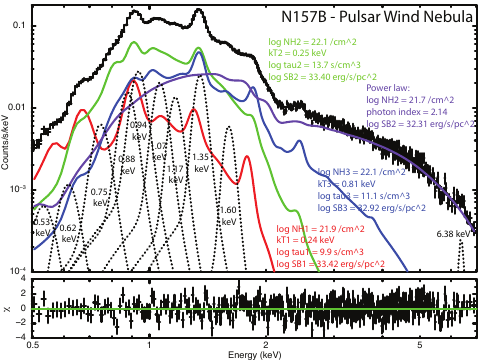}
\includegraphics[width=0.49\textwidth]{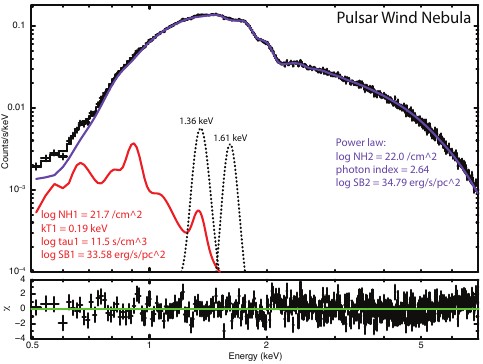}
\caption{Diffuse X-ray spectra of N157B.
(a)  The N157B SNR spectrum with the central PWN removed.
(b)  The J0537 PWN spectrum with the PSR J0537 point source removed. In these and other plots of spectral fitting presented below, the following parameters of various numbered models components are labeled: $N_H$ (absorbing hydrogen column density), $kT$ (plasma temperature), tau (timescale), and SB (surface brightness, corrected for absorption NH).
\label{N157Bspectra.fig}}
\end{figure}

We present spectral fits to N157B and the pulsar wind nebula in Figure~\ref{N157Bspectra.fig}.  N157B required a complex model consisting of three thermal plasma components plus a power law of comparable brightness. The two thermal plasma components used to model the soft end of the diffuse spectrum have $kT1\approx kT2=0.25$~keV, and \citet{Chen23-30DorB} recently concluded that the combination of the soft X-ray plasma and the physical extent of N157B require at least one more supernova preceded the one, 5000 years ago, that created PSR~J0537.

Our model also includes 10 Gaussian line components, many of which correspond to energies of emission lines that show CX enhancement in the Cygnus Loop SNR \citep{Katsuda11,Uchida19} or starburst galaxies \citep{Ranalli08,Liu12,Zhang22}. The \ion{O}{7} triplet falls near 0.53~keV, while \ion{O}{8} 
can produce lines near 0.61~keV and 0.75 keV. Highly-ionized iron can produce the lines near 0.88 keV (\ion{Fe}{17} or \ion{Fe}{18}) and 0.94 keV (\ion{Fe}{19}), but the latter could also be Ne IX. Other ionized neon and/or magnesium lines could produce the components at 1.07 keV (\ion{Ne}{10}), 1.17 keV (\ion{Ne}{10} or Mg), 1.35 keV (\ion{Mg}{11}), and 1.60 keV (possibly Mg). 

The two components attributable to magnesium CX lines also appear in the modeled spectrum of the pulsar wind nebula, which otherwise presents a comparatively simple, power-law spectrum requiring only one additional NEI plasma component to model the soft tail of the spectrum. This model resembles that of the combined ACIS-S spectrum of N157B and the pulsar wind nebula presented by \citet{Chen06}.

Unique among the T-ReX diffuse spectra modeled thus far, the fit to the hard tail of the N157B was significantly improved by the addition of a Gaussian component at 6.38 keV corresponding to the Fe K$\alpha$ fluorescence line. This is a common feature in AGN and some X-ray binaries, arising in dense accretion disks at high temperature. It is unlikely that this emission arises in the low-density SNR, but it could be produced by massive, colliding-wind binaries present in the massive NGC 2060 stellar cluster that hosted the supernova progenitor \citep[see, e.g.,][]{Walborn14}. This line is obvious in the X-ray spectrum of the $\eta$ Car system, for example \citep{Kashi21}. The average X-ray luminosity of the strongly variable $\eta$ Car is $6.33\times 10^{34}$~erg~s$^{-1}$, which would rank it among the brightest T-ReX point sources but still an order of magnitude fainter than PSR J0537-69, the sole point source detected in the central regions of N157B. 

\subsection{Spectral Fits to Global Diffuse Emission}
\begin{figure}[htb]
\centering
\includegraphics[width=0.8\textwidth]{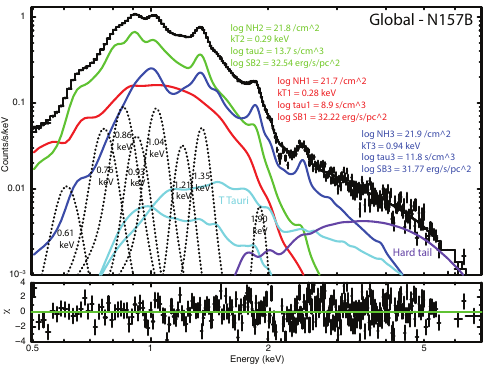}
\caption{The global diffuse X-ray spectrum of 30~Dor,  with the N157B SNR and its PWN removed.  The ``Global'' diffuse region is defined in Figure~\ref{tessellates.fig} below.
\label{globalspectrum.fig}}
\end{figure}

The spectrum ``Global $-$ N157B'' presented in Figure~\ref{globalspectrum.fig} includes everything inside the outer tessellate boundary (red outline with the approximate shape of the state of Texas) shown in Figure~\ref{tessellates.fig}, but excluding the very bright N157B supernova remnant and its pulsar wind nebula (see the previous subsection). The global boundary was chosen based on a constant level of soft-band, smoothed photon flux that enclosed the brightest diffuse emission associated with 30 Dor. This integrated diffuse X-ray emission spectrum has $1.7\times 10^6$ net counts.

The 8 brightest T-ReX point sources (those with \texttt{NetCounts\_t}$>10^4$ in Table~\ref{tbl:xray_properties}) total $2.55\times 10^6$ X-ray counts, however most of this contribution comes from outside the area enclosed by the global boundary, in the off-axis regions and PSR J0537. The fainter point sources in the 30 Dor field provide an order of magnitude fewer aggregate net counts, $2.2\times 10^5$, or 13\% of the global diffuse counts.\footnote{T-ReX has enabled a significant improvement in faint point-source extraction compared to our early analysis of the first, shallow observations of 30 Dor, for which extracted point sources represented 9\% of the integrated diffuse emission \citep{Townsley06a,Townsley06b}.} The spectral model component representing unresolved T Tauri stars represents ${<}1\%$ of the diffuse emission (Figure~\ref{globalspectrum.fig}). 

The fits to the global spectrum in Figure~\ref{globalspectrum.fig} included 8 Gaussians with energies between 0.5 and 2~keV, some of which could plausibly represent CX-enhanced line emission and several of which were also used in the fit to the spectrum of N157B (see Figure~\ref{N157Bspectra.fig}). Absent from the global spectrum are Gaussian components at the energies of the magnesium lines, seen in both the SNR and pulsar wind nebula spectrum. One new Gaussian component appears at 1.90~keV and could represent the \ion{Si}{13} emission observed in starburst galaxies \citep{Ranalli08,Katsuda11}.

\subsection{Sample Tessellate Spectral Fits \label{sec:tess}}
   
\begin{figure}[htb]
\centering
\includegraphics[width=0.95\textwidth]{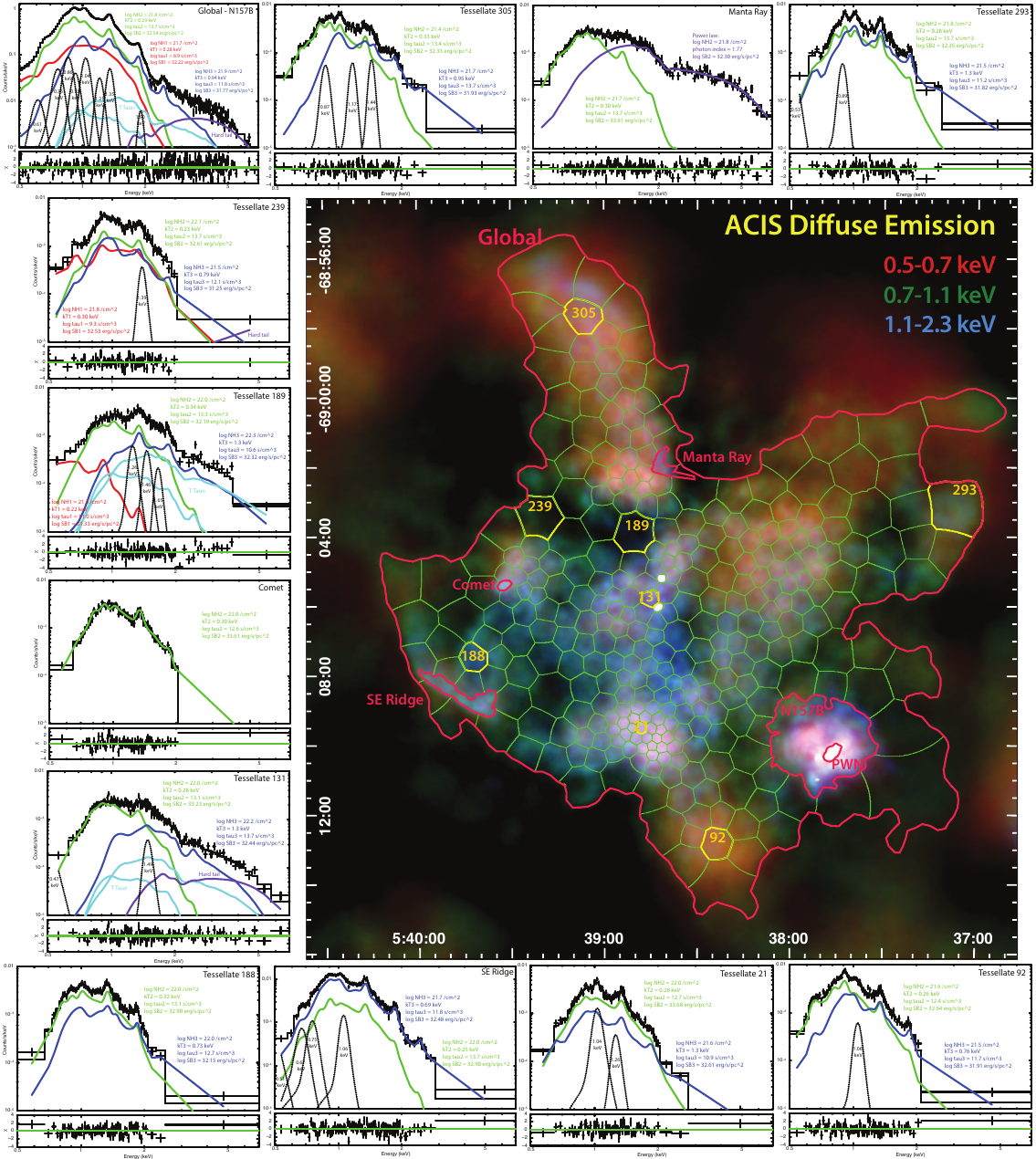}
\caption{
Tesselation used for spatially-resolved spectral fitting analysis of the ACIS diffuse emission. Spectral fits to 11 sample extractions representing a range of locations and regions of interest (labeled in red and yellow in the soft diffuse image) illustrate the spectral complexity and diversity of the diffuse X-ray emission in 30~Dor.  The outer boundary defines the ``global'' diffuse spectrum (Figure~\ref{globalspectrum.fig}), plotted in the upper-left panel to facilitate comparison with the smaller extraction regions.  Global and the other named regions outlined in red are based on surface brightness contours: N157B and PWN (Figure~\ref{N157Bspectra.fig}), Manta Ray, Comet, and SE Ridge.  The 308 tessellates (in green and yellow) come from the {\em WVT Binning} code by \citet{Diehl06}.  
\label{tessellates.fig}}
\end{figure}


Employing the strategy developed for CCCP \citep{Townsley11b}, we subdivided the 30 Dor diffuse emission into tessellates using the {\it WVT binning} algorithm \citep{Diehl06}. This method allows for higher spatial resolution in regions of brighter emission while maintaining a constant SNR across all extracted spectra.
Using SNR=70, we needed 308 tessellates to tile the field enclosed by the ``Global'' contour (Figure~\ref{tessellates.fig}).
The starting point for spectral fits to each extracted tessellate spectrum was the global model with no Gaussian components.  Unneeded components were then removed, fits optimized, and Gaussians added last if they improved the fit. 

Many tessellates (e.g. 188, and, with the addition of Gaussian lines, 305, 293, 92, 21, and SE Ridge) can be modeled with just {\em vpshock2}, a soft CIE plasma that usually dominates the surface brightness, and {\em vpshock3}, a harder plasma that may be NEI or CIE.  Of course the absorptions, temperatures, and timescales of these plasmas vary across the complex, creating 30 Dor’s rich X-ray appearance. 

The soft {\em vpshock1} component usually has the same temperature as {\em vpshock2} but a much shorter ionization timescale, implying an NEI plasma.  Thus tessellates requiring {\em vpshock1} may indicate the presence of a plasma transitioning from NEI to CIE. This component is needed only in  tesselates 189 and 239, which have the lowest apparent surface brightnesses of all sample tessellates due to high absorption north of R136. Tesselate 189 additionally required the brightest unresolved TTS components among the sample tesselates, representing the embedded stellar population north of R136 \citep[][]{Sabbi16}.

Tesselate 131 provides a sample of the diffuse plasma nearest to R136. Unsurprisingly, it includes relatively strong components modeling the unresolved T Tauri population. The absorption to all components is higher than the Global average, and the diffuse spectrum here is significantly harder, requiring both a CIE thermal plasma with $kT3 = 1.3$~keV and an hard tail. Tesselates 21 and 189, the other two samples of the central regions of 30 Dor (Figure~\ref{tessellates.fig}), have hot thermal plasma components with the same temperature but shorter timescales, implying NEI.

 Comet and SE Ridge are names given to two hand-drawn tesselate boundaries encircling areas of unusually bright, apparently hard diffuse emission.  Comet was the only sample tessellate fit with single CIE plasma component, while SE Ridge required two CIE plasmas and several Gaussian lines. The hotter $kT3 = 0.69$~keV component dominates SE Ridge more than in most other regions. Both tesselates show higher absorption than the Global average, suggesting regions where hot plasma encounters cold gas at the western edge of the 30 Dor superbubble.

If Comet and SE Ridge represent regions of plasma confinement and possible mixing with the cold ISM, sample tesselates 305 and 92 appear to sample regions at the northern and southern extremes, respectively, where there could be champagne flows of diffuse plasma leaking from 30 Dor. The X-ray flux appears softer in these lobes (Figures \ref{3soft_30dor.fig} and \ref{tessellates.fig}), and indeed the absorption is lower here compared to central regions and to the Global average. Figures~\ref{interfaces.fig}a and b show that the X-ray plasma in these regions fills cavities outlined by PAH emission.

\clearpage

\begin{figure}[htb]
\centering
\includegraphics[width=0.95\textwidth]{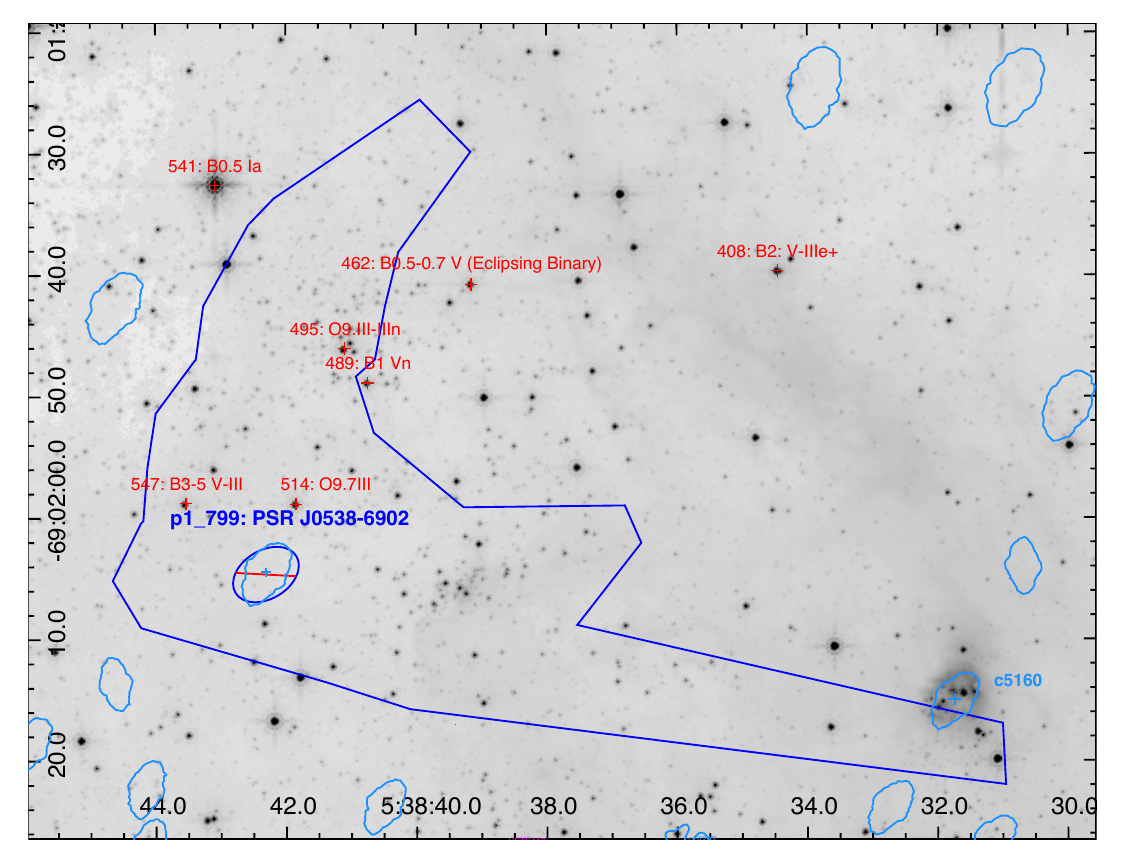}
\caption{
 HTTP F160W image \citep{Sabbi13} showing the stellar content in the region of the Manta Ray diffuse extraction tesselate (blue outline). Three stellar overdensities are visible along the trailing edges of the Manta Ray ``wings;'' the one at lower-right also shows NIR nebulosity suggesting an embedded cluster associated with the T-ReX point source c5160. The candidate X-ray pulsar (PSR J0538-6902 = T-ReX catalog source p1\_799) is marked with a blue + sign enclosed by red-blue ellipse showing mask used to exclude the source from the diffuse emission. OB stars from VFTS \citep{Walborn14,Evans15} are marked with red + symbols with spectral types labeled; the majority are evolved massive stars. Extraction apertures for regular T-ReX catalog sources are outlined with blue polygons (occasional sources are omitted).
\label{mantaray_http.fig}}
\end{figure}

\subsection{The Manta Ray: A New Candidate Pulsar Wind Nebula}

The Manta Ray (named for its morphological resemblance to the large, flat-bodied, winged fish that ``flies'' under the sea) is an unusually hard, extended X-ray source located 4\arcmin\ north of R136.  This feature is visible as a green ``V'' shape in the event data displayed in Figure~\ref{acisevents2.fig}. It required a custom, hand-drawn  aperture to extract its diffuse emission. We fit the spectrum of the Manta Ray using a similar model as with N157B and its PWN (see Figure~\ref{N157Bspectra.fig}) and found that it needed only two components (Figure~\ref{tessellates.fig}). The soft portion of the spectrum was fit with a CIE plasma with $kT = 0.30$, $\log{N_H/{\rm cm^{-2}}}=21.7$, and $\log{\rm SB/(erg~s^{-1}~pc^{-2})} = 33.61$. The hard spectrum required a power-law with $\alpha_{\rm PWN}=1.77$, $\log{N_H/{\rm cm^{-2}}}=21.8$, and $\log{\rm SB/(erg~s^{-1}~pc^{-2})} = 32.30$. 

We used a power-law with LMC absorption ({\it TBvarabs*pow}) to fit the spectrum of the T-ReX point source p1\_799 (585 net counts), which is located near the center of the Manta Ray but has no apparent counterpart at visible/NIR wavelengths (Figure~\ref{mantaray_http.fig}). The spectral fits returned $N_{H}=2.71_{-0.9}^{+1.1}\times 10^{22}$~cm$^{-2}$ (for $Z=0.4$), power-law index $\alpha_{\rm PSR}=1.5 \pm 0.3$, and $\log{L_{tc}} = 33.42$ (0.5--8~keV total-band luminosity). 
Comparing the power-law indices of the Manta Ray diffuse emission ($\alpha_{\rm PWN}$) with its point source ($\alpha_{\rm PSR}$) reveals very good agreement with the empirical relationship between the spectra of young X-ray pulsars and their wind nebulae derived by \citet{Gotthelf03}. These $\alpha$ values imply a spin-down luminosity of $L_{\rm sd} \sim 0.8$--$1.5\times 10^{37}$~erg~s$^{-1}$ for the driving pulsar. Based on the broad correlation between $L_X$ and $L_{\rm SD}$ reported by \citet[][their Fig.~2]{Possenti02}, we would expect $L_X\sim 5\times 10^{33}$~erg~s$^{-1}$ for the pulsar, in very good agreement with the absorption-corrected $L_{tc}$ for p1\_799. We therefore propose that p1\_799 be designated PSR~J0538-6902, the candidate pulsar responsible for generating the Manta Ray PWN.

Additional lines of circumstantial evidence also support the conclusion that the Manta Ray is a PWN. Since the diffuse absorption was modeled assuming solar metallicity using {\it TBabs}, doubling its value yields $N_H=1.26\times 10^{22}$~cm$^{-2}$ for the power-law nebular component at LMC metallicity, which is similar to the absorption of the point source PSR J0538-6902. There are several evolved, massive stars and three compact star clusters in the vicinity of the Manta Ray (Figure~\ref{mantaray_http.fig}), which could trace a massive stellar association that hosted the progenitor of the pulsar. The arc-shaped morphology of the PWN suggests a bow shock resulting from a high space velocity for the pulsar in an easterly direction. Assuming, for example, that the progenitor resided in the nearest HTTP cluster (${\sim}15\arcsec=3.6$~pc away), and also assuming a typical neutron star birth velocity of ${\sim}200$~km/s \citep{Sartore10}, PSR~J0538-6902 could have reached its present-day location ${\sim}20,000$~yr post-supernova. An age of a few $\times 10^4$ would be consistent with the X-ray luminosity of PSR~J0538-6902 \citep{Possenti02}, making it ${\sim}10$ times older and hence ${\sim}1000$ times less luminous than PSR J0537-6910 in N157B \citep{Chen06}.

   \subsection{Physical Properties of the Global Diffuse X-ray Emission in 30 Dor \label{sec:global}}

\input{plasma_properties.tex}

To estimate the physical conditions of the diffuse X-ray plasma within 30 Dor, we follow the approach used in previous work for M17 \citep{Townsley03} and the Carina \citep{Townsley11b} to produce Table~\ref{tbl:physics}.  
These three massive star-forming complexes span more than an order of magnitude in bolometric luminosity and physical size. Both 30 Dor and Carina clearly host multiple generations of massive star formation over the past ${\sim}10$~Myr \citep{Povich19,HTTP2}. M17 is the least luminous and most compact of the three, with the youngest rich cluster, the 0.5-Myr-old NGC 6618---but here, too there is evidence for a more evolved and spatially distributed massive stellar population \citep{Povich09,Getman14}.

Compared to previous work on much shallower \Chandra\ observations, our Global model now includes a softer (0.28 keV) NEI component  in 30 Dor, similar to both M17 and Carina \citep{Townsley11c}. The hardest, $kT3$ component is hotter in 30 Dor (0.94~keV) than in M17 or Carina (both 0.6~keV). Both 30 Dor and Carina have similar X-ray plasma densities ($n_{e,x}\sim 10^{-2}$~cm$^{-3}$) and pressures ($P_x/k\sim 10^5$~K~cm$^{-3}$), while the corresponding densities and pressures in M17 are higher by an order of magnitude. This difference seems to indicate that the hot plasma is more confined in M17 compared to the two larger, more evolved complexes.

 Because both the three-dimensional geometry and filling factor of the X-ray emitting plasma are unknown, we introduce a correction factor $\eta$ that can range from zero to unity that applies to all physical quantities related to the volume of the hot plasma (see Table~\ref{tbl:physics}).
The assumed volume ($\eta V_x$) of the hot plasma region in 30 Dor is about 50 times greater than that of Carina, while the total-band, absorption-corrected luminosity ($L_{tc}$) is ${\sim}120$ times greater \citep{Townsley11b}. 
 Both the total energy and mass ($E_x$ and $M_x$) of the hot plasma scale as $(L_{tc}\eta V_x)^{1/2}$. From \citet{Townsley03}, $L_{tc}=\Lambda n_{e,x}^2\eta V_x$, where $\Lambda=5\times 10^{-23}$~erg~cm$^{3}$~s$^{-1}$ is the bolometric X-ray emissivity \citep{LandiLandini99}. Then the total thermal energy can be written as
\begin{equation}
    E_x = \frac{3}{2}P_x\eta V_x \propto n_{e,x}T_x \eta V_x
    \propto \left(\frac{L_{tc}}{\Lambda \eta V_x} \right)^{1/2}\eta V_x T_x = (L_{tc}\eta V_x/\Lambda)^{1/2}T_x. 
\end{equation}
The total mass $M_x = \mu_x m_p n_{e}\eta V_x$ (denoting the mean particle mass as $\mu_x m_p$) has the same dependence on number density and emitting volume as the total energy, and hence also scales with luminosity as shown in Equation 2.
Based on the above scaling relations and the very similar observed plasma temperatures,  we would expect the total energy and mass of X-ray emitting plasam to be roughly 10 times greater in 30 Dor than in Carina, but instead we find they are 65--75 times greater. 

We can also compare the luminosity of the diffuse X-ray emission to the combined bolometric lumnosity of the massive stars in each region. For 30 Dor, $L_{\rm bol}=3\times 10^{41}$ \citep{Lopez11} yields $\log{L_{tc}/L_{\rm bol}}=-4.15$. Adopting $L_{tc}$ from \citet{Townsley11c} and $L_{\rm bol}$ from \citep{BP18}, we have $\log{L_{tc}/L_{\rm bol}}=-5.70$, $-$5.52, and $-$6.04 for Carina, NGC~3603, and M17, respectively, in the Milky Way. 

Based on the above simple comparisons, the superlative mass, energy, and luminosity of the X-ray-emitting plasma in 30 Dor require  enhanced mass-loss and high-energy feedback for this starburst region, exceeding expectations for scaled-up versions of the Galactic regions.\footnote{While a smaller filling factor $\eta$ for the 30 Dor plasma compared to the Galactic regions is plausible, its value would need to be vanishingly small to explain the differences in mass and energy in these comparisons, and $\eta$ does not factor into the luminosity calculation at all.} Multiple supernova explosions have likely contributed to the diffuse plasma. A SN shock propagating at $v_{\rm shock}{>}100$~km~s$^{-1}$ \citep[e.g.,][]{Chen23-30DorB} would cross the diameter of the T-ReX field in ${<}0.5$~Myr, and we have already discussed the abundance of evidence for recent SNe in the form of pulsars (+wind nebulae such as the Manta Ray) associated with evolved massive clusters. 

The brightest diffuse emission, however, is centered on R136, which is too young to have experienced SNe. This points to a wind origin, driven by the higher mass-loss rates of the numerous Wolf-Rayet and other supermassive stars in this extraordinarily rich cluster \citep[][C22]{NL00,VdK05}. 
Example tesselates in outer regions of 30 Dor that could plausibly be dominated by SN-driven bubbles (92, 293, and 305; see Figure~\ref{tessellates.fig}) have 15\%--25\% the absorption-corrected surface brightness of tessellate 131, which samples diffuse plasma closest to R136, hence 70--85\% of the diffuse X-ray luminosity in the central regions could be wind-generated.

\section{Legacy Value of T--ReX}\label{sec:legacy}
In this overview of the T-ReX project, we have provided numerous data products, including a point-source catalog, diffuse emission images and spatially-resolved spectra (Section~\ref{sec:repository}). The analysis presented in this work only scratches the surface of this unprecedentedly deep and rich X-ray dataset.
T-ReX provides our best opportunity to dissect the high-energy emission from a starburst region. 
These galactic building blocks all emit X-rays, but they blur together in distant starburst galaxies.  With T-ReX, we have quantified the relative contributions of the diverse sources that constitute the X-ray emission of a starburst.  The stars themselves (single very massive stars, colliding-wind binaries, and lower-mass pre-MS stars) and their stellar remnants (pulsars, X-ray binaries) are generally resolved and hence can be separated from truly diffuse shock emission (from massive stellar winds, individual young SNRs, thermalized cavity SNRs, and PWNe). T-ReX provides the largest number of diffuse X-ray emission counts ever detected from a massive star-forming region, surpassing Carina \citep{Townsley11c} and Cygnus OB2 \citep{CygOB2Xdiffuse}.

A rich history of multiwavelength studies leverages 30 Dor as a bridge from Galactic to extragalactic studies. T-ReX emerges as a singularly powerful dataset for study of diffuse emission and feedback, but it has also revealed the limits of current X-ray instrumentation in studying resolved stellar populations beyond the Milky Way.  While point sources become fainter with distance, X-ray surface brightness is a conserved quantity, and we have shown that the exceptionally bright diffuse emission in 30 Dor adversely affects the point-source sensitivity limits, which was not the case in Carina and other Galactic \GHIIR s \citep{Townsley11c}. Even \Chandra's unmatched spatial resolution is no match for the level of crowding and source confusion at the center of the R136 cluster (see also C22). 

T-ReX will remain the best dataset available for the foreseeable future for the analysis of diffuse X-ray emission in active star-forming regions, including evidence for charge exchange processes at hot/cold ISM interfaces. Gratings spectroscopy with current instruments, while successfully applied to young Galactic supernova remnants and more distant starburst galaxies, cannot probe the lower surface brightness of plasmas generated by stellar winds and cavity supernovae. There is need for still better spatial resolution to resolve out contributions from bright point sources that outshine the diffuse emission in more distant galaxies. Further analysis of the T-ReX data should map the variations in X-ray absorption-corrected surface brightness, plasma temperature, density, pressure, and metallicity across 30~Dor, better quantifying the effects of feedback from massive stars on the evolution of star-forming regions and their host galaxies.

\begin{acknowledgements} 
We appreciate the time and effort donated by our anonymous referee to provide helpful and supportive comments on this paper.
We thank Breanna Binder and Anna Rosen for helpful discussions, offering their perspectives as, respectively, an extragalactic high-energy observer and a theoretician of massive stellar feedback processes.

This work was supported by the {\em Chandra X-ray Observatory} Guest Observer grants
GO5-6080X (PI:  L.\ Townsley)    
and 
GO4-15131X (PI:  L.\ Townsley)   
and by the Penn State ACIS Instrument Team Contract SV4-74018.  
Both of these were issued by the \Chandra\ X-ray Center, which is operated by the Smithsonian Astrophysical Observatory for and on behalf of NASA under contract NAS8-03060.
M. Povich acknowledges support from the National Science Foundation under award AST-2108349 (RUI).
The ACIS Guaranteed Time Observations included here were selected by the ACIS Instrument Principal Investigator, Gordon P.\ Garmire, of the Huntingdon Institute for X-ray Astronomy, LLC, which is under contract to the Smithsonian Astrophysical Observatory; Contract SV2-82024.

This research used data products from the \Chandra\ Data Archive, software provided by the \Chandra\ X-ray Center in the application package {\it CIAO}, and the {\it SAOImage DS9} software developed by the Smithsonian Astrophysical Observatory.  

This research also used data products from the {\em Spitzer Space Telescope}, operated by the Jet Propulsion Laboratory (California Institute of Technology) 
under a contract with NASA.

This work used data products from the NASA/ESA Hubble Space Telescope, obtained at the Space Telescope Science Institute, which is operated by AURA Inc., under NASA contract NAS 5-26555.

This work also used data from the ESA mission {\it Gaia} (\url{https://www.cosmos.esa.int/gaia}), processed by the {\it Gaia} Data Processing and Analysis Consortium (DPAC,
\url{https://www.cosmos.esa.int/web/gaia/dpac/consortium}); funding for the DPAC has been provided by national institutions, in particular the institutions participating in the {\it Gaia} Multilateral Agreement.  

This research has also made use of NASA's Astrophysics Data System Bibliographic Services, and the SIMBAD and VizieR databases operated at the Centre de Donn\'ees Astronomique de Strasbourg. 
\end{acknowledgements}

\vspace{5mm}
\facilities{CXO (ACIS), 
Gaia,
ESO:VISTA (VIRCAM),
VLT:Kueyen (FLAMES-ARGUS),
HST (WFC3, ACS),
Spitzer (IRAC, MIPS), WISE 
Herschel (SPIRE)}.

\software{
        {\em ACIS Extract} \citep{Broos10,AE12,AE16},
        \CIAO\ \citep{Fruscione06},
        \DSnine\ \citep{Joye03},
        \MARX\ \citep{Davis12}
        \XSPEC\ \citep{Arnaud96}.
}

\appendix

\section{Possible Optical/IR Counterparts to X-ray Point Sources}\label{app:counterparts}

Identifying optical-infrared (OIR) counterparts to \Chandra\ sources (``matching'' X-ray and OIR catalogs) is often a challenging task.
The dynamic range of flux sensitivity for any OIR catalog will typically be narrower than the apparent OIR fluxes of the objects detected by \Chandra. For example, when the absorbing column and astrophysical classification of an X-ray source are unknown, its apparent magnitude could span a range ${\ga}15$ in $J$ \citep{Getman11,Broos13}. 
The difference in spatial resolution of an OIR catalog compared to the ${\sim}0.5\arcsec$ on-axis resolution of ACIS complicates the notion of a simple, one-to-one match, as in cases of large resolution departures (beyond a factor of 2) a single source in the lower-resolution catalog will frequently overlap with multiple sources in the other.
The population of X-ray sources without detected OIR counterparts can significantly degrade the reliability of most X-ray/OIR matching algorithms, because every such X-ray source is an opportunity for generating a spurious match declaration (arising from the random proximity of unrelated astrophysical objects in the two catalogs).

When matching a \Chandra\ catalog to an OIR catalog, we use simulations to estimate the fraction of \Chandra\ sources that have true counterparts and to statistically characterize the performance of the matching run \citep{Broos07,Broos11}.
One useful product of those simulations is a quality metric for the run---the fraction of declared matches expected to be spurious.
That match quality depends on the fraction of X-ray sources with true counterparts, on the size of the \Chandra\ and OIR position uncertainties, and on the density of the catalogs.

We identified possible counterparts in the six OIR catalogs listed in Table~\ref{tbl:OIR_catalogs}:
\begin{enumerate}
\item {The Gaia catalog reports optical sources across the entire T-ReX field to a limiting magnitude of $G<21$, with effective spatial resolution comparable to ACIS \citep{Gaia18}.}

\item {The VISTA Magellanic Cloud survey catalog\citep[VMC;]{Irwin04,Hambly08,Cross12}
reports near-infrared  sources across the entire T-ReX field to limiting magnitudes of $Y = 21.1$, $J = 21.3$, and $K_S=20.7$ with comparable spatial resolution to ACIS on-axis. We have used data from the 3rd data release.}

\item {The VLT-FLAMES Tarantula Survey (VFTS) catalog \citep{Evans11} is a list of 930 candidate massive stars, many subsequently confirmed via spectroscopic classification in the 30 Doradus region of the Large Magellanic Cloud \citep{Walborn14,Evans15}.}

\item {\citet{Doran13} published a comprehensive list of known and candidate massive stars in the 30 Doradus region, with cross-references to the VFTS catalog.}

\item {The {\it Hubble} Tarantula Treasury Project (HTTP) catalog reports point-source photometry in eight visible through NIR filters for a subset of the T-ReX field \citep{Sabbi16}. Each filter reaches a sensitivity of 25th magnitude or better, with bright-source limits ranging between magnitudes 15 and 17. The spatial resolution of {\it HST} is 5 to 10 times sharper than ACIS on-axis.}

\item {The {\it Spitzer} Surveying the Agents of Galaxy Evolution \citep[SAGE][]{SAGE} program cataloged mid-infrared point sources (in five filters ranging from 3/6 to 24 $\mu$m) across the LMC, including the entire T-ReX field. Point-source sensitivity ranged from $[3.6] \la 18$ to $[24]\la 11$, with saturation limiting the bright sources to ${\ga}4$ to 6 magnitudes. The angular resolution of {\it Spitzer} at 3.6~$\mu$m was about five times poorer than ACIS on-axis.}
\end{enumerate}

We performed simple experiments by which we introduced small, random positional offsets between the T-ReX catalog and OIR catalogs to assess the fraction of longer-wavelength counterparts consistent with random chance. These simulations demonstrated that within the T-ReX field the fraction of \Chandra\ sources with true, one-to-one counterparts in the OIR catalogs is very low, hence producing a counterpart list with a high level of quality is very challenging.
We employed two important strategies for improving  counterpart quality.

The first strategy was to improve the published positions in the OIR catalogs.
As shown in Table~\ref{tbl:OIR_catalogs}, we shifted the coordinate frames of the VMC, SAGE, and \Chandra\ catalogs to match the Gaia frame.\footnote
{
Gaia positions were propagated to the mean epoch of the \Chandra\ data (Y2014.5)
}.
Each subgroup of massive star coordinates we obtained from \citet{Doran13}
was shifted to optimally align with the Gaia frame.
Individual source positions in the VFTS and HTTP catalogs were significantly improved by constructing field distortion maps.
%
Single-axis ($\alpha,\delta$) position uncertainties (column PosErr in Table~\ref{tbl:OIR_catalogs}) were estimated for the VMC, VFTS, Doran, and HTTP catalogs by examining the distribution of offsets along each coordinate axis between those catalogs and Gaia.

The second counterpart quality improvement strategy was to reduce the density of the HTTP and \Chandra\ catalogs.
Among the X-ray emitting objects likely to be detected by HTTP, those of most interest to us are massive stars and T-Tauri stars.
Thus, we made two photometry selections designed to select those objects: a magnitude cut ($F555 < 19$) to select massive stars, and  a color-magnitude selection in the F555W vs F555W-F775W diagram approximating the locus of T-Tauri stars in Figure 20 of \citet{Ksoll18}.   
Those selections reduced the size of the HTTP catalog from ${>}8\times10^5$ to 29,445 rows.

We found that counterpart matching quality could also be significantly improved by trimming the \Chandra\ catalog using any of several X-ray properties, including detection significance, position uncertainty, off-axis angle, observed photon flux, and median energy of the extracted X-ray events. 
We chose to trim the \Chandra\ catalog independently for matching to each OIR catalog, using the X-ray detection significance thresholds reported in footnote $h$ of Table~\ref{tbl:OIR_catalogs}. 

\begin{deluxetable}{@{}lrrrcrrr@{}}
\tablecaption{Catalog Matching Summary \label{tbl:OIR_catalogs}}       
\tablehead{
 \multicolumn{2}{c}{Catalog} &
 \multicolumn{2}{c}{Position $(\alpha,\delta)$} &
 \multicolumn{2}{c}{Rows Retained} &
 \multicolumn{2}{c}{Matches} 
\\[-5pt]
 \multicolumn{2}{c}{\hrulefill} &
 \multicolumn{2}{c}{\hrulefill} &
 \multicolumn{2}{c}{\hrulefill} &
\\ 
 \colhead{Name} &
 \colhead{Citation} &
 \colhead{Shift} &
 \colhead{Error} &
 \colhead{OIR} &
 \colhead{\Chandra} & 
 \colhead{Gross} &
 \colhead{Correct} 
\\[-1pt]
 \colhead{} &
 \colhead{} &
 \colhead{mas} &
 \colhead{mas} &
 \colhead{} &
 \colhead{} &
 \colhead{matches} &
 \colhead{\% of gross}
\\
 \colhead{(1)} & \colhead{(2)} & \colhead{(3)} & \colhead{(4)} & \colhead{(5)} & \colhead{(6)} & \colhead{(7)} & \colhead{(8)} 
}
\startdata
Gaia      & \citet{Gaia16,Gaia18}   & \tnm{a} & published& all   &497\tnm{h,i}   & 114\tnm{k} & 78\%\tnm{k} \\
VMC       & \citet{Cioni11}         & (23,15)  & (41,44)   & all   & 497\tnm{h,i}   & 114\tnm{k}    & 78\% \tnm{k}    \\
VFTS      & \citet{Evans11}         &\tnm{b} & (15,13)   & all   & 1910\tnm{h}    &     $>$74 &\nodata  \\
Doran     & \citet{Doran13}         &\tnm{c} & \tnm{d}  &\tnm{f} & 1910\tnm{h}    &     $>$32 &\nodata  \\
HTTP      & \citet{Sabbi16}         &\tnm{b} & (2.7,4.3) &\tnm{g}& 1013\tnm{h,i,j} &       124 & 63\%  \\ 
SAGE      & \citet{SAGE}        & (64,17)  & (134,110) & all   & 851\tnm{h,i}   &        79 & 90\%  \\
\tableline
T-ReX & \nodata                 & (17,23)  & \tnm{e}  &\nodata&\nodata    &\nodata  &\nodata  
\enddata
\tablecomments{
Column ``Shift'' reports the shift applied to published coordinates to align with Gaia.\\
``Rows Retained'' reports the subset of the published catalog used for counterpart identification.\\
Column ``Gross'' reports the number of possible counterparts declared.\\
Column ``Correct'' reports the fraction of gross counterpart declarations that are likely to be true astrophysical counterparts, estimated via simulations.
}
\tablenotetext{a}{
Gaia positions were precessed to the mean epoch of the \Chandra\ data (Y2014.5).}
\tablenotetext{b}{
Field distortions in the coordinates reported by the HTTP and VFTS catalogs were reduced by constructing smooth field distortion maps derived from matching those catalogs to Gaia.
The median (maximum) HTTP correction was 67 (1200) mas.
The median (maximum) VFTS correction was 81 (160) mas.}
\tablenotetext{c}{
Each subgroup of massive star coordinates we obtained from \citet{Doran13}---originally from the CTIO, Parker, Selman, WFI, Walborn, Zaritsky, and De Marchi catalogs---was shifted to optimally align with the Gaia catalog.}
\tablenotetext{d}{
Position errors were estimated for each published catalog collated by \citet{Doran13}: 0.025\arcsec for dM, $\sim$0.1\arcsec for W, S, C, P catalogs.
}
\tablenotetext{e}{
For matching, a systematic error (0.08\arcsec) was added in quadrature to published \Chandra\ position errors.
}
\tablenotetext{f}{
Excluded Doran sources with a published VFTS identification \citep{Doran13}, and
the following Doran sources that we believe are missing VFTS cross-identifications:
VFTS416=Doran390,
VFTS562=Doran778,
VFTS1011=Doran533,
VFTS1024=Doran683,
VFTS641=Doran884,
VFTS522=Doran578,
VFTS468=Doran451.
}
\tablenotetext{g}{
For matching, we selected 29,445 candidate massive stars ($F555 < 19$ mag) and candidate T Tauri stars (a region of the F555W vs F555W-F775W color-magnitude diagram approximating the locus of T Tauri stars in Figure 20 of \citet{Ksoll18}).}
\tablenotetext{h}{
To achieve reasonable match quality (Column 8), matching used only high-significance X-ray detections defined by 
\texttt{ProbNoSrc\_MostValid} $< 1\times10^{-5}$ for Gaia and VMC,
\texttt{ProbNoSrc\_MostValid} $< 1\times10^{-3}$ for VFTS and Doran,
\texttt{ProbNoSrc\_MostValid} $< 4\times10^{-4}$ for HTTP,
\texttt{ProbNoSrc\_MostValid} $< 1.2\times10^{-4}$ for SAGE.
}
\tablenotetext{i}{
\Chandra\ sources declared to have VFTS/Doran counterparts were excluded from automated matching with Gaia, VMC, HTTP, and SAGE (Table~\ref{tbl:other_matches}).
}
\tablenotetext{j}{
\Chandra\ sources were cropped to the HTTP field of view.
}
\tablenotetext{k}{
\Chandra\ sources were matched to the union of the Gaia and VMC catalogs, rather than to each independently.
}
\end{deluxetable}


\subsection{Massive Stars}
Our list of known and candidate massive stars in the T-ReX field consists of the 930 bright stars in the VFTS catalog and the additional ${>}500$ massive stars that \citet{Doran13} collected from pre-existing catalogs.
Due to crowding in R136, we chose to identify X-ray counterparts in our list of known and candidate massive stars via visual review  and human judgement, rather than by algorithmic catalog matching.
In that visual review positions of OIR sources (VFTS, Doran, and HTTP positions, corrected as described above) and position error rectangles for our X-ray sources were displayed on an $HST$ image \citep{Sabbi16}.
In most cases, we were able to identify the VFTS/Doran object in the Gaia catalog and make our counterpart decision using the Gaia position for the massive star. The median separation between OIR (often Gaia) and X-ray positions was 0.15 arcsec.

Table~\ref{tbl:massive_matches} lists the known and candidate massive stars that we judge to have been detected as individual T-ReX point sources (with the significance cut shown in Table~\ref{tbl:OIR_catalogs}, $ProbNoSrc\_MostValid < 1\times10^{-3}$). Table~\ref{tbl:massive_blended} reports instances of close pairs or multiples of massive stars that were judged to be detected but unresolved by ACIS.
 These datasets were used by C22 for an in-depth study of X-ray emission from early-type stars in 30 Dor. 

An additional 37 \Chandra\ sources are located within 1--2\arcsec\ of a VFTS/Doran source, making it tempting to believe that these massive stars were also detected in X-rays. However, 
%
we judged that the separations between these X-ray sources and the nearby massive stars were inconsistent with their position uncertainties.
%

\subsection{Additional Counterparts}
Counterparts to X-ray sources that are not suspected to be massive stars were identified via automated catalog matching, using the Gaia, VMC, trimmed HTTP, and SAGE catalogs described above.
The matching criterion is defined such that, if the actual position uncertainties are as declared, then 99\% of the true counterparts will be identified.
Table~\ref{tbl:OIR_catalogs} reports the X-ray detection significance thresholds used to trim the X-ray catalog for each matching exercise, the gross number of matches declared, and the expected quality of the declared matches (fraction of declared expected to be spurious) estimated from simulations.
Table~\ref{tbl:other_matches} lists the declared matches.




    \include{massive_matches_stub}
    \include{massive_blended}

\include{other_counterparts_stub}

\section{High-Quality Variable Sources}\label{app:variability}

In this Appendix we briefly highlight several striking, high-quality variable sources. This is by no means an exhaustive list, and we would expect the T-ReX catalog to provide a resource for various future variability studies.

\subsection{Periodic Variation}

Binary stellar systems can exhibit periodic modulation of the unresolved X-ray emission from the system, due to time-varying accretion, eclipses, or time-variation in the physical state of colliding winds.
The long-baseline T-ReX observations (Table~\ref{tbl:obslog}) provide several high-quality examples of periodic variation.

\begin{figure}[htb]
\centering
\includegraphics[width=0.45\textwidth]{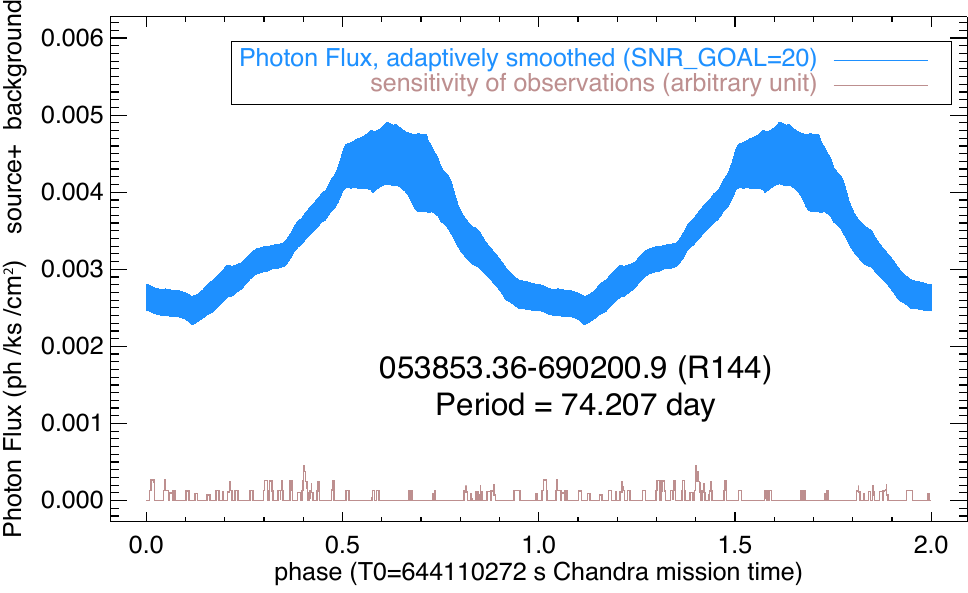}
\includegraphics[width=0.45\textwidth]{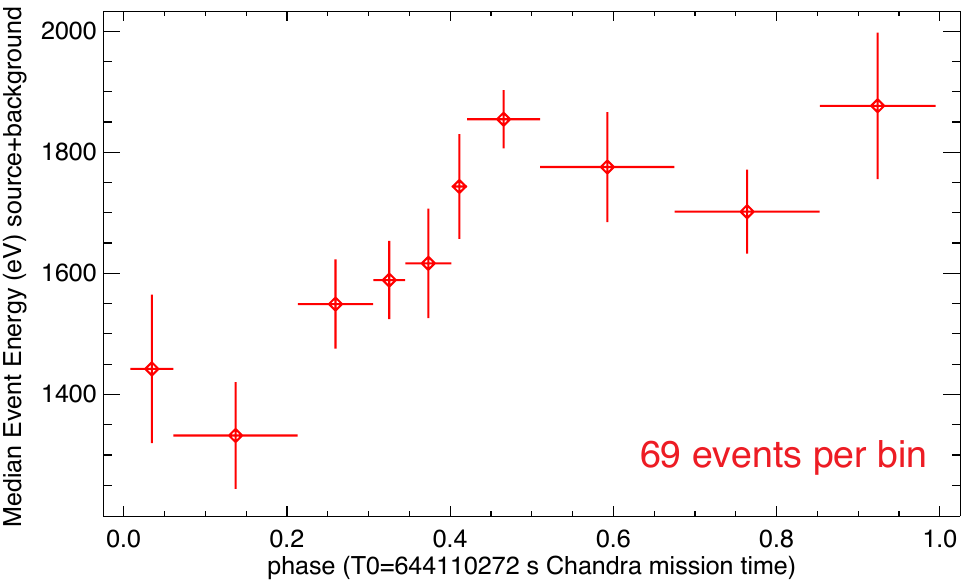}\\
\includegraphics[width=0.45\textwidth]{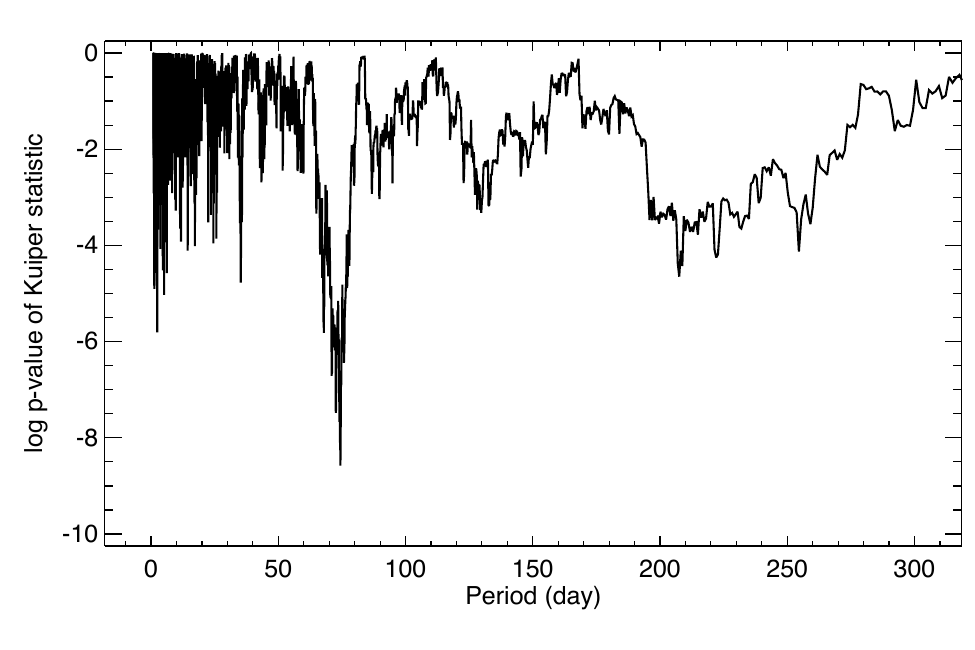}
\caption{Phase-dependent X-ray flux ({\it top left}) and median X-ray event energy ({\em top right}) for R144 (053853.36-690200.9, p1\_1194, Doran~916).
Phase is calculated for a published optical period (74.2074 day) and T0 \citep[periastron, MJD 58268.98;][]{Shenar21}.
A similar period estimate ($\sim$74.4 day) was obtained from the X-ray data ({\it bottom}).
\label{R144.fig}}
\end{figure}

The multipanel figure sets below (Figures~\ref{R144.fig}, \ref{R145(2).fig}, and \ref{R139.fig}) depict X-ray lightcurves (representing a 68\% confidence interval versus phase),
 phase-folded median X-ray energy, and periodograms.
The phase-folded folded lightcurves were constructed for the specified period by converting X-ray event timestamps (with barycentric correction) to phase values, calculating the phase distribution of those phased events, calculating the phase distribution of all the time periods in which the source was observed (using \CIAO's Good Time Interval data product), and then adaptively smoothing the event and exposure distributions to achieve a signal-to-noise goal.
The result is a smooth estimate of gross (star plus background) photon flux (in units of photon s$^{-1}~$cm$^{-2}$) with uncertainties. The median energy time-series  were constructed by simply binning the phased events, calculating the median energy of the events in each bin, and estimating uncertainties (employing a binomial distribution).


The periodograms plot $p$-values for the Kuiper statistic, under the null hypothesis of a constant flux, computed on the phased event distribution and the phased exposure distribution \citep{Paltani2004} for a large number of trial periods.
Sharp local minima in the periodogram suggest periodic variability, however interpretation requires caution because interactions between the observational sampling and aperiodic variability of the source can also produce small $p$-values.


\noindent \\ {\bf R144}\\
The binary Wolf-Rayet star R144 (053853.36-690200.9, p1\_1194, Doran~916) has a published optical period of 74.2074 day \citep{Shenar21}, which is consistent with the period we estimate from the T-ReX data ($\sim$74.4 day, Fig.~\ref{R144.fig}).
When folded on that optical period, R144 exhibits $\sim$50\% X-ray flux modulation \citep[][Fig.~17]{Shenar21} and clear modulation of the  apparent spectral hardness (Fig.~\ref{R144.fig}).

Before modeling the phase-resolved X-ray spectrum of R144,
we first identified a reasonable model for the complete spectrum (full ${\sim}2$~Ms exposure), consisting of two thermal plasmas with independent absorbers (in \XSPEC, the model {\it tbvarabs1*vapec1 + tbvarabs2*vapec2}).
That fit indicated plasma temperatures of 0.74 and 2.1 keV.
One possible interpretation of this model would be a soft plasma produced by each of the stars plus a hard plasma produced by a colliding wind shock.

%

\begin{figure}[htb]
\centering
\includegraphics[width=0.4\textwidth]{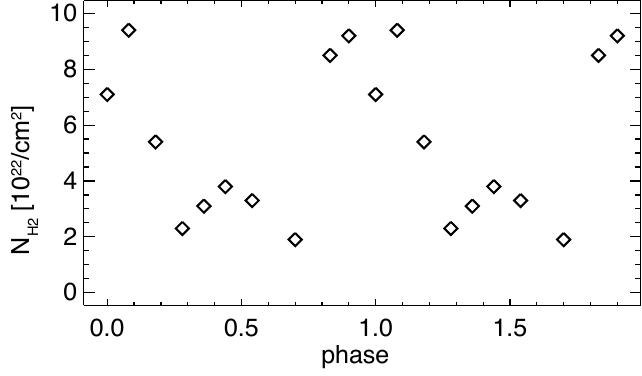}
\includegraphics[width=0.4\textwidth]{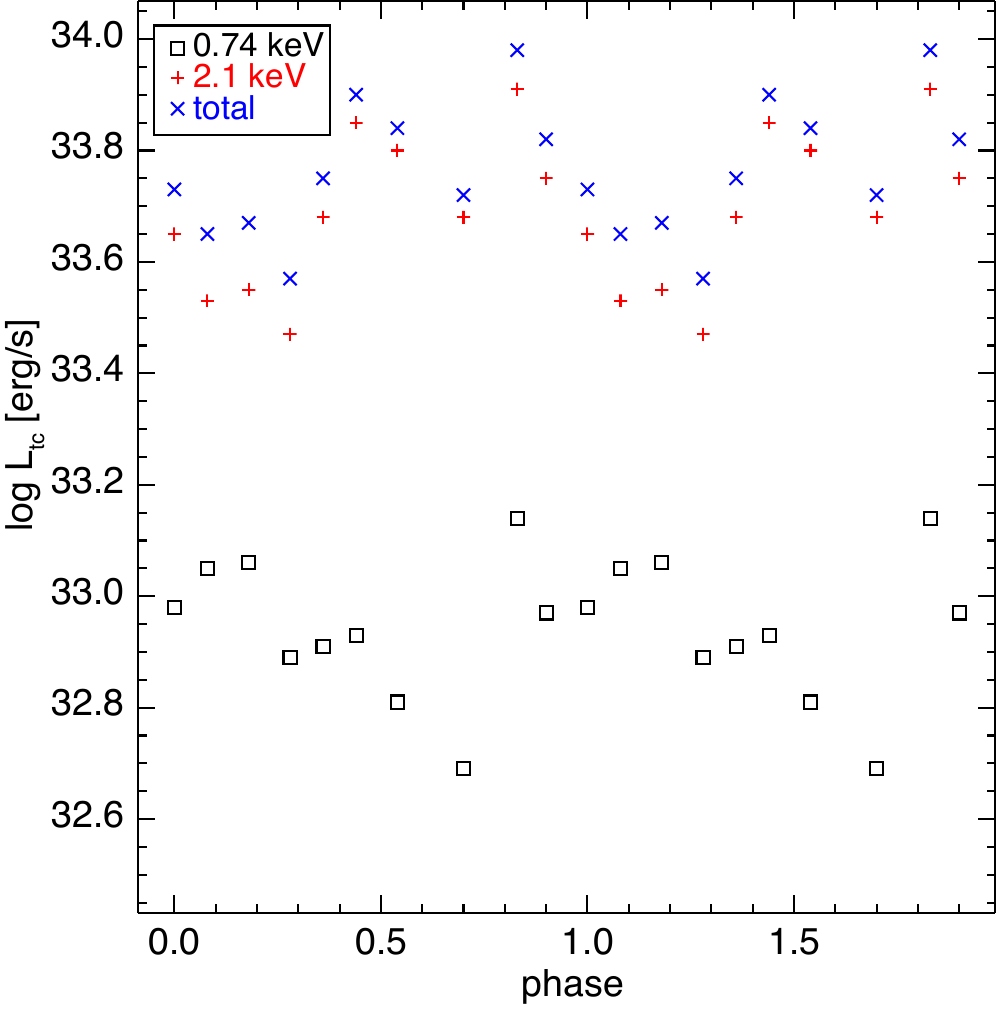}
\caption{Phase modulation of the hot plasma absorption (left) and  absorption-corrected luminosity (right) in a model of the R144 spectrum.
\label{R144_fitting.fig}}
\end{figure}

\begin{figure}[htb]
\centering
\includegraphics[width=0.45\textwidth]{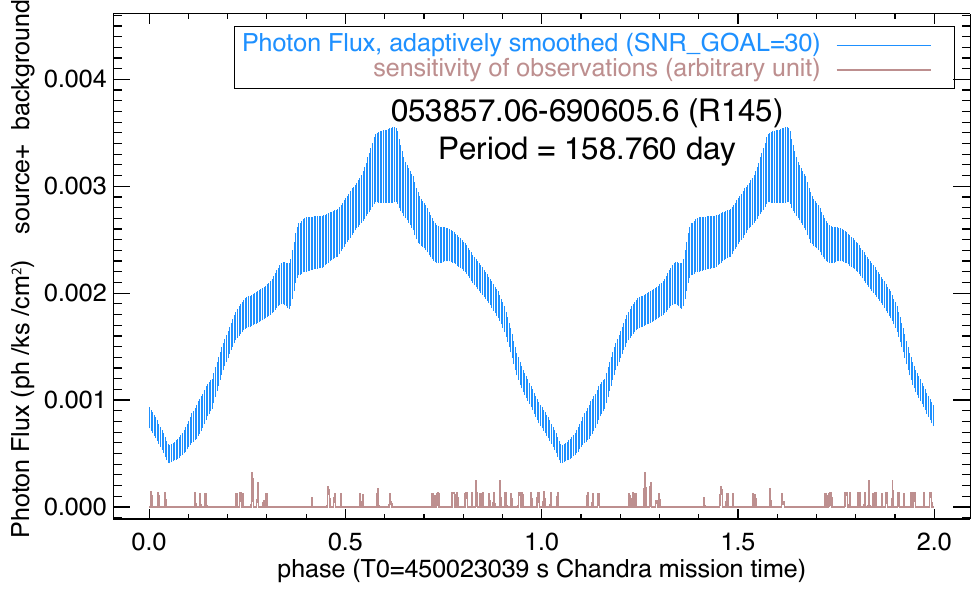}
\includegraphics[width=0.45\textwidth]{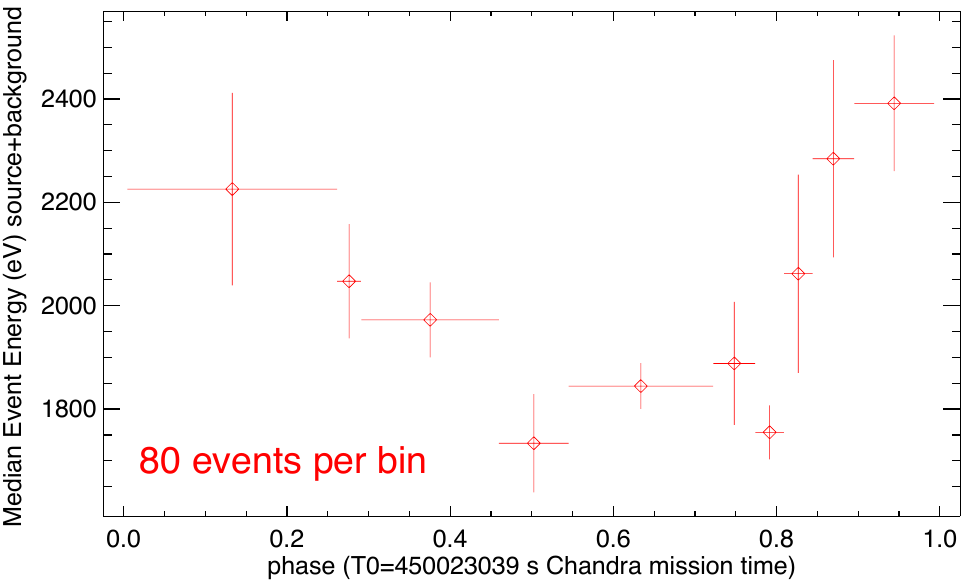}
\caption{Phase-dependent X-ray flux (left) and median X-ray event energy (right) for R145 (053857.06-690605.6, p1\_1256, VFTS~695).
Phase is calculated for a published optical period (158.76 day) and T0 (MJD 56022.6) \citep{Schnurr09,Shenar17}.
\label{R145(2).fig}}
\end{figure}

\begin{figure}[htb]
\centering
\includegraphics[width=0.45\textwidth]{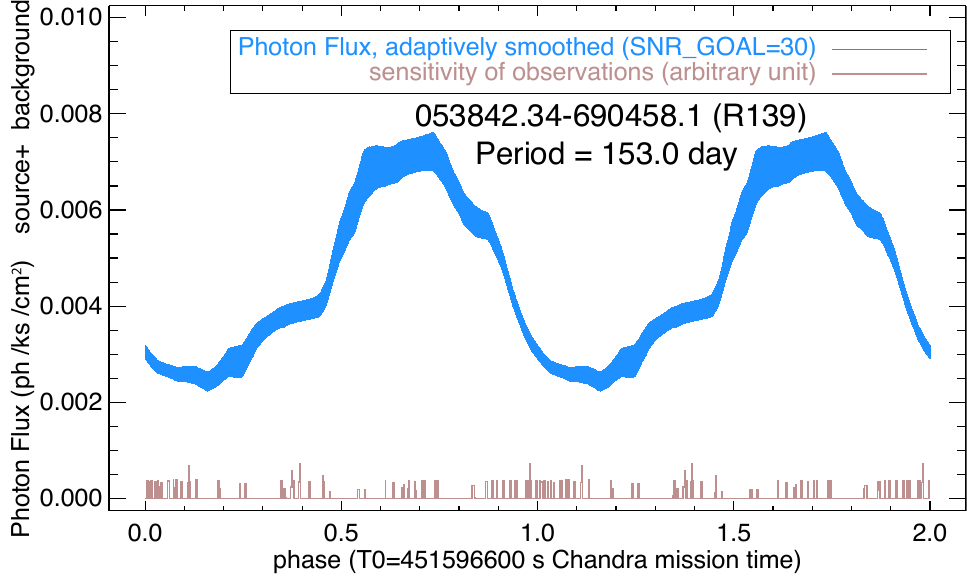}
\includegraphics[width=0.45\textwidth]{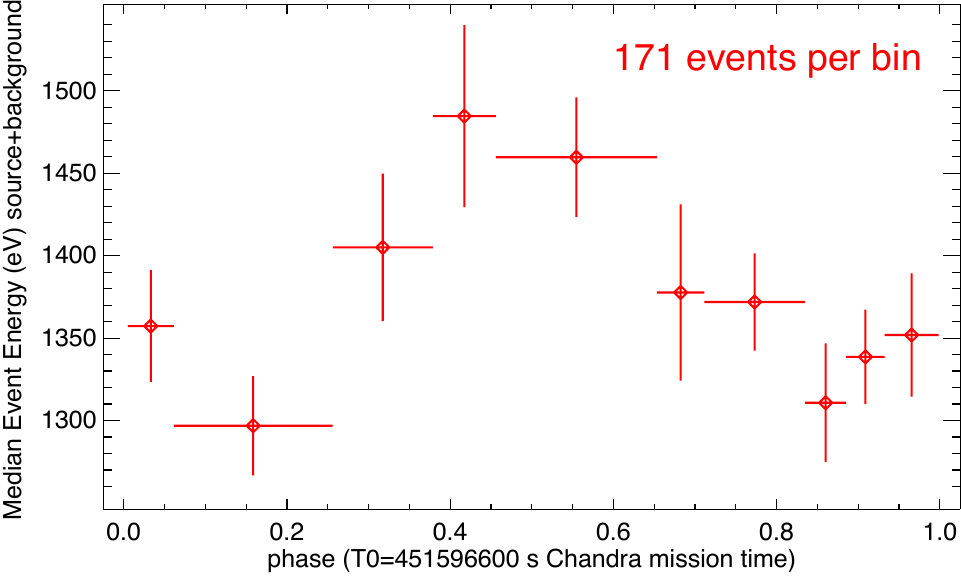}\\
\includegraphics[width=0.45\textwidth]{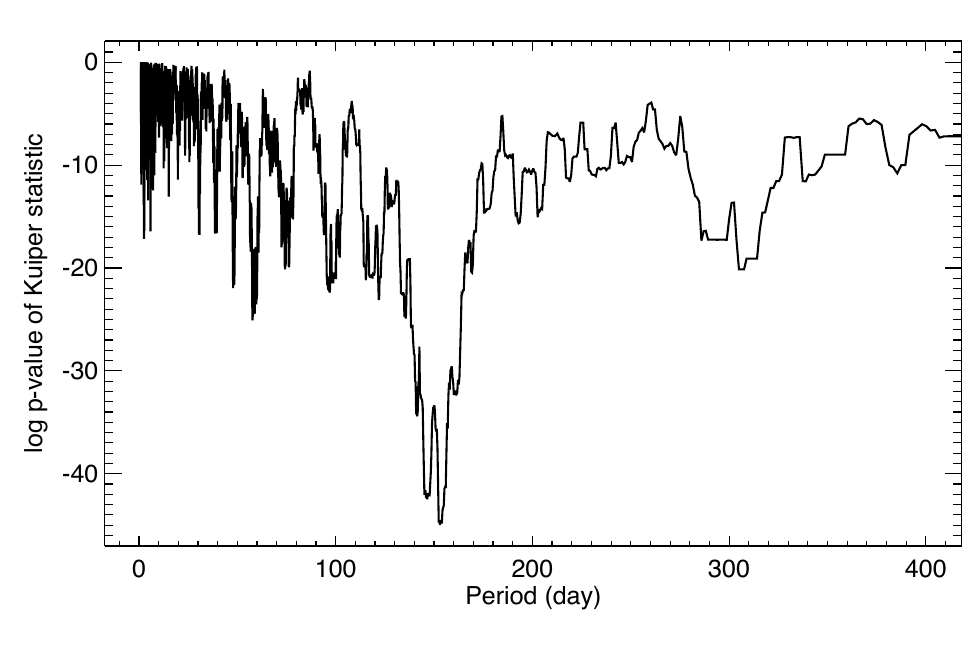}
\caption{Phase-dependent X-ray flux (left) and median X-ray event energy (right) for R139 (053842.34-690458.1, p1\_830, VFTS~527).
Phase is calculated for the period we estimate from the T-ReX data ($\sim$153.0 day, bottom).  T0 is derived from optical data (MJD 56041.3125, Hugues Sana, private communication). 
\label{R139.fig}}
\end{figure}

We then fit spectra obtained from ten groups of observations with similar phase (Table~\ref{tbl:R144_thermal_spectroscopy}). Each group contains 3--10 ObsIDs. We investigated the simple hypothesis that the observed variation was dominated by only one phase-dependent parameter, the absorption ($N_{H2}$) of the colliding wind shock emission ($kT_2$).
Specifically, each of the ten phase-resolved spectra was fit with $kT_1$,$kT_2$ and $N_{H1}$ frozen to the values obstained from the models of the complete spectrum.  $N_{H2}$ and the plasma normalizations remained free parameters.
Most of the phase-interval spectra were well-fit by this model (see $\chi_{\nu}^2$ in Table~\ref{tbl:R144_thermal_spectroscopy}).
Under these model assumptions, the phase modulations of $N_{H2}$ and the absorption-corrected luminosities are plotted in Figure~\ref{R144_fitting.fig}.


\setlength{\tabcolsep}{0.5mm}


 \begin{deluxetable}{@{}hhhlhhr@{\hspace{2em}}cccc@{\hspace{2em}}cccchhhhhhcc@{\hspace{2em}}cc@{\hspace{2em}}cchh@{}}   

\tabletypesize{\scriptsize} 

\tablecaption{R144 ACIS X-ray Spectroscopy for Ten Phase Intervals
\label{tbl:R144_thermal_spectroscopy}}

\tablehead{
\multicolumn{6}{c}{phase\tablenotemark{a}} &
\multicolumn{15}{c}{Spectral Fit\tablenotemark{b}} &   
\multicolumn{6}{c}{X-ray Luminosities\tablenotemark{c}} &
\nocolhead{} 
\\[-8pt] 
\multicolumn{6}{c}{\hrulefill} &
\multicolumn{15}{c}{\hrulefill} &    
\multicolumn{6}{c}{\hrulefill} \\
\nocolhead{\#} & \nocolhead{Label} & \nocolhead{CXOU J} & \colhead{} & \nocolhead{NetCts} & \nocolhead{SNR} &
\colhead{$\chi_{\nu}^2$} &
\colhead{$N_{H1}$} & \colhead{$A_{V1}$} & \colhead{$kT_1$} & \colhead{$EM1$} & \colhead{$N_{H2}$} & \colhead{$A_{V2}$} &\colhead{$kT_2$} & \colhead{$EM2$} &  
\nocolhead{O} & \nocolhead{Ne} & \nocolhead{Mg} & \nocolhead{Si} & \nocolhead{S} & \nocolhead{Fe} & 
%
\colhead{$L_{1t}$} & \colhead{$L_{1tc}$} &
\colhead{$L_{2t}$} & \colhead{$L_{2tc}$} &
\colhead{$L_t$}    & \colhead{$L_{tc}$}  & 
\colhead{}  \\
\colhead{} & \colhead{} & \colhead{} & \colhead{} & \nocolhead{(counts)} & \nocolhead{} & \nocolhead{} &
\colhead{($10^{22}$cm$^{-2}$)} & \colhead{(mag)} & \colhead{ (keV) } &  \colhead{(log cm$^{-3}$)} &  
\colhead{($10^{22}$cm$^{-2}$)} & \colhead{(mag)} & \colhead{ (keV) } &  \colhead{(log cm$^{-3}$)} &  
\multicolumn{6}{c}{} &                    
\multicolumn{6}{c}{(log erg s$^{-1}$)} &
\colhead{}
}

\startdata
&&&\multicolumn{10}{l}{R144, D=50.0 kpc}\\
   1 & p1\_1194 & 053853.36$-$690200.9 & {\bf p00} &    90 &     8.5 &  1.11 & $0.452\phd\!*$ &     0.7 & $0.74\phd\!*$ & $ 56.1\phd_{-0.3}^{+0.2}$   & {\boldmath $7.1\phd_{-3.4}^{+4.7}$} &    10.9 & $2.1\phd\!*$ & $ 56.7\phd_{-0.2}^{+0.2}$   & & & & & & &    32.73 &   32.98 &   33.14 &   33.65 &   33.28 &   33.73 &                grp1.4\_T+T:LMC,LMC &                        HandFit \\[1ex] 
   2 & p1\_1194 & 053853.36$-$690200.9 & {\bf p08} &    95 &     8.7 &  1.59 &$0.452\phd\!*$ &     0.7 & $0.74\phd\!*$ & $ 56.1\phd_{-0.2}^{+0.1}$   & {\boldmath $9.4\phd_{-3.9}^{+8.8}$} &    14.4 & $2.1\phd\!*$ & $ 56.6\phd_{-0.2}^{+0.2}$   & & & & & & &    32.80 &   33.05 &   32.95 &   33.53 &   33.18 &   33.65 &                grp1.5\_T+T:LMC,LMC &                        HandFit \\[1ex] 
   3 & p1\_1194 & 053853.36$-$690200.9 & {\bf p18} &   106 &     9.2 &  0.58 &$0.452\phd\!*$ &     0.7 & $0.74\phd\!*$ & $ 56.2\phd_{-0.2}^{+0.1}$   & {\boldmath $5.4\phd_{-3.1}^{+4.9}$} &     8.4 & $2.1\phd\!*$ & $ 56.6\phd_{-0.2}^{+0.2}$   & & & & & & &    32.81 &   33.06 &   33.08 &   33.55 &   33.27 &   33.67 &                grp1.6\_T+T:LMC,LMC &                        HandFit \\[1ex] 
   4 & p1\_1194 & 053853.36$-$690200.9 & {\bf p28} &   148 &    11.0 &  1.04 &$0.452\phd\!*$ &     0.7 & $0.74\phd\!*$ & $ 56.0\phd_{-0.5}^{+0.2}$   & {\boldmath $2.3\phd_{-1.0}^{+1.5}$} &     3.5 & $2.1\phd\!*$ & $ 56.6\phd_{-0.1}^{+0.1}$   & & & & & & &    32.63 &   32.89 &   33.15 &   33.47 &   33.27 &   33.57 &                grp2.0\_T+T:LMC,LMC &                        HandFit \\[1ex] 
   5 & p1\_1194 & 053853.36$-$690200.9 & {\bf p36} &   263 &    15.1 &  0.40 &$0.452\phd\!*$ &     0.7 & $0.74\phd\!*$ & $ 56.0\phd_{-0.3}^{+0.2}$   & {\boldmath $3.1\phd_{-0.9}^{+1.1}$} &     4.7 & $2.1\phd\!*$ & $ 56.8\phd_{-0.09}^{+0.08}$ & & & & & & &    32.66 &   32.91 &   33.32 &   33.68 &   33.40 &   33.75 &                grp3.0\_T+T:LMC,LMC &                        HandFit \\[1ex] 
   6 & p1\_1194 & 053853.36$-$690200.9 & {\bf p44} &   157 &    11.7 &  1.15 &$0.452\phd\!*$ &     0.7 & $0.74\phd\!*$ & $ 56.0\phd_{-0.6}^{+0.2}$   & {\boldmath $3.8\phd_{-1.3}^{+1.6}$} &     5.8 & $2.1\phd\!*$ & $ 56.9\phd_{-0.11}^{+0.09}$ & & & & & & &    32.68 &   32.93 &   33.45 &   33.85 &   33.52 &   33.90 &                grp2.2\_T+T:LMC,LMC &                        HandFit \\[1ex] 
   7 & p1\_1194 & 053853.36$-$690200.9 & {\bf p54} &   106 &     9.6 &  1.23 &$0.452\phd\!*$ &     0.7 & $0.74\phd\!*$ & $ 55.9\phd_{\cdots}^{+0.3}$ & {\boldmath $3.3\phd_{-1.5}^{+2.0}$} &     5.1 & $2.1\phd\!*$ & $ 56.9\phd_{-0.1}^{+0.1}$   & & & & & & &    32.55 &   32.81 &   33.42 &   33.80 &   33.48 &   33.84 &                grp1.6\_T+T:LMC,LMC &                        HandFit \\[1ex] 
   8 & p1\_1194 & 053853.36$-$690200.9 & {\bf p70} &    67 &     7.9 &  0.86 &$0.452\phd\!*$ &     0.7 & $0.74\phd\!*$ & $ 55.8\phd_{\cdots}^{+0.4}$ & {\boldmath $1.9\phd_{-1.3}^{+2.2}$} &     2.9 & $2.1\phd\!*$ & $ 56.8\phd_{-0.2}^{+0.2}$   & & & & & & &    32.43 &   32.69 &   33.39 &   33.68 &   33.44 &   33.72 &                grp1.2\_T+T:LMC,LMC &                        HandFit \\[1ex] 
   9 & p1\_1194 & 053853.36$-$690200.9 & {\bf p83} &    97 &     9.3 &  1.15 &$0.452\phd\!*$ &     0.7 & $0.74\phd\!*$ & $ 56.2\phd_{-0.2}^{+0.1}$   & {\boldmath $8.5\phd_{-4.4}^{+7.2}$} &    13.1 & $2.1\phd\!*$ & $ 57.0\phd_{\cdots}^{+0.2}$ & & & & & & &    32.88 &   33.14 &   33.36 &   33.91 &   33.49 &   33.98 &                grp1.5\_T+T:LMC,LMC &                        HandFit \\[1ex] 
  10 & p1\_1194 & 053853.36$-$690200.9 & {\bf p90} &   104 &     9.1 &  0.94 &$0.452\phd\!*$ &     0.7 & $0.74\phd\!*$ & $ 56.1\phd_{-0.2}^{+0.1}$   & {\boldmath $9.2\phd_{-3.4}^{+5.2}$} &    14.2 & $2.1\phd\!*$ & $ 56.8\phd_{-0.2}^{+0.1}$   & & & & & & &    32.71 &   32.97 &   33.18 &   33.75 &   33.31 &   33.82 &                grp1.6\_T+T:LMC,LMC &                        HandFit \\[1ex] 
\enddata


\tablenotetext{a}{Phase groups are labeled, for example, with ``p36'' representing the spectrum obtained by merging a group of observations with phase $\sim$0.36.}

\tablenotetext{b}{
%
%
All fits used the thermal plasma model \mbox{\it{tbvarabs*vapec+tbvarabs*vapec}} with abundances for both absorption and plasma components as adopted by \citet{Tehrani19}.
%
%
Best-fit values for the extinction are shown both as a column density and as $A_V = N_H / 6.5 \times 10^{21}$~cm$^{-2}$. 
Quantities marked with an asterisk (*) were frozen in the fit.
Uncertainties represent 90\% confidence intervals.
Uncertainties are missing when \XSPEC\ was unable to compute them or when their values were so large that the parameter is effectively unconstrained.  
}

\tablenotetext{c}{ X-ray luminosities in the total ($t$) band (0.5--8 keV) are derived from the model spectrum, assuming a distance of 50~kpc.
Absorption-corrected luminosities are subscripted with a $c$.
Model luminosities may underestimate the true X-ray luminosity for sources with high absorption because an additional soft thermal plasma could be present but unmeasurable in the X-ray spectrum.
}

\end{deluxetable}

\noindent \\ {\bf R145}\\
The binary Wolf-Rayet star R145 (053857.06-690605.6, p1\_1256, VFTS~695) has a published optical period of 158.76 day \citep{Schnurr09,Shenar17}, which is consistent with the T-ReX data \citep[][]{Tehrani19}.
When folded on that period, we find that R145 exhibits $\sim$80\% X-ray flux modulation and clear modulation of the  apparent spectral hardness (Fig.~\ref{R145(2).fig}).

\noindent \\ {\bf R139}\\
The T-ReX data for the O6.5I+O6I binary 
R139 (053842.34-690458.1, p1\_830, VFTS~527) exhibit $\sim$65\% X-ray flux modulation and clear modulation of the apparent spectrum when folded on the period we estimate from the T-ReX data ($\sim$153.0 day, Fig.~\ref{R139.fig}).
This period is very similar to the 153.90--day period derived from optical radial velocity measurements \citep{Taylor11}.
%
%

\noindent \\ {\bf Mk33Na}\\
Periodic variation of the O3If \citep{Massey98}  
star Mk33Na (053844.33-690554.7, p1\_1000, Doran~775, Hunter~16, MUSE~1943) has been detected in both the T-ReX data (P$\sim$18.3 day) and in optical spectra obtained over a short baseline
\citep[P=$18.32 \pm 0.14$ day][]{Bestenlehner22}.
When folded on that period, the T-ReX data exhibit flux modulation of $\sim$50\% and modulation of the apparent spectrum consistent with a phase-dependent absorbing column density \citep{Bestenlehner22}.

\noindent \\ {\bf VFTS~399}\\
\citet{Clark15} discovered that the massive (O9) star VFTS~399 (053833.43-691159.0, p1\_610) probably has an accreting neutron star companion, based in part on the $\sim$2567 s periodic modulation found in its T-ReX lightcurve.

\noindent \\ {\bf Mk34}\\
\citet{Pollock18} find that Mk34 (053844.25-690605.9, p1\_995, VFTS~770) is an eccentric colliding-wind Wolf-Rayet binary system with a period of $\sim$155 days, based on its remarkable T-ReX lightcurve.

\noindent \\ {\bf Mk39}\\
\citet[][Chapter 4]{Tehrani19} estimate the period ($\sim$640 day) of the binary Wolf-Rayet star Mk39 (053840.22-690559.8, p1\_698, VFTS~482) using the T-ReX data.

\subsection{``Flares''}

\begin{figure}[htb]
\centering
\includegraphics[width=0.7\textwidth]{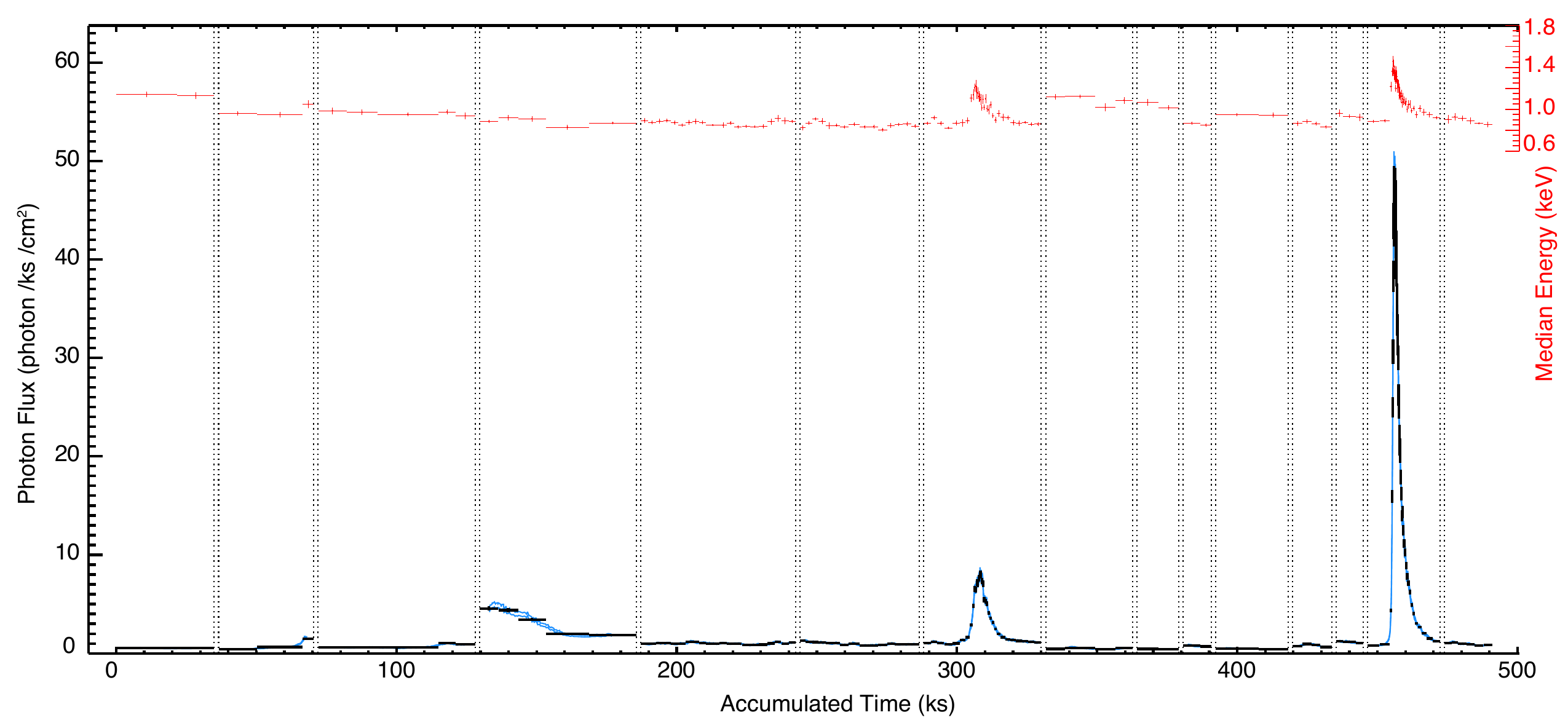}
\caption{HD~269921 (053834.66-685306.1, pass52\_4) lightcurve (blue, with grouped events in black) and median energy timeseries (red).
Two complete ``flare'' events exhibit spectral hardening. Vertical dotted lines mark the boundaries of individual ObsIDs.
\label{HD269921.fig}}
\end{figure}

Fast-rise, slow-decay ``flares'' were seen in at least two foreground stars.
The low-mass star HD~269921\footnote{
94 pc, Gaia DR2 4657783116463742336 (0.47\arcsec separation)
}
(053834.66-685306.1, pass52\_4) shows variability by a factor of $\sim$100 (Figure~\ref{HD269921.fig}).
\citet{Moor13} report a spectral type of G7V, and assert membership in the Columba moving group.
These data (85,000 X-ray events) should allow time-resolved spectroscopy of the two complete ``flare'' events observed, which exhibit spectral hardening (red curve). 

\begin{figure}[htb]
\centering
\includegraphics[width=0.9\textwidth]{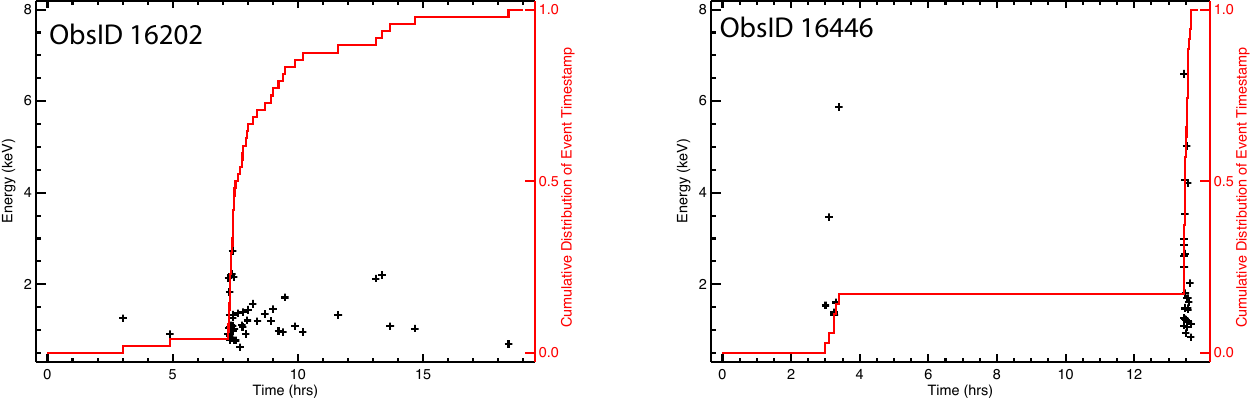}
\caption{Flares within two single ObsIDs of foreground source 053935.54-690438.2 (p1\_1521). Black + symbols depict the arrival time and energy of individual events, while the red curves give the cumulative distributions of arrival times.
The {\em p}-values for the null hypothesis (constant flux) are  $1\times10^{-5}$ and $3\times10^{-21}$ for ObsIDs 16202 and 16446, respectively.
\label{CXOUJ053935.54-690438.2.fig}}
\end{figure}

The foreground star VFTS 2077 \footnote{
220 pc, Gaia DR2 4657682923488239360 (0.2\arcsec separation)
} (053935.54-690438.2 , p1\_1521)
shows  dramatic variability in two of its 54 observations (Figure~\ref{CXOUJ053935.54-690438.2.fig}).
The peak count rate at the end of ObsID~16446 (0.038 ct/s = 29 counts in a 770 s interval) is 460 times the mean count rate ($8\times10^{-5}$ ct/s = 169 net counts in 2 Ms). 



\begin{figure}[htb]
\centering
\includegraphics[width=0.7\textwidth]{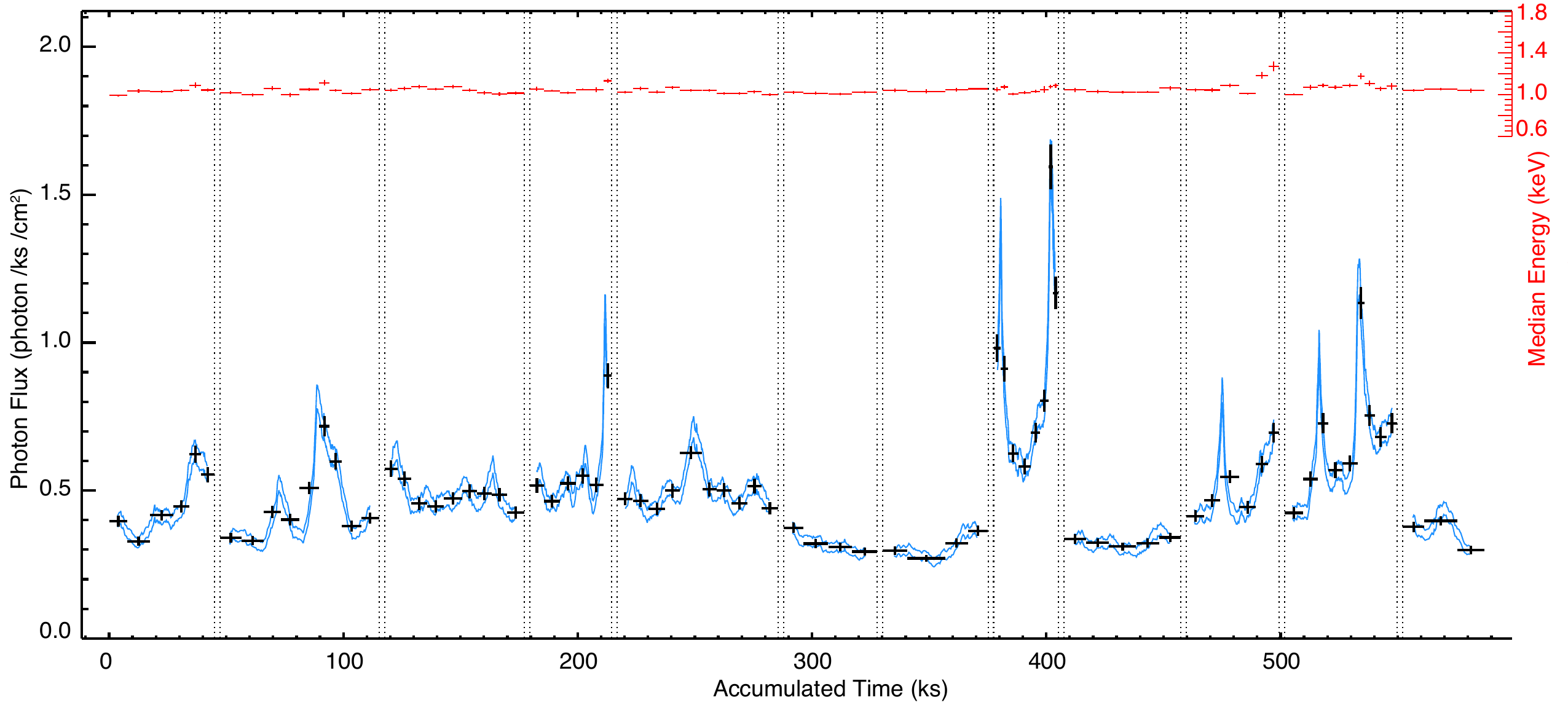}
\caption{CAL 69 (053816.28-692331.1, c4107) lightcurve, plotted with symbols and axes as in Figure~\ref{HD269921.fig}.
Frequent, low-amplitude flares are observed, with evidence for spectral hardening.
\label{CAL69.fig}}
\end{figure}

Finally, the foreground low-mass star CAL 69\footnote{53 pc, Gaia DR3 4657651793560005632 (1.33\arcsec separation at large off-axis angle)} (053816.28-692331.1, c4107) is an early M dwarf \citep{Riaz06} exhibiting complex, frequent flaring activity (Figure~\ref{CAL69.fig}). Obtaining this long-baseline lightcurve for this star was serendipitous, as it happened to fall within the extreme southern portion of the T-ReX field (Figure~\ref{acisevents_fullfield.fig}). X-ray activity in low-mass stars provides an important constraint on exoplanet habitability models.

\clearpage

\input{bib.tex}

\end{document}

%% file: observing_log.tex
\startlongtable
\begin{deluxetable}{hrrrccrhhhlhC}
\centering 
\tabletypesize{\footnotesize} \tablewidth{0pt}

\tablecaption{ Log of {\em Chandra} Observations 
 \label{tbl:obslog}}

\tablehead{
\nocolhead{Target} & 
\colhead{ObsID} & 
\colhead{Start Date} & 
\colhead{Exposure\tablenotemark{a}} & 
\multicolumn{2}{c}{On-axis Position\tablenotemark{b}} & 
\colhead{Roll} & 
\nocolhead{Config\tablenotemark{c}} & 
\nocolhead{Mode\tablenotemark{d}} &
\nocolhead{PI}  &
\colhead{TGAIN\tablenotemark{c}} &
\nocolhead{OBF\tablenotemark{f}} &
\colhead{Shift\tablenotemark{d}}  \\
\cline{5-6}
\colhead{} & 
\colhead{} & 
\colhead{(UT)} & 
\colhead{(s)} & 
\colhead{$\alpha_{\rm J2000}$} & 
\colhead{$\delta_{\rm J2000}$} & 
\colhead{(\arcdeg)} & 
\colhead{} &
\colhead{} &
\colhead{} &
\colhead{} &
\colhead{} &
\colhead{(SKY pixel)} 
}
\decimalcolnumbers
\startdata
               30 Doradus & \dataset[  5906]{\doi{10.25574/05906}} &       2006-01-21 &   12317 & 05:38:43.92 & -69:06:16.3 &  323 &   I (012367) &  TE-VF & Leisa Townsley & \path{ 2005-11-01N6 } & N0010 & (-0.285, -0.315) \\ 
               30 Doradus & \dataset[  7263]{\doi{10.25574/07263}} &       2006-01-22 &   42528 & 05:38:43.96 & -69:06:16.4 &  323 &   I (012367) &  TE-VF & Leisa Townsley & \path{ 2005-11-01N6 } & N0010 & (-0.120, -0.254) \\ 
               30 Doradus & \dataset[  7264]{\doi{10.25574/07264}} &       2006-01-30 &   37593 & 05:38:43.54 & -69:06:17.4 &  314 &   I (012367) &  TE-VF & Leisa Townsley & \path{ 2005-11-01N6 } & N0010 & (-0.076, -0.417) \\ 
               30 Doradus & \dataset[ 16192]{\doi{10.25574/16192}} &       2014-05-03 &   93761 & 05:38:39.69 & -69:06:28.1 &  222 &   I (012367) &  TE-VF & Leisa Townsley & \path{ 2014-05-01N6_revA } & N0010 & (+0.305, -1.001) \\ 
               30 Doradus & \dataset[ 16193]{\doi{10.25574/16193}} &       2014-05-08 &   75993 & 05:38:39.31 & -69:06:26.8 &  218 &   I (012367) &  TE-VF & Leisa Townsley & \path{ 2014-05-01N6_revA } & N0010 & (+0.119, -0.957) \\ 
               30 Doradus & \dataset[ 16612]{\doi{10.25574/16612}} &       2014-05-11 &   22672 & 05:38:39.31 & -69:06:26.8 &  218 &   I (012367) &  TE-VF & Leisa Townsley & \path{ 2014-05-01N6_revA } & N0010 & (-0.048, -0.662) \\ 
               30 Doradus & \dataset[ 16194]{\doi{10.25574/16194}} &       2014-05-12 &   31333 & 05:38:38.87 & -69:06:24.9 &  212 &   I (012367) &  TE-VF & Leisa Townsley & \path{ 2014-05-01N6_revA } & N0010 & (+0.480, -1.665) \\ 
               30 Doradus & \dataset[ 16615]{\doi{10.25574/16615}} &       2014-05-15 &   45170 & 05:38:38.86 & -69:06:25.0 &  212 &   I (012367) &  TE-VF & Leisa Townsley & \path{ 2014-05-01N6_revA } & N0010 & (-0.193, -0.727) \\ 
               30 Doradus & \dataset[ 16195]{\doi{10.25574/16195}} &       2014-05-24 &   44405 & 05:38:38.20 & -69:06:21.4 &  202 &   I (012367) &  TE-VF & Leisa Townsley & \path{ 2014-05-01N6_revA } & N0010 & (+0.928, -1.448) \\ 
               30 Doradus & \dataset[ 16196]{\doi{10.25574/16196}} &       2014-05-30 &   67109 & 05:38:38.19 & -69:06:21.4 &  202 &   I (012367) &  TE-VF & Leisa Townsley & \path{ 2014-05-01N6_revA } & N0010 & (+0.323, -0.942) \\ 
               30 Doradus & \dataset[ 16617]{\doi{10.25574/16617}} &       2014-05-31 &   58860 & 05:38:38.19 & -69:06:21.3 &  202 &   I (012367) &  TE-VF & Leisa Townsley & \path{ 2014-05-01N6_revA } & N0010 & (+1.063, -1.710) \\ 
               30 Doradus & \dataset[ 16616]{\doi{10.25574/16616}} &       2014-06-03 &   34530 & 05:38:38.19 & -69:06:21.3 &  202 &   I (012367) &  TE-VF & Leisa Townsley & \path{ 2014-05-01N6_revA } & N0010 & (+0.008, -0.746) \\ 
               30 Doradus & \dataset[ 16197]{\doi{10.25574/16197}} &       2014-06-06 &   67790 & 05:38:38.19 & -69:06:21.3 &  202 &   I (012367) &  TE-VF & Leisa Townsley & \path{ 2014-05-01N6_revA } & N0010 & (+0.911, -1.483) \\ 
               30 Doradus & \dataset[ 16198]{\doi{10.25574/16198}} &       2014-06-11 &   39465 & 05:38:37.45 & -69:06:14.9 &  187 &   I (012367) &  TE-VF & Leisa Townsley & \path{ 2014-05-01N6_revA } & N0010 & (+1.397, -1.247) \\ 
               30 Doradus & \dataset[ 16621]{\doi{10.25574/16621}} &       2014-06-14 &   44400 & 05:38:37.44 & -69:06:14.9 &  187 &   I (012367) &  TE-VF & Leisa Townsley & \path{ 2014-05-01N6_revA } & N0010 & (+1.211, -1.110) \\ 
               30 Doradus & \dataset[ 16200]{\doi{10.25574/16200}} &       2014-06-26 &   27360 & 05:38:37.01 & -69:06:06.8 &  170 &   I (012367) &  TE-VF & Leisa Townsley & \path{ 2014-05-01N6_revA } & N0010 & (+1.720, -0.801) \\ 
               30 Doradus & \dataset[ 16201]{\doi{10.25574/16201}} &       2014-07-21 &   58390 & 05:38:37.14 & -69:05:54.0 &  145 &   I (012367) &  TE-VF & Leisa Townsley & \path{ 2014-05-01N6_revA } & N0010 & (+0.158, -0.038) \\ 
               30 Doradus & \dataset[ 16640]{\doi{10.25574/16640}} &       2014-07-24 &   61679 & 05:38:37.15 & -69:05:54.0 &  145 &   I (012367) &  TE-VF & Leisa Townsley & \path{ 2014-05-01N6_revA } & N0010 & (+0.594, -0.135) \\ 
               30 Doradus & \dataset[ 16202]{\doi{10.25574/16202}} &       2014-08-19 &   65128 & 05:38:38.53 & -69:05:41.8 &  117 &   I (012367) &  TE-VF & Leisa Townsley & \path{ 2014-08-01N6_revA } & N0010 & (+0.561, +0.312) \\ 
               30 Doradus & \dataset[ 17312]{\doi{10.25574/17312}} &       2014-08-22 &   44895 & 05:38:38.53 & -69:05:41.8 &  117 &   I (012367) &  TE-VF & Leisa Townsley & \path{ 2014-08-01N6_revA } & N0010 & (+0.481, +0.346) \\ 
               30 Doradus & \dataset[ 16203]{\doi{10.25574/16203}} &       2014-09-02 &   41422 & 05:38:39.63 & -69:05:37.4 &  102 &   I (012367) &  TE-VF & Leisa Townsley & \path{ 2014-08-01N6_revA } & N0010 & (+0.243, +0.248) \\ 
               30 Doradus & \dataset[ 17413]{\doi{10.25574/17413}} &       2014-09-08 &   24650 & 05:38:39.63 & -69:05:37.3 &  102 &   I (012367) &  TE-VF & Leisa Townsley & \path{ 2014-08-01N6_revA } & N0010 & (+0.718, +0.789) \\ 
               30 Doradus & \dataset[ 17414]{\doi{10.25574/17414}} &       2014-09-13 &   17317 & 05:38:40.32 & -69:05:35.5 &   94 &   I (012367) &  TE-VF & Leisa Townsley & \path{ 2014-08-01N6_revA } & N0010 & (+0.356, -0.146) \\ 
               30 Doradus & \dataset[ 16442]{\doi{10.25574/16442}} &       2014-10-25 &   48350 & 05:38:44.49 & -69:05:35.6 &   50 &   I (012367) &  TE-VF & Leisa Townsley & \path{ 2014-08-01N6_revA } & N0010 & (-0.138, -0.336) \\ 
               30 Doradus & \dataset[ 17545]{\doi{10.25574/17545}} &       2014-10-28 &   34530 & 05:38:44.50 & -69:05:35.5 &   50 &   I (012367) &  TE-VF & Leisa Townsley & \path{ 2014-08-01N6_revA } & N0010 & (-0.071, -0.008) \\ 
               30 Doradus & \dataset[ 17544]{\doi{10.25574/17544}} &       2014-11-01 &   25642 & 05:38:44.49 & -69:05:35.7 &   50 &   I (012367) &  TE-VF & Leisa Townsley & \path{ 2014-11-01N6 } & N0010 & (+0.015, -0.032) \\ 
               30 Doradus & \dataset[ 16443]{\doi{10.25574/16443}} &       2014-11-14 &   34530 & 05:38:45.51 & -69:05:38.5 &   38 &   I (012367) &  TE-VF & Leisa Townsley & \path{ 2014-11-01N6 } & N0010 & (-0.014, +0.069) \\ 
               30 Doradus & \dataset[ 17486]{\doi{10.25574/17486}} &       2014-12-04 &   33541 & 05:38:47.19 & -69:05:48.3 &   12 &   I (012367) &  TE-VF & Leisa Townsley & \path{ 2014-11-01N6 } & N0010 & (+0.257, -0.702) \\ 
               30 Doradus & \dataset[ 17555]{\doi{10.25574/17555}} &       2014-12-06 &   55246 & 05:38:47.18 & -69:05:48.3 &   12 &   I (012367) &  TE-VF & Leisa Townsley & \path{ 2014-11-01N6 } & N0010 & (+0.557, -0.761) \\ 
               30 Doradus & \dataset[ 17561]{\doi{10.25574/17561}} &       2014-12-20 &   54567 & 05:38:47.86 & -69:05:58.9 &  350 &   I (012367) &  TE-VF & Leisa Townsley & \path{ 2014-11-01N6 } & N0010 & (+0.707, -0.514) \\ 
               30 Doradus & \dataset[ 17562]{\doi{10.25574/17562}} &       2014-12-25 &   42031 & 05:38:47.85 & -69:05:58.9 &  350 &   I (012367) &  TE-VF & Leisa Townsley & \path{ 2014-11-01N6 } & N0010 & (+0.602, -0.204) \\ 
               30 Doradus & \dataset[ 16444]{\doi{10.25574/16444}} &       2014-12-27 &   41440 & 05:38:47.85 & -69:05:58.9 &  350 &   I (012367) &  TE-VF & Leisa Townsley & \path{ 2014-11-01N6 } & N0010 & (+0.270, -0.188) \\ 
               30 Doradus & \dataset[ 16448]{\doi{10.25574/16448}} &       2015-02-14 &   34599 & 05:38:46.12 & -69:06:24.6 &  295 &   I (012367) &  TE-VF & Leisa Townsley & \path{ 2015-02-01N6 } & N0010 & (+0.722, -0.198) \\ 
               30 Doradus & \dataset[ 17602]{\doi{10.25574/17602}} &       2015-02-19 &   51705 & 05:38:46.12 & -69:06:24.5 &  295 &   I (012367) &  TE-VF & Leisa Townsley & \path{ 2015-02-01N6 } & N0010 & (+1.587, +0.330) \\ 
               30 Doradus & \dataset[ 16447]{\doi{10.25574/16447}} &       2015-03-26 &   26868 & 05:38:43.31 & -69:06:31.8 &  262 &   I (012367) &  TE-VF & Leisa Townsley & \path{ 2015-02-01N6 } & N0010 & (+0.243, +0.421) \\ 
               30 Doradus & \dataset[ 16199]{\doi{10.25574/16199}} &       2015-03-27 &   39461 & 05:38:43.31 & -69:06:31.9 &  262 &   I (012367) &  TE-VF & Leisa Townsley & \path{ 2015-02-01N6 } & N0010 & (+0.287, +0.669) \\ 
               30 Doradus & \dataset[ 17640]{\doi{10.25574/17640}} &       2015-03-31 &   26318 & 05:38:43.31 & -69:06:31.9 &  262 &   I (012367) &  TE-VF & Leisa Townsley & \path{ 2015-02-01N6 } & N0010 & (+0.326, +0.421) \\ 
               30 Doradus & \dataset[ 17641]{\doi{10.25574/17641}} &       2015-04-04 &   24638 & 05:38:43.32 & -69:06:31.9 &  262 &   I (012367) &  TE-VF & Leisa Townsley & \path{ 2015-02-01N6 } & N0010 & (+0.969, +1.656) \\ 
               30 Doradus & \dataset[ 16445]{\doi{10.25574/16445}} &       2015-05-27 &   49310 & 05:38:37.81 & -69:06:19.1 &  196 &   I (012367) &  TE-VF & Leisa Townsley & \path{ 2015-05-01N6 } & N0010 & (+0.286, +0.166) \\ 
               30 Doradus & \dataset[ 17660]{\doi{10.25574/17660}} &       2015-05-29 &   38956 & 05:38:37.81 & -69:06:19.1 &  196 &   I (012367) &  TE-VF & Leisa Townsley & \path{ 2015-05-01N6 } & N0010 & (-0.023, +0.124) \\ 
               30 Doradus & \dataset[ 16446]{\doi{10.25574/16446}} &       2015-06-02 &   47547 & 05:38:37.68 & -69:06:18.1 &  193 &   I (012367) &  TE-VF & Leisa Townsley & \path{ 2015-05-01N6 } & N0010 & (+0.117, +0.072) \\ 
               30 Doradus & \dataset[ 17642]{\doi{10.25574/17642}} &       2015-06-08 &   34438 & 05:38:37.44 & -69:06:15.6 &  188 &   I (012367) &  TE-VF & Leisa Townsley & \path{ 2015-05-01N6 } & N0010 & (+0.182, -0.065) \\ 
               30 Doradus & \dataset[ 16449]{\doi{10.25574/16449}} &       2015-09-28 &   24628 & 05:38:41.66 & -69:05:33.6 &   80 &   I (012367) &  TE-VF & Leisa Townsley & \path{ 2015-08-01N6 } & N0010 & (-1.468, -0.353) \\ 
               30 Doradus & \dataset[ 18672]{\doi{10.25574/18672}} &       2015-11-08 &   30574 & 05:38:45.82 & -69:05:39.8 &   34 &   I (012367) &  TE-VF & Leisa Townsley & \path{ 2015-11-01N6 } & N0010 & (-0.480, -1.424) \\ 
               30 Doradus & \dataset[ 18706]{\doi{10.25574/18706}} &       2015-11-10 &   14776 & 05:38:45.80 & -69:05:39.7 &   34 &   I (012367) &  TE-VF & Leisa Townsley & \path{ 2015-11-01N6 } & N0010 & (-0.889, -1.345) \\ 
               30 Doradus & \dataset[ 18720]{\doi{10.25574/18720}} &       2015-12-02 &    9832 & 05:38:45.30 & -69:05:58.0 &  357 &   I (012367) &  TE-VF & Leisa Townsley & \path{ 2015-11-01N6 } & N0010 & (-0.563, -1.301) \\ 
               30 Doradus & \dataset[ 18721]{\doi{10.25574/18721}} &       2015-12-08 &   25598 & 05:38:44.95 & -69:05:54.0 &   12 &   I (012367) &  TE-VF & Leisa Townsley & \path{ 2015-11-01N6 } & N0010 & (+0.040, -1.745) \\ 
               30 Doradus & \dataset[ 17603]{\doi{10.25574/17603}} &       2015-12-09 &   13778 & 05:38:45.02 & -69:05:55.1 &    8 &   I (012367) &  TE-VF & Leisa Townsley & \path{ 2015-11-01N6 } & N0010 & (+0.155, -1.927) \\ 
               30 Doradus & \dataset[ 18722]{\doi{10.25574/18722}} &       2015-12-11 &    9826 & 05:38:45.16 & -69:05:56.0 &    3 &   I (012367) &  TE-VF & Leisa Townsley & \path{ 2015-11-01N6 } & N0010 & (-0.510, -1.237) \\ 
               30 Doradus & \dataset[ 18671]{\doi{10.25574/18671}} &       2015-12-13 &   25617 & 05:38:45.17 & -69:05:56.1 &    3 &   I (012367) &  TE-VF & Leisa Townsley & \path{ 2015-11-01N6 } & N0010 & (-0.465, -1.112) \\ 
               30 Doradus & \dataset[ 18729]{\doi{10.25574/18729}} &       2015-12-21 &   16742 & 05:38:45.18 & -69:05:56.1 &    3 &   I (012367) &  TE-VF & Leisa Townsley & \path{ 2015-11-01N6 } & N0010 & (+0.500, -1.643) \\ 
               30 Doradus & \dataset[ 18750]{\doi{10.25574/18750}} &       2016-01-20 &   48318 & 05:38:45.44 & -69:06:07.5 &  324 &   I (012367) &  TE-VF & Leisa Townsley & \path{ 2015-11-01N6 } & N0010 & (+1.496, -0.875) \\ 
               30 Doradus & \dataset[ 18670]{\doi{10.25574/18670}} &       2016-01-21 &   14565 & 05:38:45.43 & -69:06:07.5 &  324 &   I (012367) &  TE-VF & Leisa Townsley & \path{ 2015-11-01N6 } & N0010 & (+0.905, -0.902) \\ 
               30 Doradus & \dataset[ 18749]{\doi{10.25574/18749}} &       2016-01-22 &   22153 & 05:38:45.42 & -69:06:07.4 &  324 &   I (012367) &  TE-VF & Leisa Townsley & \path{ 2015-11-01N6 } & N0010 & (+0.730, -1.058) \\ 
\enddata

\tablenotetextbox{a}{0.95\textwidth}{Exposure times are the net usable times after various filtering steps are applied in the data reduction process. 
No exposure time needed to be removed during periods of high instrumental background. 
The time variability of the ACIS background is discussed in \S6.16.3 of the \anchorparen{http://asc.harvard.edu/proposer/POG/}{\Chandra\ Proposers' Observatory Guide} and in the ACIS Background Memos at \url{http://asc.harvard.edu/cal/Acis/Cal_prods/bkgrnd/current/}.
}

\tablenotetextbox{b}{0.95\textwidth}{The On-axis Position is the \anchorparen{http://cxc.harvard.edu/ciao/faq/nomtargpnt.html}{time-averaged location of the optical axis (CIAO parameters RA\_PNT,DEC\_PNT)}.
Units of right ascension ($\alpha$) are hours, minutes, and seconds; units of declination ($\delta$) are degrees, arcminutes, and arcseconds.}



\tablenotetextbox{c}{0.95\textwidth}{Abbreviated name of the ACIS Time-Dependent Gain file used for calibration of event energies, e.g. ``2002-02-01N6'' = ``acisD2002-02-01t\_gainN0006.fits''.}


\tablenotetextbox{d}{0.95\textwidth}{The shift (in RA and Dec) applied to the ObsID's aspect file (via the {\em CIAO} tool {\em wcs\_update}) to achieve astrometric alignment, expressed as (dx,dy) in the \anchorparen{http://asc.harvard.edu/ciao/ahelp/coords.html}{Chandra ``SKY'' coordinate system}; 1 SKY pixel = 0.492 arcsecond.  Note that astrometric shifts are correlated in time.}

\end{deluxetable}

%% file: xray_column_labels.tex

%



\startlongtable
\begin{deluxetable*}{llp{4.0in}}

\tablecaption{T-ReX X-ray Source Properties \label{tbl:xray_properties}
}
\tablewidth{6.5in}
\tabletypesize{\scriptsize}
\tablecolumns{3}

\tablehead{   
  \colhead{Column Label} & 
  \colhead{Units} & 
  \colhead{Description} 
}
\colnumbers
\startdata
\cutinhead{Source identifiers}
 Name                       & \nodata              & \parbox[t]{4.0in}{X-ray source name in IAU format; prefix is CXOU~J} \\
 Label                      & \nodata              & X-ray source name used within the project\tablenotemark{a} \\
\cutinhead{Source position, derived from the set of ObsIDs that minimized position uncertainty \citep[][\S6.2 and 7]{Broos10}}
 RAdeg                      & deg                  & right ascension (ICRS)\tablenotemark{b} \\
 DEdeg                      & deg                  & declination (ICRS)\tablenotemark{b} \\
 PosErr                     & arcsec               & 1-$\sigma$ error circle around (RAdeg,DEdeg) \\
 PosType                    & \nodata              & algorithm used to estimate position \citep[][\S7.1]{Broos10}  \\
\cutinhead{Validity metrics, derived from a pre-defined set of ObsID combinations \citep[][\S2.2.3]{Townsley18} }
 ProbNoSrc\_MostValid       & \nodata              & smallest of ProbNoSrc\_t, ProbNoSrc\_s, ProbNoSrc\_h, ProbNoSrc\_v (\citealp[][\S4.3]{Broos10} and \citealp[][Appendix~A2]{Weisskopf07})\tablenotemark{c} \\
 ProbNoSrc\_t               & \nodata              & \parbox[t]{4.0in}{smallest {\em p}-value under the no-source (0.5--7~keV) null hypothesis among validation merges} \\
 ProbNoSrc\_s               & \nodata              & \parbox[t]{4.0in}{smallest {\em p}-value under the no-source (0.5--2~keV) null hypothesis among validation merges}\\
 ProbNoSrc\_h               & \nodata              & \parbox[t]{4.0in}{smallest {\em p}-value under the no-source (2--7~keV) null hypothesis among validation merges}\\
 ProbNoSrc\_v               & \nodata              & \parbox[t]{4.0in}{smallest {\em p}-value under the no-source (4--7~keV) null hypothesis among validation merges}\\
IsOccasional                & boolean              & \parbox[t]{4.0in}{flag indicating that source validation failed in all multi-ObsID merges; source validation comes from a single ObsID \citep[\S3.1]{Townsley19}.}\\
\cutinhead{Variability indicators, derived from all ObsIDs}
 ProbKS\_single      & \nodata              & \parbox[t]{4.0in}{$\min(P_1,P_2, \ldots P_{N\_KS\_single})$, where each $P_i$ is the {\em p}-value arising from a Kolmogorov--Smirnov test of event arrive times under the null hypothesis of no variability within a single observation} \\
 N\_KS\_single      & observations         & \parbox[t]{4.0in}{number of {\em p}-values compared to produce ProbKS\_single}\\
 ProbKS\_merge       & \nodata              & \parbox[t]{4.0in}{{\em p}-value from a Kolmogorov--Smirnov test on the combined events from all observations; the null hypothesis is no variability on any time scale} \\
 ProbChisq\_PhotonFlux & \nodata            & \parbox[t]{4.0in}{{\em p}-value from a $\chi^2$ test on the single-ObsID measurements of PhotonFlux\_t; the null hypothesis is no inter-ObsID variability} \\
\cutinhead{Observation details and photometric quantities, derived from {\bf all} observations of the source}
 ExposureTimeNominal        & s                    & total exposure time in merged ObsIDs \\
 ExposureFraction           & \nodata              & fraction of ExposureTimeNominal that source was observed\tablenotemark{d} \\
 RateIn3x3Cell              &  count/frame         & 0.5:8 keV, in 3$\times$3 CCD pixel cell\tablenotemark{e}  \\
 NumObsIDs                  & observations         & total number of ObsIDs extracted \\
 NumMerged                  & observations         & \parbox[t]{4.0in}{number of ObsIDs merged to estimate photometry properties} \\
 MergeBias                  & \nodata              & fraction of exposure discarded in merge \\
 Theta\_Lo                  & arcmin               & smallest off-axis angle for merged ObsIDs \\
 Theta                      & arcmin               &  average off-axis angle for merged ObsIDs \\
 Theta\_Hi                  & arcmin               &  largest off-axis angle for merged ObsIDs \\
 PsfFraction                & \nodata              & average PSF fraction (at 1.5 keV) for merged ObsIDs \\
 SrcArea                    & (0.492 arcsec)$^2$   & average aperture area for merged ObsIDs \\
 AfterglowFraction          & \nodata              & suspected afterglow fraction\tablenotemark{f} \\
 SrcCounts\_t               & count                & observed counts (0.5--8~keV) in merged apertures \\
 SrcCounts\_s               & count                & observed counts (0.5--2~keV) in merged apertures \\
 SrcCounts\_h               & count                & observed counts (2--8~keV) in merged apertures \\
 BkgScaling                 & \nodata              & scaling of the background extraction \citep[][\S5.4]{Broos10} \\
 BkgCounts\_t               & count                & observed counts (0.5--8~keV) in merged background regions \\
 BkgCounts\_s               & count                & observed counts (0.5--2~keV) in merged background regions \\
 BkgCounts\_h               & count                & observed counts (2--8~keV) in merged background regions \\
 NetCounts\_t               & count                & net counts (0.5--8~keV) in merged apertures \\
 NetCounts\_s               & count                & net counts (0.5--2~keV) in merged apertures \\
 NetCounts\_h               & count                & net counts (2--8~keV) in merged apertures \\
 NetCounts\_Lo\_t           & count                & 1-$\sigma$ lower bound on NetCounts\_t\tablenotemark{g}  \\
 NetCounts\_Hi\_t           & count                & 1-$\sigma$ upper bound on NetCounts\_t\tablenotemark{g}  \\
 NetCounts\_Lo\_s           & count                & 1-$\sigma$ lower bound on NetCounts\_s\tablenotemark{g}  \\
 NetCounts\_Hi\_s           & count                & 1-$\sigma$ upper bound on NetCounts\_s\tablenotemark{g}  \\
 NetCounts\_Lo\_h           & count                & 1-$\sigma$ lower bound on NetCounts\_h\tablenotemark{g}  \\
 NetCounts\_Hi\_h           & count                & 1-$\sigma$ upper bound on NetCounts\_h\tablenotemark{g}  \\  
 MeanEffectiveArea\_t       & cm$^2$~count~photon$^{-1}$ & mean ARF value (0.5--8~keV)\tablenotemark{h} \\
 MeanEffectiveArea\_s       & cm$^2$~count~photon$^{-1}$ & mean ARF value (0.5--2~keV)\tablenotemark{h} \\
 MeanEffectiveArea\_h       & cm$^2$~count~photon$^{-1}$ & mean ARF value (2--8~keV)\tablenotemark{h} \\
 MedianEnergy\_t            & keV                  & median energy, observed spectrum (0.5--8 keV)\tablenotemark{i} \\
 MedianEnergy\_s            & keV                  & median energy, observed spectrum (0.5--2 keV)\tablenotemark{i} \\
 MedianEnergy\_h            & keV                  & median energy, observed spectrum (2--8 keV)\tablenotemark{i} \\
 PhotonFlux\_t              & photon~cm$^{-2}$~s$^{-1}$     & apparent photon flux (0.5--8~keV)\tablenotemark{j} \\
 PhotonFlux\_s              & photon~cm$^{-2}$~s$^{-1}$     & apparent photon flux (0.5--2~keV)\tablenotemark{j} \\
 PhotonFlux\_h              & photon~cm$^{-2}$~s$^{-1}$     & apparent photon flux (2--8~keV)\tablenotemark{j} \\
 EnergyFlux\_t              & erg~cm$^{-2}$~s$^{-1}$ & max(EnergyFlux\_s,0) + max(EnergyFlux\_h,0) \\
 EnergyFlux\_s              & erg~cm$^{-2}$~s$^{-1}$ & apparent energy flux (0.5--2~keV)\tablenotemark{j} \\
 EnergyFlux\_h              & erg~cm$^{-2}$~s$^{-1}$ & apparent energy flux (0.5--2~keV)\tablenotemark{j}  
\enddata
                                                                       
\tablecomments{
These X-ray columns are produced by the
{\em ACIS Extract} \citep[\AEacro,][]{Broos10,AE12,AE16} software package.  Similar column labels were previously published by the CCCP \citep{Broos11}, MOXC1, MOXC2, and MOXC3.
\AEacro\ and its User's Guide are publicly available from the Astrophysics Source Code Library, from Zenodo, and at \href{http://personal.psu.edu/psb6/TARA/ae_users_guide.html}{\url{http://personal.psu.edu/psb6/TARA/ae_users_guide.html}}. 
}

\tablenotetext{a}{Source ``labels'' identify each source during data analysis, as the source position (and thus the Name) is subject to change.}

\tablenotetext{b}{\ACIS\ ObsIDs are shifted to align with our astrometric reference catalog, composed of 20 stars chosen from the VISTA Magellanic Survey \citep[VMC]{Cioni11} and Two Micron All Sky Survey \citep[2MASS,][]{Skrutskie06}.} 

\tablenotetext{c}{The suffixes ``\_t'', ``\_s'', ``\_h'', and ``\_v'' on source validation {\em p}-values (``ProbNoSrc\_*'') designate the 0.5--7~keV, 0.5--2~keV,  2--7~keV, and 4--7~keV energy bands.} 

\tablenotetext{d}{Due to dithering over inactive portions of the focal plane, a \Chandra source often is not observed during some fraction of the nominal exposure time (\url{http://cxc.harvard.edu/ciao/why/dither.html}).  We report here the \CIAO\ quantity ``FRACEXPO'' produced by the tool {\em mkarf}.}

\tablenotetext{e}{
Source properties in this table are {\em not} corrected for pile-up effects.
RateIn3x3Cell is an estimate of the observed count rate falling on an event detection cell of size 3$\times$3 \ACIS\ pixels, centered on the source position.
When RateIn3x3Cell $>0.05$ (count/frame), the reported source properties may be biased by pile-up effects.
See \S~\ref{sec:pileup} for a list of source extractions confirmed to have significant pile-up.
}

\tablenotetext{f}{Some background events arising from an instrumental effect known as ``afterglow'' (\url{http://cxc.harvard.edu/ciao/why/afterglow.html}) may contaminate source extractions. 
After extraction, we attempt to identify afterglow events using the \AEacro\ tool {\em ae\_afterglow\_report}, and report the fraction of extracted events attributed to afterglow; see the \AEacro\ manual.}

\tablenotetext{g}{Confidence intervals (68\%) for NetCounts quantities are estimated by the \CIAO\ tool {\em aprates} (\url{http://asc.harvard.edu/ciao/ahelp/aprates.html}).}

\tablenotetext{h}{The Ancillary Response File (ARF) in \ACIS\ data analysis represents both the effective area of the Observatory and the fraction of the observation time for which data were actually collected for the source (column ExposureFraction).}

\tablenotetext{i}{MedianEnergy is the median energy of extracted events, corrected for background but dependent on the shape of the Effective Area curves for the extractions of the source\citep[][\S7.3]{Broos10}. 
} 

\tablenotetext{j}{PhotonFlux = (NetCounts / MeanEffectiveArea / ExposureTimeNominal) \citep[][\S7.4]{Broos10}. 
EnergyFlux = $1.602 \times 10^{-9} {\rm (erg/keV)} \times$ PhotonFlux $\times$ MedianEnergy \citep[][\S2.2]{Getman10}. 
Because MeanEffectiveArea depends on the shape of the Effective Area curves for the extractions of the source, PhotonFlux and EnergyFlux exhibit source-dependent calibration errors.
}

\end{deluxetable*}

%% file: plasma_properties.tex

\begin{deluxetable}{@{\hspace{2em}}lccccc@{}}
\centering  \tabletypesize{\footnotesize} \tablewidth{0pt}

\tablecaption{Physical Properties of the Diffuse Plasma Components \label{tbl:physics}}

\tablehead{
\colhead{Parameter} & \colhead{~Scale factor~~} & 
\multicolumn{3}{c}{T-ReX ``Global'' Region}\\[-7pt]
\colhead{}{\hrulefill} & \colhead{}{\hrulefill} &
\multicolumn{3}{c}{\hrulefill}\\[-7pt]
\colhead{} & \colhead{} &
\colhead{vpshock1} & \colhead{vpshock2} & \colhead{vpshock3}
}
\startdata
\multicolumn{5}{l}{Observed X-ray properties} \\
\hline
median $kT_x$ (keV)      & \nodata     &  0.28                &  0.29                & 0.94                 \\
$L_{tc}$ (erg~s$^{-1}$)  & \nodata     & $6.2 \times 10^{36}$ &$12.9 \times 10^{36}$ & $2.2 \times 10^{36}$ \\
$V_x$ (cm$^3$)           & $\eta$      & $5.5 \times 10^{61}$ & $5.5 \times 10^{61}$ & $5.5 \times 10^{61}$ \\
                         &             &                      &                      &                      \\
\hline
\multicolumn{5}{l}{Derived X-ray plasma properties} \\
\hline
$T_x$ (K)                & \nodata     & $3.2 \times 10^6$    & $3.4 \times 10^6$    &$10.9 \times 10^6$    \\
$n_{e,x}$ (cm$^{-3}$)    &$\eta^{-1/2}$&  0.05                &  0.07                &  0.03                \\
$P_x/k$ (K~cm$^{-3}$)  	 &$\eta^{-1/2}$& $3.4 \times 10^5$    & $5.1 \times 10^5$    & $6.8 \times 10^5$    \\
$E_x$ (erg)              &$\eta^{1/2}$ & $3.9 \times 10^{51}$ & $5.8 \times 10^{51}$ & $7.7 \times 10^{51}$ \\
$\tau_{cool}$ (Myr)      & $\eta^{1/2}$& 20                   & 14                   & 111                  \\
$M_x$ (M$_\odot$)        & $\eta^{1/2}$& 1358                 & 1964                 &  809                 \\
\enddata

\tablecomments{Equations detailing how the derived properties were obtained from the observed properties are given in \citet{Townsley03}.  The volume is assumed to be the area of the ``Global'' diffuse region multiplied by a depth of 50~pc.  The quantity $\eta$ is a ``filling factor,'' $0 < \eta < 1$, accounting for partial filling of the ``Global'' volume with the X-ray-emitting plasmas.  The parameters in the table should be multiplied by the appropriate scale factor (Column 2) to account for this filling factor.  Derived plasma properties are proportional to $\eta^{1/2}$ and are thus only weakly sensitive to this correction.}

\end{deluxetable}

%% file: massive_matches_stub.tex
\setlength{\tabcolsep}{1mm}

\begin{longrotatetable}
\startlongtable
\begin{deluxetable}{@{}lrrrrllllllcchhhp{1.0in}@{}}
\movetabledown=0.5in
\tablewidth{10in}
\tabletypesize{\tiny}
\tablecolumns{14}
\tablecaption{Candidate and Known Massive Stars with T-ReX Detections \label{tbl:massive_matches}}       
\tablehead{
 \multicolumn{4}{c}{X-ray Source} &
 \multicolumn{4}{c}{VFTS/Doran ID, Adopted Coordinates\tablenotemark{a}} &
 \colhead{} &
 \multicolumn{3}{c}{Other Counterpart Names} &
\\[-5pt]
 \multicolumn{4}{c}{\hrulefill} &
 \multicolumn{4}{c}{\hrulefill} &
 \colhead{} &
 \multicolumn{3}{c}{\hrulefill} &
\\[-8pt]
 \colhead{Label} &     
 \colhead{NetCt} & 
 \colhead{PhotonFlux\_t} & 
 \colhead{$E_{med}$} & 
 \colhead{ID} &
 \colhead{RA} &
 \colhead{DEC} & 
 \colhead{Src} &
 \colhead{Sep.\tablenotemark{b}} & 
 \colhead{Gaia} &
 \colhead{VMC} &
 \colhead{HTTP\tablenotemark{c}} &
 \colhead{See also\tablenotemark{d}} &
 \nocolhead{Other} &  
 \nocolhead{Type} &
 \nocolhead{Ref.} &
 \colhead{Notes}
\\[-12pt]
 \colhead{} &
 \colhead{ct} &
 \colhead{ph~cm$^{-2}$~s$^{-1}$} &
 \colhead{keV} &
 \colhead{} &
 \colhead{} &
 \colhead{} &
 \colhead{} &
 \colhead{\arcsec} &
 \colhead{\#46576} &
 \colhead{\#55834} &
 \colhead{} &
 \colhead{} &
 \colhead{} 
}
\startdata
\multicolumn{12}{l}{\bf\footnotesize Declared Counterparts in VFTS Catalog}\\
p1\_610 &22077& 4.93E-5& 3.2& 399        & 05:38:33.413 & -69:11:59.02 &G  & 0.13 &  77456003693440 & 9048274 &053833.417-691159.00 &C22 C15       & See C15. & O9IIIn & Walborn & X-ray binary\\ 
1\_698  &7988& 1.69E-5& 1.6& 482        & 05:38:40.221 & -69:05:59.91 &G  & 0.03 &  79659312960256 &         &053840.214-690559.86 &C22 T19       & Brey 78, Mk 39, P93-0767, S99-014. See T19. & O2.5If*/WN6 & CW11 & \\ 
p1\_830  &1680& 4.19E-6& 1.4& 527        & 05:38:42.359 & -69:04:58.19 &G  & 0.06 &  85564856502528 & 9019909 &053842.351-690458.19 &C22           & R139, Brey 86, P93-0952, S99-002 , BAT99-107 & O6.5Iafc+O6Iaf & T11 & \\ 
p1\_893  &6955& 1.69E-5& 1.8&1025        & 05:38:42.903 & -69:06:04.96 &G  & 0.02 &  85534828237440 &         &\nodata              &C22 T19       & R136c; Blend of S99 027 \& 063. See T19. & WN5h & CD98 & \\ 
p1\_1088 & 693& 1.74E-6& 1.4& 617        & 05:38:47.522 & -69:00:25.28 &G  & 0.23 &  87733851287040 & 8944472 &053847.533-690025.23 &C22 T19       & HDE 269926, R146, Sk-69 245, Brey 88, BAT99-117, P93-9033. See T19. & WN5ha & Evans & \\ 
p1\_1256 & 784& 1.87E-6& 2.0& 695        & 05:38:57.072 & -69:06:05.65 &G  & 0.06 &  85500468567040 & 9018590 &053857.072-690605.58 &C22 T19       & HDE 269928, R145, Sk-69 248, Brey 90, P93-1788, BAT99-119. See T19. & WN6h+? & Evans & \\ 
\hline
\hline
\multicolumn{12}{l}{\bf\footnotesize Declared Counterparts in Doran13 Catalog}\\
p1\_296 &1174 & 2.80E-6& 1.9& 185  & 05:37:49.039 & -69:05:08.21 &G   & 0.12 & 92338057418624 & 8986916 &053749.028-690508.33&C22 T19       & RMC 130. See T19.& WN/C+B1I & BAT99 & \\ 
p1\_812 & 933 & 2.02E-6& 1.5& 580  & 05:38:42.112 & -69:05:55.33 &G   & 0.06 & 85534828234496 &         &053842.104-690555.29&C22           & Mk 42            & O2If* & CW11 & 3 other Doran srcs (573, 577, 587) in ACIS aperture, but this is definitely the best match\\ 
p1\_867 & 522 & 1.61E-6& 1.4& 682  & 05:38:42.658 & -69:06:02.91 & D+ & 0.02 &                &         &\nodata             &C22           &                  & O3If* & MH98 & Doran682 uses dM position\\ 
p1\_979 &2180 & 4.70E-6& 1.4& 765  & 05:38:44.129 & -69:05:56.71 &G   & 0.03 & 85534828260608 &         &053844.123-690556.63&C22           & Mk 33 Sa         & O3III(f*) & MH98 & \\ 
p1\_995 &74445& 1.48E-4& 1.8& 770  & 05:38:44.256 & -69:06:06.00 &G   & 0.04 & 85534828257792 & 9020003 &053844.252-690605.93&C22 P18       & MK 34. See P18.  & WN5h & CD98 & ACIS pile-up \\ 
p1\_1000&1865 & 4.09E-6& 1.5& 775  & 05:38:44.334 & -69:05:54.72 &G   & 0.02 & 85534828252672 &         &053844.329-690554.66&C22           & Mk 33 Na         & O3If* & MH98 & \\ 
p1\_1194&1233 & 2.90E-6& 1.8& 916  & 05:38:53.383 & -69:02:00.88 &G   & 0.1  & 86222022977920 & 8959289 &053853.363-690200.88&C22 T19       & RMC 144. See T19.& WN6h & BAT99 & \\ 
  \enddata
\tablecomments{
Sources with ${>}500$ net counts are displayed above; the full table is available in the online journal. Reported properties of ACIS sources (NetCounts\_t, PhotonFlux\_t, MedianEnergy\_t from Table~\ref{tbl:xray_properties}) are derived by combining all observations.  In some cases the source detection is most significant in a single observation.
NetCounts suggests the quality of the extracted spectrum, but is unsuitable for photometry.  Apparent Photon Flux (NetCounts / MeanEffectiveArea / ExposureTimeNominal) is a rough scaling of NetCounts to account for source-to-source variations in exposure time, effective area, and aperture size \citep[][\S7.4]{Broos10}.
}
\tablenotetext{a}{
The coordinates published in the VFTS and Doran catalogs were not directly used to assess positional coincidence between X-ray sources and candidate/known massive stars.
When the star could be identified in the Gaia catalog (Src ``G''), Gaia coordinates (propagated to the mean epoch of the \Chandra\ data, Y2014.5) and uncertainties were adopted.
Otherwise, when the star could be identified in the VMC catalog (Src ``V''), VMC coordinates were adopted and uncertainties were estimated by the authors.
Otherwise, pubished VFTS coordinates (``V+ '') were corrected using a field distortion map derived using Gaia and uncertainties were estimated by the authors.
Otherwise, published Doran coordinates (``D+ '') were corrected using frame shifts derived using Gaia and uncertainties were estimated by the authors.\\
}
\tablenotetext{b}{
Separation between ACIS source and star's Adopted Coordinates; median is 0.15 arcsec.
}
\tablenotetext{c}{
Source label in the HTTP catalog.  Due to space limitations, our corrected HTTP positions are not reported.
}
\tablenotetext{d}{
Modeling of the T-ReX spectrum is described by 
\citet[C22]{Crowther22},
\citet[T19]{Tehrani19},
\citet[P18]{Pollock18},
\citet[C15]{Clark15}.
}

\end{deluxetable}
\end{longrotatetable}

%% file: massive_blended.tex
%
%
\setlength{\tabcolsep}{1mm}

\begin{deluxetable}{@{}lrcclllchhhp{1.0in}@{}}
\movetabledown=0.5in
\tablewidth{10in}
\tabletypesize{\tiny}
\tablecolumns{14}
\tablecaption{Candidate and Known Massive Stars Associated with ACIS Source Blends In R136 \label{tbl:massive_blended}}       
\tablehead{
 \multicolumn{4}{c}{X-ray Source} &
 \multicolumn{2}{c}{VFTS/Doran} &
 \colhead{} &
\\[-5pt]
 \multicolumn{4}{c}{\hrulefill} &
 \multicolumn{2}{c}{\hrulefill} &
 \colhead{} &
\\[-8pt]
 \colhead{Label} &     
 \colhead{NetCt} & 
 \colhead{PhotonFlux\_t} & 
 \colhead{$E_{med}$} & 
 \colhead{ID} &
 \colhead{Src} &
 \colhead{Sep.} & 
\colhead{See also} &
 \nocolhead{Other} &  
 \nocolhead{Type} &
 \nocolhead{Ref.} &
 \colhead{Notes}
\\[-12pt]
 \colhead{} &
 \colhead{ct} &
 \colhead{ph~cm$^{-2}$~s$^{-1}$} &
 \colhead{keV} &
 \colhead{} &
 \colhead{} &
 \colhead{} &
 \colhead{} &
 \colhead{} &
 \colhead{} &
 \colhead{} 
}
\startdata
  p1\_987 &   84 &1.96E-7&1.2& V1034\tnm{a}  &G  & 0.2                    &C22       &Mk 32, S99-021& O8III(f)  & MH98 & Both could contribute to ACIS light\\ 
          &      &       &   &    D772      &D+ & 0.3                     &          &              & 06V       & MH98 & \\
\hline
  p1\_866 &   62 &1.50E-7&1.4&     D685     &G  & 0.25                    &C22       &              &\nodata    && Both could contribute to ACIS light\\ 
          &      &       &   &     D671     &D+ & 0.5                     &          &              &\nodata    && \\
\hline
  p1\_832 & 4452 &1.15E-5&1.5&     D633     &D+ &                         &C22 T19   &R136a1/a2/a4/a5/a7/a8. See T19.& WN5h      & CD98 & Mushroom aperture at ACIS aimpoint; hopelessly confused at R136 core\\ 
          &      &       &   &     D630     &D+ &                         &          &              & WN5h      & CD98 & \\
          &      &       &   &     D642     &D+ &                         &          &              & O2If*     & CW11 & \\
          &      &       &   &     D638     &D+ &                         &          &              & O3III(f*) & MH98 & \\                                 
\hline
  p1\_766 &   31 &9.70E-8&1.6& V513\tnm{b}   &G  & 0.2                     &C22       & S99-266      & O6-7II(f) &Walborn& Both could contribute to ACIS light\\ 
          &      &       &   &       V515   &V+ & 0.4                     &          & S99-434      &\nodata    && \\
\hline
  cc4970  &  496 &2.39E-6&1.7&    D613      &D+ &                         &C22 T19   & R136a3/a6. See T19.    & WN5h      & CD98 & $\sim$8 Doran srcs in ACIS aperture\\ 
          &      &       &   &    D621      &D+ &                         &          &              &\nodata    && \\
\hline
  c7182   &  258 &6.18E-7&1.4&    D657      &D+ &                         &C22       &              & O3V       & MH98 & Many Doran sources in ACIS aperture\\ 
          &      &       &   &    D652      &D+ &                         &          &              &\nodata    && \\
          &      &       &   &    D649      &D+ &                         &          &              &\nodata    && \\
          &      &       &   &    D659      &D+ &                         &          &              &\nodata    && \\
\hline
  c7157   &  345 &8.16E-7&1.3&    D648      &D+ &                         &C22       &              & O3V       & MH98 & Many Doran sources in ACIS aperture\\ 
          &      &       &   &    D624      &D+ &                         &          &              & O3V       & MH98 & \\
\hline
  c7018   &   45 &2.20E-7&1.5&    D589      &D+ &                         &C22       &              &\nodata    && Many Doran sources in ACIS aperture\\ 
          &      &       &   &    D583      &D+ &                         &          &              &\nodata    && 
\enddata
\tablecomments{See footnotes to Table~\ref{tbl:massive_matches}.}
\tablenotetext{a}{
We suspect that the identifiers VFTS1034 and Doran769 refer to the same star.
}
\tablenotetext{b}{
We suspect that the identifiers VFTS513 and Doran558 refer to the same star.
}
\end{deluxetable}

%% file: other_counterparts_stub.tex
\setlength{\tabcolsep}{1mm}

\begin{longrotatetable}
\startlongtable
\begin{deluxetable}{@{}lcccccccrccccp{2.0in}@{}}
\movetabledown=0.5in
\tablewidth{10in}
\tabletypesize{\tiny}
\tablecolumns{14}
\tablecaption{Counterparts To X-ray Sources Beyond the Massive Stellar Population \label{tbl:other_matches}}       
\tablehead{
 \multicolumn{2}{c}{X-ray Source} &
 \multicolumn{4}{c}{Gaia+VMC} &
 \multicolumn{4}{c}{HTTP\tablenotemark{a}} &
 \multicolumn{3}{c}{SAGE} &
\\[-5pt]
 \multicolumn{2}{c}{\hrulefill} &
 \multicolumn{4}{c}{\hrulefill} &
 \multicolumn{4}{c}{\hrulefill} &
 \multicolumn{3}{c}{\hrulefill}
\\[-8pt]
 \colhead{Label} &
 \colhead{PosErr} &
 \colhead{Dist.\tablenotemark{b}} &
 \colhead{Gaia ID} &
 \colhead{VMC ID} &
 \colhead{K} &
 \colhead{Dist.\tablenotemark{b}} &
 \colhead{ID} &
 \colhead{Shift} &
 \colhead{F555} &
 \colhead{Dist.\tablenotemark{b}} &
 \colhead{SAGE ID} &
 \colhead{3.6\micron} &
 \colhead{SIMBAD\tablenotemark{c}}   
\\[-7pt]
 \colhead{} &
 \colhead{\arcsec} &
 \colhead{\arcsec} &
 \colhead{\#4657} &
 \colhead{\#55834} &
 \colhead{mag} &
 \colhead{\arcsec} &
 \colhead{} &
 \colhead{mas} &
 \colhead{mag} &
 \colhead{\arcsec} &
 \colhead{} &
 \colhead{mag} 
}
\startdata
  p1\_255\tnm{d}     &     &     &                  &         &     &     &                     &              &     &     &                       &      & Pulsar (PSR~J0537-6910) \\
  pass52\_4\tnm{d}   &     & 0.47& 783116463742336  & 8880341 & 10  &     &                       &            &     &     &                       &      & Rotationally variable Star (HD~269921) \\
p1\_1548     & 0.2 & 0.4 &                  & 9712401 & 19  & 0.2 &  053955.602-690846.62 &   130,688  & 25  &     &                       &      &  \\
p1\_1529     & 0.2 & 0.1 &  684980764259456 & 8944793 & 16  &     &                       &            &     & 0.2 &   J053939.44-690029.7 & 14   &  AGN ([GC2009] J053939.48-690030.0)\\
 c11125       & 0.3 & 0.2 &                  & 8919503 & 15  &     &                       &            &     & 0.1 &   J053938.25-685740.2 & 14   &  YSO (2MASS J05393821-6857404)\\
 p1\_1521     & 0.1 & 0.2 &  682923488239360 & 9014587 & 13  & 0.3 &  053935.478-690438.90 &   204,381  & 17  & 0.4 &   J053935.50-690438.6 & 12   &  \\
 p1\_1517     & 0.2 & 0.3 &                  & 9778551 & 20  &     &                       &            &     & 0.5 &   J053933.97-691130.4 & 16   &  \\
 c10740       & 0.3 & 0.5 &                  & 9713682 & 18  & 0.2 &  053925.779-691031.35 &   -28,-54  & 23  &     &                       &      &  \\
 c10422       & 0.2 & 0.2 &                  & 9768044 & 18  &     &                       &            &     & 0.2 &   J053915.91-685752.2 & 17   &  \\
 p1\_1420     & 0.2 & 0.3 &                  & 8939594 & 16  &     &                       &            &     & 0.6 &   J053915.91-685950.0 & 15   &  \\
 c10393       & 0.2 & 0.1 &  679281354744320 & 9023693 & 14  & 0.1 &  053914.961-690923.81 &   -20,-139 & 19  & 0.1 &   J053914.96-690923.7 & 14   &  Long-period variable star (2MASS J05391496-6909238) \\
 c10370       & 0.2 & 0.1 &                  & 9767609 & 19  &     &                       &            &     & 0.1 &   J053913.96-685728.6 & 17   &  \\
 c10266       & 0.1 & 0.2 &  679384433817472 & 9014737 &     & 0.2 &  053910.764-690756.78 &   10,-129  & 18  & 0.4 &   J053910.79-690756.7 & 15   &  \\
p1\_1368     & 0.1 & 0.1 &                  & 9706871 & 18  &     &                       &            &     & 0.1 &   J053908.45-685958.6 & 16   &  \\
p1\_1317     & 0.0 & 0.0 &                  & 9774684 & 21  & 0.0 &  053903.176-690703.87 &   24,-125  & 24  &     &                       &      &  \\
c9590        & 0.2 & 0.3 &                  & 8938310 & 18  &     &                       &            &     & 0.4 &   J053859.69-685949.3 & 17   &  \\
 c9506        & 0.4 & 0.3 &                  & 8849680 & 17  &     &                       &            &     & 0.3 &   J053859.31-684959.5 & 15   &  \\
p1\_1246     & 0.2 & 0.4 &  688416745861760 & 8912946 & 17  &     &                       &            &     & 0.3 &   J053856.83-685636.3 & 15   &  \\
 c9209        & 0.1 & 0.1 &                  & 8978864 &     & 0.1 &  053856.710-690426.29 &   42,-84   & 22  &     &                       &      &  \\
 c8790        & 0.1 & 0.2 &                  & 9707745 & 18  &     &                       &            &     & 0.2 &   J053852.25-690126.9 & 17   &  \\
cc5713       & 0.1 &     &                  &         &     & 0.4 &  053848.720-690715.56 &   27,-94   & 23  & 0.4 &   J053848.73-690715.6 & 15   &  \\
 c8315        & 0.1 & 0.3 &                  & 9770921 & 18  &     &                       &            &     & 0.4 &   J053848.05-690210.8 & 16   &  \\
 c8216        & 0.1 &     &                  &         &     & 0.2 &  053846.871-690505.58 &   38,-73   & 24  & 0.2 &   J053846.84-690505.4 & 10   &  \\
 p1\_1056     & 0.2 & 0.2 &                  & 8921036 & 18  &     &                       &            &     & 0.5 &   J053845.63-685759.7 &      &  \\
 c7568        & 0.1 & 0.1 &                  & 8996929 & 17  & 0.2 &  053843.546-690628.82 &   33,-69   & 18  & 0.4 &   J053843.50-690629.0 & 14   &  YSO (OGLE LMC-ECL-21448)\\
 c6236        & 0.1 & 0.2 &                  & 9774528 & 17  & 0.2 &  053839.963-690652.52 &   33,-54   & 26  &     &                       &      &  \\
 c6150        & 0.1 & 0.0 &  685534828103552 &         &     & 0.0 &  053839.633-690532.25 &   39,-48   & 17  &     &                       &      &  \\
  c5995        & 0.1 & 0.2 &  679659312977664 &         &     & 0.2 &  053838.549-690612.85 &   35,-45   & 17  &     &                       &      &  \\
   c5836        & 0.3 & 0.5 &                  & 9705089 & 17  &     &                       &            &     & 0.7 &   J053837.20-685655.0 & 16   &  \\
 p1\_641      & 0.2 & 0.2 &                  & 9767536 & 19  &     &                       &            &     & 0.3 &   J053836.65-685725.9 & 18   &  \\
p1\_639      & 0.1 & 0.2 &                  & 9038525 &     & 0.2 &  053836.577-691056.66 &   -14,-38  & 22  &     &                       &      &  \\
p1\_609      & 0.3 & 0.7 &                  & 8923829 & 18  &     &                       &            &     & 0.3 &   J053833.49-685817.9 & 17   &  \\
 c5204        & 0.2 & 0.0 &                  & 8968459 &     & 0.1 &  053832.322-690314.89 &   14,-9    & 22  &     &                       &      &  \\
 p1\_574      & 0.1 & 0.2 &                  & 9707046 & 18  &     &                       &            &     & 0.4 &   J053826.21-690012.9 & 16   &  \\
 c4405        & 0.4 & 0.7 &  783490104306304 & 8855976 & 10  &     &                       &            &     & 0.9 &   J053821.25-685034.0 & 10   &  \\
  c3916        & 0.2 & 0.3 &  686698750750848 & 8976411 &     & 0.3 &  053810.121-690224.70 &   14,80    & 19  &     &                       &      &  \\
 p1\_464      & 0.2 & 0.3 &  688863422482304 & 8912592 & 16  &     &                       &            &     & 0.2 &   J053809.91-685657.7 & 15   &  \\
 c3775        & 0.2 &     &                  &         &     & 0.6 &  053806.273-691310.00 &   -37,46   & 24  & 0.8 &   J053806.27-691310.1 & 15   &  \\
  c3766        & 0.2 & 0.2 &  679968549406848 & 9021508 & 13  & 0.2 &  053805.644-690909.77 &   13,70    & 19  & 0.3 &   J053805.61-690909.6 & 11   &  YSO (2MASS6X J05380564-6909098)\\
 p1\_450      & 0.0 & 0.1 &                  & 9708672 & 19  & 0.1 &  053802.741-690259.15 &   24,105   & 22  & 0.1 &   J053802.73-690258.9 & 16   &  \\
 c3526        & 0.2 & 0.6 &                  & 8872736 & 17  &     &                       &            &     & 0.7 &   J053757.84-685238.2 & 16   &  \\
p1\_410      & 0.1 & 0.3 &                  & 9710167 & 16  & 0.2 &  053754.561-690516.57 &   23,107   & 26  &     &                       &      &  \\
p1\_406      & 0.1 & 0.1 &                  & 9771021 & 20  &     &                       &            &     & 0.0 &   J053754.32-690222.1 & 17   &  \\
c2202        & 0.5 & 0.5 &  701198560089088 & 8877930 & 17  &     &                       &            &     & 0.5 &   J053745.77-685311.3 & 15   & AGN (MQS J053745.77-685311.4) \\
 c1348        & 0.5 & 0.6 &                  & 9705614 & 16  &     &                       &            &     & 0.5 &   J053738.09-685753.1 & 15   &  \\
  c1167        & 0.2 & 0.5 &  692024485433088 & 8982580 & 16  &     &                       &            &     & 0.2 &   J053729.61-690451.8 & 16   &  \\
 p1\_11       & 0.2 & 0.2 &                  & 9711726 & 17  & 0.2 &  053725.617-690738.37 &   -12,29   & 25  & 0.4 &   J053725.55-690738.2 & 16   &  \\
c1042        & 0.3 &     &                  &         &     & 0.3 &  053721.382-690621.28 &   -13,11   & 26  & 0.6 &   J053721.45-690621.4 & 17   &  \\
c979         & 1.1 & 1.9 &  663849561921024 & 9131446 &     &     &                       &            &     & 1.5 &   J053714.70-692021.0 & 17   &  \\
 c919         & 0.3 & 0.2 &  690001595436160 & 9033573 & 15  & 0.2 &  053710.840-691025.19 &   49,51    & 18  & 0.4 &   J053710.87-691025.1 & 14   &  \\
p1\_5        & 0.3 & 0.4 &  690070295111040 & 9025732 & 18  &     &                       &            &     & 0.5 &   J053703.80-690942.6 & 17   &  \\
c749         & 0.4 & 0.7 &  693433288582784 & 8962789 & 16  &     &                       &            &     & 0.6 &   J053658.57-690234.2 & 16   &  \\
pass53\_1    & 0.2 & 0.3 &  666323458738560 & 9061522 & 15  & 0.3 &  053657.042-691328.57 &   309,290  & 21  & 0.1 &   J053657.12-691328.2 & 14   &  \\
c373         & 0.9 & 0.8 &  702989558190336 & 8894623 & 14  &     &                       &            &     & 0.8 &   J053612.44-685455.9 & 14   &  \\
 c73          & 0.2 & 0.5 &                  & 9090770 & 15  &     &                       &            &     &     &                       &      &  SNR 1987A\\
\enddata
\tablecomments{A portion of this table, highlighting sources with counterparts in more than one optical/IR catalog, is provided here to illustrate format and content. The full table is available in the online journal.}
\tablenotetext{a}{
Column ``ID'' is the published label for a detection in the HTTP catalog.  Column ``Correction'' reports the shift we applied to the published position.
}
\tablenotetext{b}{
Separation between ACIS source and Adopted Coordinates.
}
\tablenotetext{c}{
Identifiers reported by SIMBAD.
}
\tablenotetext{d}{
Using our judgement, we assert two associations that were not found by automated methods:
Pulsar PSR~J0537-6910 == T-Rex 053747.40-691020.2 (p1\_255);
the low-mass star HD~269921 == T-ReX  053834.66-685306.1 (pass52\_4).
}
%
\end{deluxetable}
\end{longrotatetable}